\documentclass[a4paper,11pt]{article}

\usepackage{classeJBArxive}

\usepackage{arydshln}
\usepackage{algorithm}
\usepackage[noend]{algpseudocode}

\newcommand*\unit[1]{\bigl[\, \mathsf{#1} \,\bigr]}

\newcommand{\apr}{\mathrm{apr}}
\newcommand{\Fo}{\mathrm{Fo}}

\newcommand{\Tinf}{T_{\,\infty}}
\newcommand{\Tref}{T_{\,\reff}}
\newcommand{\tref}{t_{\,\reff}}
\newcommand{\tini}{t_{\,\mathrm{ini}}}

\newcommand{\apriori}{\mathrm{apr}}

\newcommand{\fin}{\mathrm{f}}
\newcommand{\ini}{\mathrm{ini}}

\newcommand{\oed}{\mathrm{oed}}

\newcommand{\reff}{\mathrm{ref}}

\newcommand{\mA}{\boldsymbol{A}}
\newcommand{\mB}{\boldsymbol{B}}
\newcommand{\mC}{\boldsymbol{C}}
\newcommand{\mF}{\boldsymbol{F}}
\newcommand{\mG}{\boldsymbol{G}}
\newcommand{\mH}{\boldsymbol{H}}
\newcommand{\mM}{\boldsymbol{M}}
\newcommand{\mQ }{\boldsymbol{Q}}
\newcommand{\mU}{\boldsymbol{U}}
\newcommand{\mX}{\boldsymbol{X}}
\newcommand{\mY}{\boldsymbol{Y}}
\newcommand{\mTheta}{\boldsymbol{\Theta}}

\newcommand{\RK}{\textsc{Runge}--\textsc{Kutta}}

\title{Estimation of the thermal properties of an historic building wall by combining Modal Identification Method and Optimal Experiment Design
}

\author{Julien Berger\textsuperscript{a}$^{\ast}$, Benjamin Kadoch\textsuperscript{b}\\
\date{\today\vspace{-0.5cm}}}

\begin{document}

\maketitle

\begin{center}
\small
\textsuperscript{a}  Laboratoire des Sciences de l’Ingénieur pour l’Environnement (LaSIE), UMR 7356 CNRS, La Rochelle Université, CNRS, 17000, La Rochelle, France \\
\textsuperscript{b} Aix Marseille Université, CNRS, IUSTI UMR 7343, 13453 Marseille, France\\
$^{\ast}$corresponding author, e-mail address : julien.berger@univ-lr.fr\\
\end{center}


\begin{abstract}

The estimation of wall thermal properties by \emph{in situ} measurement enables to increase the reliability of the model predictions for building energy efficiency. Nevertheless, retrieving the unknown parameters has an important computational cost. 
Indeed, several computations of the heat transfer problem are required to identify these thermal properties. To handle this drawback, an innovative approach is investigated. The first step is to search the optimal experiment design among the sequence of observation of several months. A reduced sequence of observations of three days is identified which guarantees to estimate the parameter with the maximum accuracy. Moreover, the inverse problem is only solved for this short sequence. To decrease further the computational efforts, a reduced order model based on the modal identification method is employed. This \emph{a posteriori} model reduction method approximates the solution with a lower degree of freedom. The whole methodology is illustrated to estimate the thermal diffusivity of an historical building that has been monitored with temperature sensors for several months. The computational efforts is cut by five. The estimated parameter improves the reliability of the predictions of the wall thermal efficiency.



\end{abstract}

\section{Introduction}

Within the environmental context, requirements on building energy efficiency becomes more and more important. In France, the building stock increases slowly \cite{Stephan_2014} with an average around $1\%\,$. Thus, there is a crucial issue in retrofitting existing building. To this end, models such as building simulation programs are employed to define the best  actions. However, there are some uncertainties on the input parameters of the model \cite{Cornaro_2016,Evangelisti_2015}. \revision{Particularly, for such old buildings, the material properties may differ strongly from one site to another \cite{Hamard_2018}. Indeed, due to the vernacular traditional architecture, the material were generally taken from locals around the building. Furthermore, the aging of materials may have changed the properties \cite{Evangelisti_2018}.} Parameter such as the thermal properties of the walls require to be known precisely since they play a crucial role on the assessment of the building energy efficiency \cite{Jumabekova_2019,Busser_2019}.

The unknown properties can be estimated by solving inverse problems or more precisely parameter estimation problem \cite{Berger_2016b} to calibrate the model. This procedure aims at minimizing a so-called cost function between the model numerical predictions and the experimental observations obtained by on-site measurements. \revision{Such approaches have been successfully applied for real walls and a detailed state-of-the-art is proposed in \cite{Rasooli_2019}. However, the determination of thermo-physical properties using \emph{in situ} dynamic measurements faces important challenge in terms of computational effort. This challenge arises from two reasons. The first one is the duration of the experimental observations. If long measurement duration are better for the accuracy of the retrieved parameter, it implies important computational costs to solve the inverse problem. Note that \emph{in situ} measurement depends also on occupants' acceptance and long period can be constraining. The standard techniques ISO 9869 requires more than a month of measurement periods \cite{Rasooli_2018}. Thus, various alternative techniques have been developed to reduced the monitoring time. In \cite{Suklje_2019} an experiment of $7 \ \mathsf{days}$ is carried. In \cite{Rodler_2019}, the accuracy of the parameter estimation is investigated according to the measurement period. For their case study, the authors suggest to carry at least three days of monitoring  to have an accurate estimation. In \cite{Rasooli_2019}, a fast estimation method is proposed based excitation pulse method and less than $1 \ \mathsf{h}$ of measurement. }

\revision{The second reason is the mathematical model to represent the physical phenomena. Indeed, the computational efforts required to solve the inverse problem are directly proportional to the so-called direct model (as well as the optimization algorithm). Thus, numbers of investigations aimed at developing direct model with reduced computational cost. Lumped model such as RC approaches have been proposed due to their very small computational cost. However, their reliability to estimate accurately the thermal diffusivity is questioned in \cite{Berger_2019}. The model has to be based on detailed heat diffusion process. In \cite{Gori_2018} an equivalent homogeneous model is proposed for the description of multi-layer walls with the issue of reducing the cost of parameter estimation as shown in \cite{Evangelisti_2018}. When possible, analytical solution are also developed as for instance in \cite{Simoes_2012}.}

\revision{Thus, the objective of this article is to answer both issues in order to propose a fast and accurate parameter estimation method. First, the length of the measurement observations is determined with the optimal experiment design methodology \cite{Ucinski,Beck_1977}. The optimal design is searched according to the conditions of the experiment. Carrying the measurement for the optimal experiment design ensure to estimate the unknown parameter with the highest accuracy. This approach has been used in \cite{Berger_2017,Berger_2018} to determine the optimal boundary conditions of laboratory experiments to retrieve material properties of heat and mass transfer. In \cite{Dalessandro_2019}, an optimal design is found to retrieve the thermal conductivity and volumetric heat capacity of a material using laboratory experiments. This approach can be extended for \emph{in-situ} measurement to determine the optimum sequence of measurement observations.}

\revision{Then, to decrease the computational effort of the direct numerical model, model reduction techniques can be employed.  Several methods are reported in the literature to model the physical phenomena in building walls \cite{Berger_2016}.} Among all, the Modal Identification Method (MIM) demonstrated successful applications for inverse problem. The primer works of \textsc{Girault}, \textsc{Petit}, and \textsc{Videcoq} introduced the MIM approach for direct simulations in \cite{Petit_1997,Girault_2003}. Then, it has been applied for the identification of boundary fluxes through sequential procedure in \cite{Girault_2004a,Girault2005a,Girault2005b} or surface temperature as in \cite{Gerardin_2017}. To our best knowledge, it has never been employed for the estimation of thermal diffusivity in building walls under climatic varying conditions. \revision{The MIM ROM is employed since in the state-space representation of the MIM model, both the field and its sensitivity to the unknown parameter are computed with a reduced computational cost. It can be combined with a gradient-based algorithm to retrieve the thermal properties of the wall.}

This article presents and evaluate the proposed methodology to estimate the thermal diffusivity of a wall monitored in an old building. First the mathematical model is described in Section~\ref{sec:math_model}. The MIM model reduction method is detailed in Section~\ref{sec:numerical_model}. Then, the search optimal experiment design combined with the solution of the parameter estimation problem is presented in Section~\ref{sec:pep}. A brief synthesis of the whole methodology and the metrics to evaluate its efficiency are given in Section~\ref{sec:synthesis_methodology}. Subsequently, a first case study is introduced to validate the MIM method for estimating the thermal diffusivity of a wall. Last, a real case study is investigated in Section~\ref{sec:real_case}.

\section{Mathematical model}
\label{sec:math_model}

\revision{First, the mathematical model is presented. It is based on two equations. The first corresponds to the diffusion one occurring in the wall. In addition, the sensitivity equation is also described. It enables to compute the sensitivity coefficient which qualifies the sensitivity of the field of temperature relative to change in the unknown diffusivity. This is required in the algorithm to solve the inverse problem.} 

\subsection{Physical formulation}

The field of interest is the temperature $T$ evolving in a building wall material \revision{according to a diffusion process}. The space domain is defined by $x \, \in \bigl[\, 0 \,\,\, L \,\bigr]\,$, where $L$ is the length of the wall. The time domain is defined by $t \, \in \, \Omega_{\,t}\,$ with $\Omega_{\,t} \, \eqdef \, \bigl[\, 0 \,,\, t_{\,\fin} \,\bigr] \,$. The temperature is computed using the heat diffusion equation \cite{Ozisik_2000}:
\begin{align}
\label{eq:heat_diffusion}
\pd{T}{t} \egal \alpha \cdot \pd{^{\,2} T}{x^{\,2}} \,,
\end{align}
where $\alpha \, \eqdef \, \displaystyle \frac{k}{c}$ is the heat diffusion, $k$ is the heat conductivity and $c$ is the volumetric heat capacity. 
At the interface between ambient air and the wall, \textsc{Dirichlet} type boundary conditions are assumed. \revision{The temperature is prescribed by the one measured:}
\begin{align*}
T \egal T_{\,\infty\,,\,L}\,\bigl(\,t\,\bigr)  \,, \qquad x \egal 0 \,, \\[4pt]
T \egal T_{\,\infty\,,\,R}\,\bigl(\,t\,\bigr)  \,, \qquad x \egal L \,,
\end{align*}
\revision{where  $\Tinf$ are the temperatures measured by sensors and depending on time:}
\begin{align*}
\Tinf \,:\, t \,\longmapsto \, \Tinf\,\bigl(\,t\,\bigr) \,.
\end{align*}
At $t \egal 0\,$, the temperature in the material is known:
\begin{align*}
T \egal T_{\,\ini}\,\bigl(\,x\,\bigr) \,, \qquad t \egal 0 \,.
\end{align*}
\revision{with $T_{\,\ini}$ a function of $x$ corresponding to the initial field in the wall:}
\begin{align*}
T_{\,\ini} \,:\, x \,\longmapsto \, T_{\,\ini}\,\bigl(\,x\,\bigr) \,.
\end{align*}
The so-called observable field $T_{\,s}$ is defined as follows:
\begin{align*}
T_{\,s} \,:\, (\, x_{\,s} \,,\, t \,) \,\longmapsto \, T\,\bigl(\,x_{\,s}\,,\,t\,\bigr) \,, \qquad s \, \in \, \bigl\{\,1 \,, \ldots \,, N_{\,s} \,\bigr\}\,,
\end{align*}
where $N_{\,s}$ is the number of points of interests. In our case, it corresponds to the sensor locations. \revision{Thus, the observable fields correspond to the temperature computed with the model at the sensor locations. It enables to compare the model predictions with the experimental observations.} The direct problem is now transformed into a dimensionless formulation to be solved by the numerical model. 

\subsection{Dimensionless formulation}

As discussed and thoroughly motivated in \cite{Nayfeh_2000,Kahan_1979}, it is of capital importance to obtain a dimensionless problem before elaborating a numerical model. \revision{It allows to define scaling parameters such as \textsc{Fourier} number that may decrease the number of unknown parameters in the inverse problem.} For this, dimensionless fields are defined: 
\begin{align*}
u & \eqdef \, \frac{T \moins T_{\,\min}}{\Tref}  \,,
&& u_{\,\infty\,,\,L} \, \eqdef \, \frac{T_{\,\infty\,,\,L}\moins T_{\,\min}}{\Tref} \,, \\[4pt]
u_{\,\infty\,,\,R} & \eqdef \, \frac{T_{\,\infty\,,\,R}\moins T_{\,\min}}{\Tref} \,, 
&& u_{\,\ini} \, \eqdef \, \frac{T_{\,\ini}\moins T_{\,\min}}{\Tref} \,,
\end{align*}
where $\Tref$ and $T_{\,\min}$ are user--defined reference temperature. The space and time coordinates are also transformed into dimensionless variables:
\begin{align*}
t^{\,\star} & \eqdef \, \frac{t}{\tref} \,,
&& x^{\,\star} \, \eqdef \, \frac{x}{L} \,.
\end{align*}
\revision{The \textsc{Fourier} dimensionless number is defined:}
\begin{align*}
\Fo & \eqdef \, \frac{\tref \cdot \alpha}{L^{\,2}} \,.
\end{align*}
\revision{It corresponds to the thermal diffusivity in the dimensionless representation and quantifies the magnitude of diffusion inside the material. A high number indicates a fast heat transfer process inside the material.} With these transformations, the dimensionless problem is written as:
\begin{align}
\label{eq:heat_diffusion_dimless}
\pd{u}{t^{\,\star}} \egal \Fo \cdot \pd{^{\,2} u}{\,x^{\,\star\,2}} 
\end{align}
with the boundary condition:
\begin{subequations}
\label{eq:bc_dimless} 
\begin{align}
u \egal u_{\,\infty\,,\,L}\,\bigl(\,t\,\bigr)  \,, && x^{\,\star} \egal 0 \\[4pt]
u \egal u_{\,\infty\,,\,R}\,\bigl(\,t\,\bigr) \,, && x^{\,\star} \egal 1 
\end{align}
\end{subequations}
and the initial condition:
\begin{align}
\label{eq:ic_dimless} 
u \egal u_{\,\ini}\,\bigl(\,x\,\bigr)  \,, \qquad t^{\,\star} \egal 0 \,.
\end{align}
To have a well-posed problem, initial and boundary conditions must be compatible. The observable field is also transformed according to $ \displaystyle u_{\,s} \, \eqdef \, \frac{T_{\,s}\moins T_{\,\min}}{\Tref} \,, \quad s \, \in \, \bigl\{\,1 \,, \ldots \,, N_{\,s} \,\bigr\}\,$.

\subsection{Sensitivity equation}

\revision{The issue is to estimate the unknown thermal diffusivity $\alpha$ of the wall or its equivalent in the dimensionless representation, the \textsc{Fourier} number $\Fo\,$. The solution of the inverse problem requires the so-called sensitivity function $\theta\,$:
\begin{align*}
\theta \,:\, (\,\Fo \,,\, x\,,\,t\,) & \longmapsto \, \pd{u}{\Fo} \,.
\end{align*} 
It quantifies the sensitivity of the field of temperature according to the unknown parameter. A small magnitude reveals that large change in the parameter induces small change in the field. Thus, the parameter cannot be estimated with accuracy. Inversely, high magnitude of sensitivity coefficient are favorable conditions to estimate the unknown parameter. One can obtain can compute the sensitivity equations by differentiating Eq.~\eqref{eq:heat_diffusion_dimless} relatively to the parameter:}
\begin{align}
\label{eq:sensitivity_dimless}
\pd{\theta}{t^{\,\star}} \egal \Fo \cdot \pd{^{\,2} \theta}{x^{\,\star\,2}} \plus \pd{^{\,2} u}{x^{\,\star\,2}} \,,
\end{align}
\revision{with the following boundary conditions obtained by differentiating Eq.~\eqref{eq:bc_dimless}:}
\begin{align*}
\theta \egal 0  \,, \qquad x^{\,\star} \egal 0 \,, \\[4pt]
\theta \egal 0  \,, \qquad x^{\,\star} \egal 1 \,,
\end{align*}
\revision{and the initial condition arising from Eq.~\eqref{eq:ic_dimless}:}
\begin{align*}
\theta \egal 0  \,, \qquad t^{\,\star} \egal 0 \,.
\end{align*}
\revision{Particularly, the sensitivity function at the location $x_{\,s}$ of the sensors are needed. It is computed by:}
\begin{align*}
\theta_{\,s} \,:\, (\, x_{\,s} \,,\, t \,) \,\longmapsto \, \theta\,\bigl(\,x_{\,s}\,,\,t\,\bigr) \,, \qquad s \, \in \, \bigl\{\,1 \,, \ldots \,, N_{\,s} \,\bigr\}\,.
\end{align*}

\section{MIM reduced order model of the field and its sensitivity}
\label{sec:numerical_model}

In this Section, the numerical models used to perform simulations are described starting first with the large original model. Then the reduced MIM approach is detailed. A uniform discretisation based on finite differences, is adopted for the space domain, with the discretisation parameter denoted by $\Delta x\,$. The discrete value of $u\,(\,x\,,\,t\,)$ and $\theta \,(\,x\,,\,t\,)$ becomes $u_{\,j} \, \eqdef \, u\,(\,x_{\,j}\,,\,t\,)$ and $\theta_{\,j} \, \eqdef \, \theta\,(\,x_{\,j}\,,\,t\,)\,$, respectively, with $j \, \in \, \bigl\{\, 1\,,\, \ldots \,,\, N \,\bigr\}\,$. Note that other discretisation methds (finite volumes, finite elements, ...) could be used and will give the same MIM formulation.

\subsection{The state space representation of the large original model}

The straightforward semi-discretization of Eqs.~\eqref{eq:heat_diffusion_dimless} and \eqref{eq:sensitivity_dimless} using central finite differences yields for $j \, \in \, \bigl\{\, 2 \,,\, \ldots \,,\, N-1 \,\bigr\}$ to:
\begin{subequations}
\label{eq:semi_discrete_equations}
\begin{align}
\pd{u}{t} & \egal \frac{\Fo}{\Delta x^{\,2}} \cdot \Bigl(\, u_{\,j+1} \moins 2 \cdot u_{\,j} \plus u_{\,j-1} \,\Bigr) \,, \\[4pt]
\pd{\theta}{t} & \egal \frac{\Fo}{\Delta x^{\,2}} \cdot \Bigl(\, \theta_{\,j+1} \moins 2 \cdot \theta_{\,j} \plus \theta_{\,j-1} \,\Bigr) \plus
 \frac{1}{\Delta x^{\,2}} \cdot \Bigl(\, u_{\,j+1} \moins 2 \cdot u_{\,j} \plus u_{\,j-1} \,\Bigr) \,.
\end{align}
\end{subequations}
The problem is governed by \textsc{Dirichlet} boundary conditions for both fields $u$ and $\theta\,$:
\begin{align*}
u_{\,1} & \egal u_{\,\infty\,,\,L}\,(\,t\,) \,,
&& u_{\,N} \egal u_{\,\infty\,,\,R}\,(\,t\,) \,,
&& \theta_{\,1} \egal 0 \,,
&& \theta_{\,N} \egal 0 \,.
\end{align*}
Thus, the semi-discrete equations~\eqref{eq:semi_discrete_equations} can be written in the matrix form:
\begin{align*}
\dot{\mU} & \egal \Fo \cdot \mA \cdot \mU \plus \Fo \cdot \mB \cdot \mQ \,, \\[4pt]
\dot{\mTheta} & \egal \Fo \cdot \mA \cdot \mTheta \plus \mA \cdot \mU \plus \mB \cdot \mQ  \,, 
\end{align*}
where $\mA \, \in \, \mathcal{M}\,\bigl(\, \mathbb{R}^{\,N\times N}\,\bigr)$ and $\mB \, \in \, \mathcal{M}\,\bigl(\, \mathbb{R}^{\,N\times 2}\,\bigr)$ are matrices defined to ensure a second order accuracy in space:
\begin{align*}
\mA \, \eqdef \,
 \frac{1}{\Delta x^{\,2}} \cdot \begin{bmatrix}
-2 & 1 & 0 & \ldots &  0 \\
1 & -2 & 1 & \ddots &   0 \\
0 & \ddots & \ddots & \ddots & 0  \\
0 & \ldots & 1 & -2 & 1 \\
0 & \ldots & 0 & 1 &  -2 
\end{bmatrix} \,, \qquad 
\mB \, \eqdef \, 
\begin{bmatrix}
1 & 0 \\
0 & 0 \\
\vdots & \vdots \\
0 & 0\\
0 & 1 \\
\end{bmatrix} \,, \qquad
\mQ \, \eqdef \, 
\begin{bmatrix} u_{\,\infty\,,\,L} \quad u_{\,\infty\,,\,T}
\end{bmatrix}^{\intercal}
\end{align*}
The computation of the observable fields $u_{\,s}$ and $\theta_{\,s}$ can also be formulated in a matrix form:
\begin{align*}
\mY_{\,u} \egal \mC \cdot \mU \,, \qquad \mY_{\,\theta} \egal \mC \cdot \mTheta \,. 
\end{align*}
where $\mC \, \in \, \mathcal{M}\,\bigl(\, \mathbb{R}^{\,N_{\,s}\times N_{\,s}}\,\bigr)\,$, so that $\mY_{\,u} \, \in \, \mathcal{M}\,\bigl(\, \mathbb{R}^{\,N_{\,s}\times 1}\,\bigr)\,$ and $\mY_{\,\theta}\, \in \, \mathcal{M}\,\bigl(\, \mathbb{R}^{\,N_{\,s}\times 1}\,\bigr)\,$ are vectors. In the end, the state space representation of the problem is formulated as:
\begin{align}
\label{eq:state_space_rep}
\begin{cases}
\dot{\mU} & \egal  \Fo \cdot \mA \cdot \mU \plus \Fo \cdot \mB \cdot \mQ  \,, \\[4pt]
\dot{\mTheta} & \egal \Fo \cdot \mA \cdot \mTheta \plus \mA \cdot \mU \plus \mB \cdot \mQ  \,, \\[4pt]
\mY_{\,u} & \egal \mC \cdot \mU \,, \\[4pt]
\mY_{\,\theta} & \egal \mC \cdot \mTheta \,. 
\end{cases}
\end{align}
The model is solved using a \RK ~order $2$ solver. It computes the field and its sensitivity at the points of interests $x_{\,s}\,$, corresponding to the sensor locations. It is denoted as Large Original Model (LOM).

\subsection{The MIM reduced order model}

The MIM reduced order model enables to compute the field $\tilde{u}_{\,s}$ and $\tilde{\theta}_{\,s}\,, \quad s \, \in \, \bigl\{\,1 \,, \ldots \,, N_{\,s} \,\bigr\}\,$. They are gathered in the vectors $\tilde{\mY}_{\,u}$ and $\tilde{\mY}_{\,\theta}$. Both are computed from the unknown reduced vectors $\mX_{\,u}\, \in \, \mathcal{M}\,\bigl(\, \mathbb{R}^{\, N_{\,r}} \times 1\,\bigr)$ and $\mX_{\,\theta}\, \in \, \mathcal{M}\,\bigl(\, \mathbb{R}^{\,N_{\,r}\times 1}\,\bigr)\,$. The reduced model equations is assumed as: 
\begin{align}
\label{eq:MIM}
\begin{cases}
\dot{\mX}_{\,u} & \egal \Fo \cdot \mF \cdot \mX_{\,u} \plus \Fo \cdot \mG \, \cdot \mQ \,, \\[4pt]
\dot{\mX}_{\,\theta} & \egal \Fo \cdot \mF \cdot \mX_{\,\theta} \plus \mF \cdot \mX_{\,u} \plus \mG \, \cdot \mQ  \,,  \\[4pt]
\tilde{\mY}_{\,u} & \egal \mH_{\,u} \cdot \mX_{\,u} \,, \\[4pt]
\tilde{\mY}_{\,\theta} & \egal \mH_{\,\theta} \cdot \mX_{\,\theta}  \,,
\end{cases}
\end{align}
where the matrix $\mF \, \in \, \mathcal{M}\,\bigl(\, \mathbb{R}^{\,N_{\,r}\times N_{\,r}}\,\bigr)$ is assumed as diagonal. And the matrices $\mG \, \in \, \mathcal{M}\,\bigl(\, \mathbb{R}^{\,N_{\,r}\times 2}\,\bigr)\,$, $\mH_{\,u}\, \in \, \mathcal{M}\,\bigl(\, \mathbb{R}^{\,N_{\,s}\times N_{\,r}}\,\bigr)$ and $\mH_{\,\theta}\, \in \, \mathcal{M}\,\bigl(\, \mathbb{R}^{\,N_{\,s}\times N_{\,r}}\,\bigr)$ are fulled. 
Indeed the general formulation of MIM (Eq. \ref{eq:MIM}) is obtained by considering the transformation $\mU= \mM \mX_{\,u}$ applied to the LOM equations (Eq. \ref{eq:state_space_rep}). It can be noted that the columns of the matrix $\mM \, \in \, \mathcal{M}\,\bigl(\, \mathbb{R}^{\,N\times N_{\,r}}\,\bigr)$ contains the eigenvectors of $\mA$ and the diagonal matrix $\mF \, \in \, \mathcal{M}\,\bigl(\, \mathbb{R}^{\,N_{\,r}\times N_{\,r}}\,\bigr)$ the $N_{\,s}$ eigenvalues of $\mA$.

The reduced model will thus give at time $t\,$ the reduced vectors $\mX_{\,u}$ and $\mX_{\,\theta}\,$. To come back to the fields at the interest points, the operations $\tilde{\mY}_{\,u} \egal \mH_{\,u} \cdot \mX_{\,u}$ and $\tilde{\mY}_{\,\theta} \egal \mH_{\,\theta} \cdot \mX_{\,\theta}$ are performed. These observable fields are obtained by the reduced model and thus denoted with the super script $\sim \,$. It is interesting to remark that knowing matrices $\mF\,$, $\mG$, we have straightforwardly the reduced model for the sensitivity $\mX_{\,\theta}\,$. Here, the MIM model is solved using a \RK ~order $2$ method.

The construction of the MIM reduced model is performed by defining the differences between the reduced model and the complete model as a square residues functional $J_{\,rom}\,$:
\begin{align*}
J_{\,rom}\,\bigl(\,N_{\,r} \,,\, \mF \,,\, & \mG \,,\, \mH_{\,u} \,,\, \mH_{\,\theta} \,\bigr) \\
& \, \eqdef \, 
\sum_{i = 1}^{N_{\,s}} \, \int_{\,0}^{\,t_{\,\fin}} \, 
\Biggl(\, \frac{ \tilde{u}_{\,s}\,(\,t\,) \moins u_{\,s}\,(\,t\,)}{u_{\,s}\,(\,t\,)} \Biggr)^{\,2} \plus
\Biggl(\, \frac{\tilde{\theta}_{\,s}\,(\,t\,) \moins \theta_{\,s}\,(\,t\,)}{\theta_{\,s}\,(\,t\,)} \Biggr)^{\,2} \, \mathrm{d}t \,.
\end{align*}
The functional $J_{\,rom}$ is minimized to determined the order of the reduced model $N_{\,r}$ and matrices $\mF\,$, $\mG\,$, $\mH_{\,u}$ and $\mH_{\,\theta}\,$. The minimization procedure is carried out preliminary to the solution of the parameter estimation problem, using any optimizer such as quasi-Newton, \revision{PSO} or genetic algorithm. This procedure is denoted as the \emph{learning step}. Here the construction is performed for several values of the unknown diffusivity $\alpha\,$. The set of thermal diffusivity used for the construction of the ROM is denoted $\Omega_{\,\alpha}\,$.

\section{Parameter estimation problem with reduced order model}
\label{sec:pep}

The reduced order model is used in the framework of parameter estimation problem. For the sake of clarity, the methodology is explained for the unknown parameter denoted by: 
\begin{align*}
p \egal \alpha \,.
\end{align*}
It is also distinguished the solution of the parameter estimation problem $p^{\,\circ}\,$. It is different from the exact solution of the problem, noted $p^{\,r}\,$, when known. The so-called \emph{a priori} parameter, used as initial guess in the algorithm to solve the inverse problem, is denoted by $p^{\,\apr}\,$. The set of searched parameter is denoted $\Omega_{\,p}\,$. The experimental observations of the field $u$ are written with the super-script $^{\,m}\,$. Thus, the vector of observation is denoted $\mY^{\,m} \, \in \, \mathcal{M}\,\bigl(\, \mathbb{R}^{\,N_{\,s}\times 1}\,\bigr)\,$.

\subsection{The optimal sequence of observations}

To use the MIM ROM in the framework of parameter estimation problem, it requires a learning step to build \emph{a posteriori} the matrices. Since this step is time consuming, the optimal experiment design (OED) methodology \cite{Berger_2018,Jumabekova_2020} is used to select a reduced measurement sequence of $3 \ \mathsf{days}$, denoted:
\begin{align*}
\Omega_{\,t}^{\,\oed}  \, \eqdef \, \bigl[\, \tini \,,\, \tini + 3 \,\bigr] \,,
\end{align*}
where $\tini \ \unit{d}$ is the beginning of the reduced sequence. Note that the reduced sequence respects the condition $\Omega_{\,t}^{\,\oed} \, \subset \, \Omega_{\,t} \,$. The measurement plan $\Pi$ is introduced:
\begin{align*}
\Pi \egal \bigl\{\, \tini \,\bigr\} \,,
\end{align*}
The search of the optimal measurement plan $\Pi^{\,\circ}$ is carried out through the maximization of the D-criteria \cite{Ucinski,Beck_1977}
\begin{align*}
\Pi^{\,\circ} \egal \arg \max_{\,\Pi} \, \Psi \,,
\end{align*}
where $\Psi$ is the determinant of the \textsc{Fisher} matrix $\mathcal{F}$:
\begin{align*}
\Psi \, \eqdef \, \det \mathcal{F}(\,\Pi\,) \,,
\end{align*}
and
\begin{align*}
\displaystyle \mathcal{F} \, \eqdef \, \frac{1}{\sigma^{\,2}} \ \sum_{i \egal 1}^{N_{\,s}} \int_{\,\Omega_{\,t}^{\,\oed}} \theta_{\,s}\,\bigl(\,t\,\bigr)^{\,2}\ \mathrm{d}t \,,
\end{align*}
with $\sigma$ is the measurement uncertainty. This methodology is performed using the LOM and the \emph{a priori} value of the unknown parameter. It enables to determine the optimal reduced sequence of measurement that maximize the parameter estimation accuracy. With this approach, the learning step of the MIM and the solution of the inverse problem are performed over a reduced sequence $\Omega_{\,t}^{\,\oed} \, \subset \, \Omega_{\,t}\,$.

\subsection{Estimating the unknown parameter}
\label{sec:inv_prob_algo}

The parameter estimation problem is solved over the reduced sequence $\Omega_{\,t}^{\,\oed}\,$ to cut the computational efforts. It aims at determining the unknown parameter $p$ verifying: 
\begin{align*}
p^{\,\star} \egal \min_{\Omega_{\,p}} J\,(\,p\,) \,,
\end{align*}
where $J$ is the cost function defined using the least square estimator and the reduced model: 
\begin{align}
\label{eq:cost_function}
J\,:\, p \, \longmapsto \, 
\sum_{i = 1}^{N_{\,s}} \, \int_{\,0}^{\,t_{\,\fin}} \, 
\biggl(\, \tilde{u}_{\,s}\,(\,p\,,\,t\,) \moins u_{\,s}^{\,m}\,(\,t\,) \biggr)^{\,2} \, \mathrm{d}t \,.
\end{align}
%
The minimisation of the cost function is realized using the \textsc{Gau}\ss ~algorithm \cite{Beck_1977,Ozisik_2000}. It assumes the following condition on $J\,$:
\begin{align}
\label{eq:grad_J}
\grad_{\,p} \, J \egal 0 \,.
\end{align}
Thus, Eq.~\eqref{eq:grad_J} is equivalent to:
\begin{align}
\label{eq:derivative_J}
\sum_{i = 1}^{N_{\,s}} \, \int_{\,0}^{\,t_{\,\fin}} \,  \tilde{\theta}_{\,s}\,(\,p\,,\,t\,) \cdot \Bigl(\, \tilde{u}_{\,s}\,(\,p\,,\,t\,) \moins u_{\,s}^{\,m}\,(\,t\,)  \,\Bigr) \egal 0 \,.
\end{align}
Using the \textsc{Taylor} expansion of $\tilde{u}$ around a parameter $p^{\,k}\,$, equation~\eqref{eq:derivative_J} gives a system of $N_{\,s}$ equations:
\begin{align*}
\Bigl(\, \tilde{u}_{\,s}\,(\,p^{\,k}\,,\,t\,) \plus \tilde{\theta}_{\,s}\,(\,p^{\,k}\,,\,t\,) \cdot \bigl(\,p \moins p^{\,k} \,\bigr)
\moins u_{\,s}^{\,m}\,\Bigr) \egal \boldsymbol{0} \,.
\end{align*}
It enables to compute the next candidate $p \egal p^{\,k+1}$ better than $p^{\,k}$ in an iterative procedure, written with the matrix formulation:
\begin{align}
\label{eq:descent_step}
p^{\,k+1} \egal p^{\,k} \plus 
\Bigl(\, Y_{\,\theta}^{\,\intercal}\,(\,p^{\,k}\,) \cdot Y_{\,\theta}\,(\,p^{\,k}\,)  \,\Bigr)^{\,-1} 
\cdot Y_{\,\theta}\,(\,p^{\,k}\,) \cdot 
\Bigl(\, Y^{\,m} \moins \tilde{\mY}_{\,u}\,(\,p^{\,k}\,) \,\Bigr) \,.
\end{align} 
In terms of computational time, the cost to compute the new candidate $p^{\,k+1}$ is strongly reduced due to the use of MIM reduced order model. Indeed, it only manipulates the reduced vectors $\mX_{\,u}$ and $\mX_{\,\theta}$ of size $N_{\,r}\, \ll \, N_{\,s}\,$. Moreover, the MIM model provides straightforwardly the sensitivity functions of the observable fields.

The Algorithm~\ref{alg:PEP_algorithm} describes the main steps to solve the inverse problem. At the iteration $k\,$, the field and its sensitivity are computed using the MIM ROM and a candidate $p_{\,k}$ for the unknown parameter. With these results, the cost function between measurement and model predictions is evaluated. Moreover, using the sensitivity $\mY_{\,\theta}\,$, the candidate $p_{\,k+1}$for the next iteration is computed. The algorithm runs until a maximum number of iterations $N_{\,k}$ is reached or one of the two following criteria is satisfied: 
\begin{align*}
\gamma_{\,1} \,\geqslant\,\eta_{\,1} \qquad \& \qquad \gamma_{\,2} \,\geqslant\,\eta_{\,2} \,,
\end{align*}
where $\eta_{\,1}$ and $\eta_{\,2}$ are user defined values. The criteria $\gamma_{\,1}$ and $\gamma_{\,2}$ evaluates the relative magnitude changes of the cost function and candidate parameter, respectively:
\begin{align*}
\gamma_{\,1} \, \eqdef \, \displaystyle \frac{\Bigl\| \, J(\,p_{\,k+1}\,) \moins J(\,p_{\,k}\,) \,\Bigr\|_{\,2}}
{\Bigl\| \, J(\,p_{\,k}\,) \,\Bigr\|_{\,2}} \,, \qquad
\gamma_{\,2} \, \eqdef \, \displaystyle \frac{\Bigl\| \, p_{\,k+1} \moins p_{\,k} \,\Bigr\|_{\,2}}
{\Bigl\| \, p_{\,k} \,\Bigr\|_{\,2}} \,.
\end{align*}

\begin{algorithm}
\caption{ \textsc{Gau}\ss ~algorithm using the MIM ROM.}\label{alg:PEP_algorithm}
\begin{algorithmic}[1]
\State Build matrices  $\mF\,$, $\mG\,$, $\mH_{\,u}$ and $\mH_{\,\theta}$ according to MIM learning step
\State Set iteration indicator $k \egal 1\,$
\State Set \emph{a priori} parameter $p_{\,k} \egal p^{\,\apr}\,$
\While{$k\,\leqslant\,N_{\,k}$ \& $\gamma_{\,1} \,\geqslant\,\eta_{\,1}$ \& $\gamma_{\,2} \,\geqslant\,\eta_{\,2}$}
\State Compute the direct problem $\mY_{\,u}$ and $\mY_{\,\theta}$ using reduced order model with Eq.~\eqref{eq:MIM} and a candidate $p_{\,k}$
\State Evaluate the cost function $J$ using Eq.~\eqref{eq:cost_function}
\State Compute the next candidate for the unknown parameter $p_{\,k+1}$ using Eq.~\eqref{eq:descent_step}
\State Increment: $k \egal k + 1\,$, $p_{\,k} \, \leftarrow \, p_{\,k+1}$
\State Compute stopping criteria $\gamma_{\,1}$ and $\gamma_{\,2}$ with Eqs.~
\EndWhile 
\State \textbf{end}
\end{algorithmic}
\end{algorithm}

\section{Synthesis of the methodology and metrics of its efficiency}
\label{sec:synthesis_methodology}

\begin{figure}[h!]
\centering
\includegraphics[width=.95\textwidth]{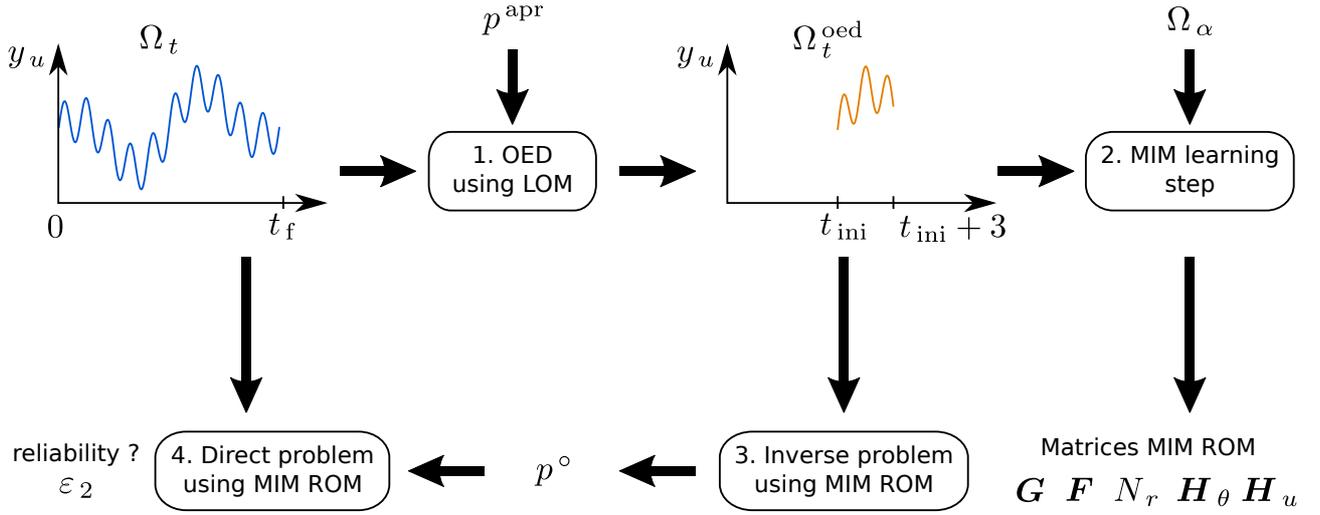}
\caption{illustration of the methodology.}
\label{fig:methodo}
\end{figure}

The global methodology to solve the parameter estimation problem of heat transfer using MIM ROM is synthesized in Fig.~\ref{fig:methodo}. First, the OED methodology is used to define an optimal reduced sequence $\Omega_{\,t}^{\,\oed}$ of the observations of the field. This results have two advantages noticeable at steps two and three. At step two, the learning step of the MIM is carried only for $3 \ \mathsf{days}$ and not for the whole signal. It saves important computational efforts to determine the matrices of the ROM MIM. Then, at step three, the inverse problem is solved for a reduced sequence. Moreover, the sequence is optimal since it ensures the maximum accuracy of the retrieved parameters. Last, when the unknown parameter is estimated, the MIM ROM is computed considering the whole signal. The error between the numerical predictions and the whole sequence of experimental observations is computed to evaluate the reliability of the model.

Several metrics are defined to evaluate the efficiency of the proposed method. The $\mathcal{L}_{\,2}$ and $\mathcal{L}_{\,\infty}$ errors are denoted by $\varepsilon_{\,2}$ and $\varepsilon_{\,\infty}\,$, respectively. They are computed to evaluate the accuracy of the predictions of the numerical models:
\begin{align*}
\varepsilon_{\,2}  & \eqdef \ \sqrt{\,\frac{1}{N_{\,s} \ N_{\,t}} \, \sum_{n\, =\, 1}^{N_{\,t}} \ \sum_{s\, =\, 1}^{N_{\,s}}\, \Bigl( \, u\, (\,x_{\,s} \,,\, t_{\,n} \,) \moins u^{\mathrm{\, ref}}\, (\,x_{\,s} \,,\, t_{\,n}\, \Bigr)^{\,2}}\,, \\[4pt]
\varepsilon_{\,\infty}  & \eqdef \ \max_{x_{\,s} \,,\, t_{\,n}} \ \Bigl| \, u\, (\,x_{\,s} \,,\, t_{\,n} \,) \moins u^{\mathrm{\, ref}}\, (\,x_{\,s} \,,\, t_{\,n}\,)\,\bigr) \Bigr|\,,
\end{align*}
where $u^{\mathrm{\, ref}}$ is a so-called reference solution, obtained generally by the LOM model, and $N_{\,t}$ the number of time step. In addition, the relative error $\varepsilon_{\,r}$ is calculated to evaluate the accuracy of the retrieved parameter:
\begin{align*}
\varepsilon_{\,r}  & \eqdef \ \frac{p^{\,\circ} \moins p^{\,r}}{p^{\,r}} \,.
\end{align*}

Another important criteria is the computational efforts evaluated in the \texttt{Matlab\texttrademark} environment with a computer equipped with \texttt{Intel} i$7$ CPU and $32$ GB of RAM. The measured computational time is denoted by $t_{\,\mathrm{cpu}} \ \unit{s}\,$. A dimensionless version is also computed: 
\begin{align*}
t_{\,\mathrm{cpu}}^{\,\star} \, \eqdef \, \frac{t_{\,\mathrm{cpu}}}{t_{\,\mathrm{cpu}\,,\,\mathrm{ref}}}\,,
\end{align*}
where $t_{\,\mathrm{cpu}\,,\,\mathrm{ref}}$ corresponds to the computational time required using the LOM.

\section{Validation of the MIM model for parameter estimation problem}
\label{sec:validation_MIM}

The construction of the MIM ROM and its use to solve  parameter estimation problem is first explained on a simple case with simulated experimental data. Since the computational costs are reduced, step one of the methodology illustrated in Figure~\ref{fig:methodo} is not carried out. Thus, next section presents the learning step to build the MIM ROM for the field and its sensitivity to the unknown parameter. Then, the model is used for parameter estimation problem. 

\subsection{Construction of the MIM ROM}
\label{sec:construction_MIM}

The MIM ROM is built during the so-called learning step. For this, Eq.~\eqref{eq:heat_diffusion} is solved for a wall of length $L \egal 0.5 \ \mathsf{m}\,$, an horizon of simulation $t_{\,\fin} \egal 24 \ \mathsf{h}$ and the following initial and right boundary condition: 
\begin{align*}
T_{\,\ini} & \egal T_{\,\infty\,,\,R} \egal 20 \ \mathsf{^{\,\circ}C}  \,.
\end{align*}
The solution is generated for four cases at five points of interests defined as:
\begin{align}
\label{eq:point_interests}
x_{\,s} \egal \bigl\{\, 0.1 \,,\, 0.15 \,,\, 0.25 \,,\, 0.3 \,,\, 0.45\,\bigr\}\,.
\end{align}
Three cases corresponds to the following diffusivity $\Omega_{\,\alpha} \egal \bigl\{\, 2.03 \,,\, 5.21 \,,\, 5.81 \,\bigr\} \cdot 10^{\,-7} \ \mathsf{m^{\,2} \,.\,s^{\,-1}}\,$, corresponding to the ones of concrete, gypsum and mineral wool, combined with the left boundary condition: 
\revision{
\begin{align}
\label{eq:learning_step_BC}
T_{\,\infty\,,\,L}^{\,1} & \egal T_{\,\ini} \plus 5 \, \sin \biggl(\, \frac{2 \, \pi}{24 \cdot 3600} \ t \,\biggr) \,.
\end{align}}
The last case is carried out for the diffusivity of concrete $\alpha \egal 5.21 \cdot 10^{\,-7} \ \mathsf{m^{\,2} \,.\,s^{\,-1}}$ and a slightly modified left boundary condition:
\revision{
\begin{align}
\label{eq:learning_step_BC_2}
T_{\,\infty\,,\,L}^{\,2} & \egal T_{\,\ini} \plus 20 \, \sin \biggl(\, \frac{6 \, \pi}{24 \cdot 3600} \ t \,\biggr)  \,.
\end{align}}
To obtain each order of the MIM, PSO is used to perform the learning step with $100$ iterations and a size of $50$ populations. It can be noted that taking $100$ iterations is more than sufficient to ensure the convergence.

\begin{figure}[h!]
\centering
\subfigure[\label{fig:e2_fN}]{\includegraphics[width=.45\textwidth]{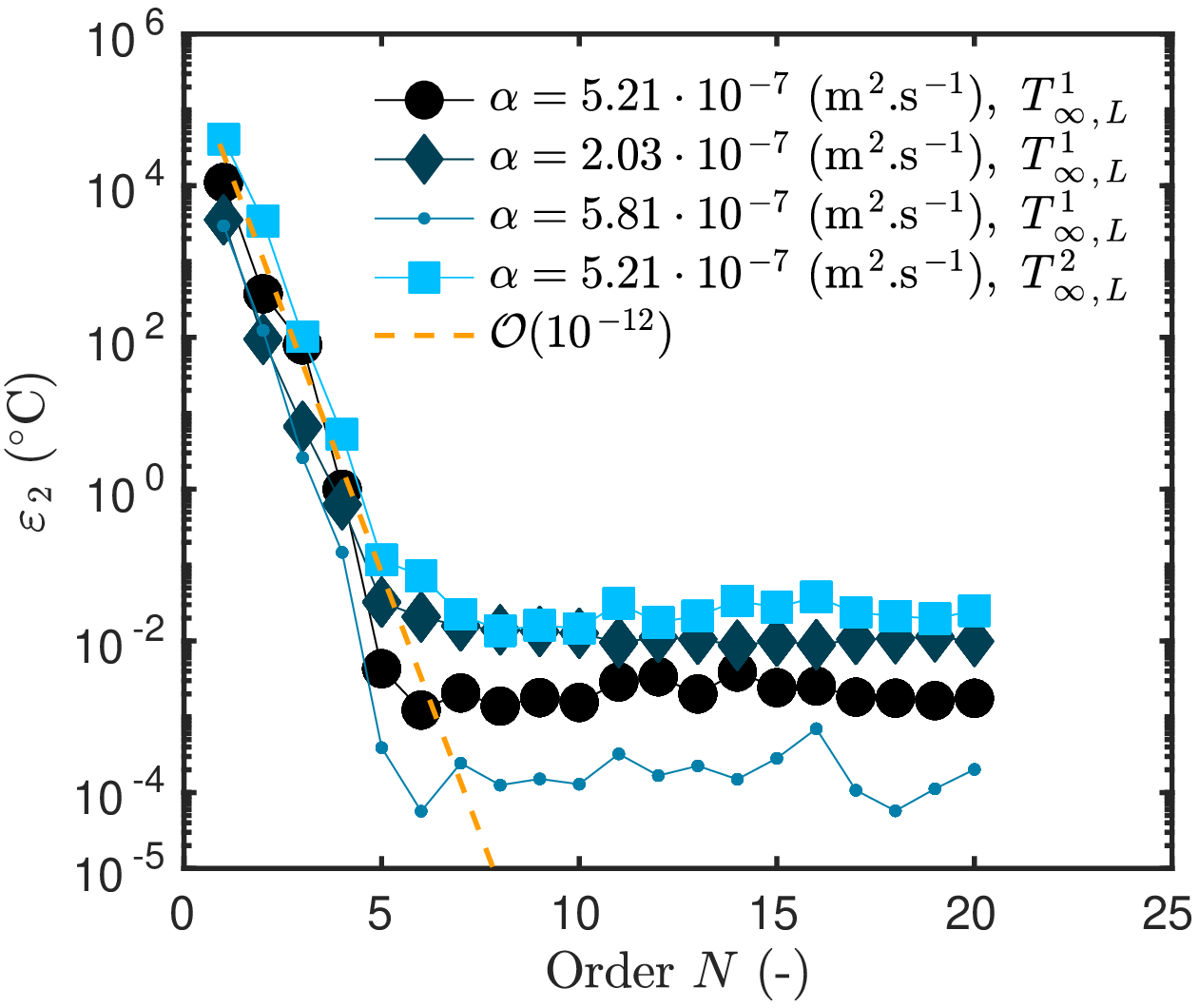}} \hspace{0.2cm}
\subfigure[\label{fig:einf_fN}]{\includegraphics[width=.45\textwidth]{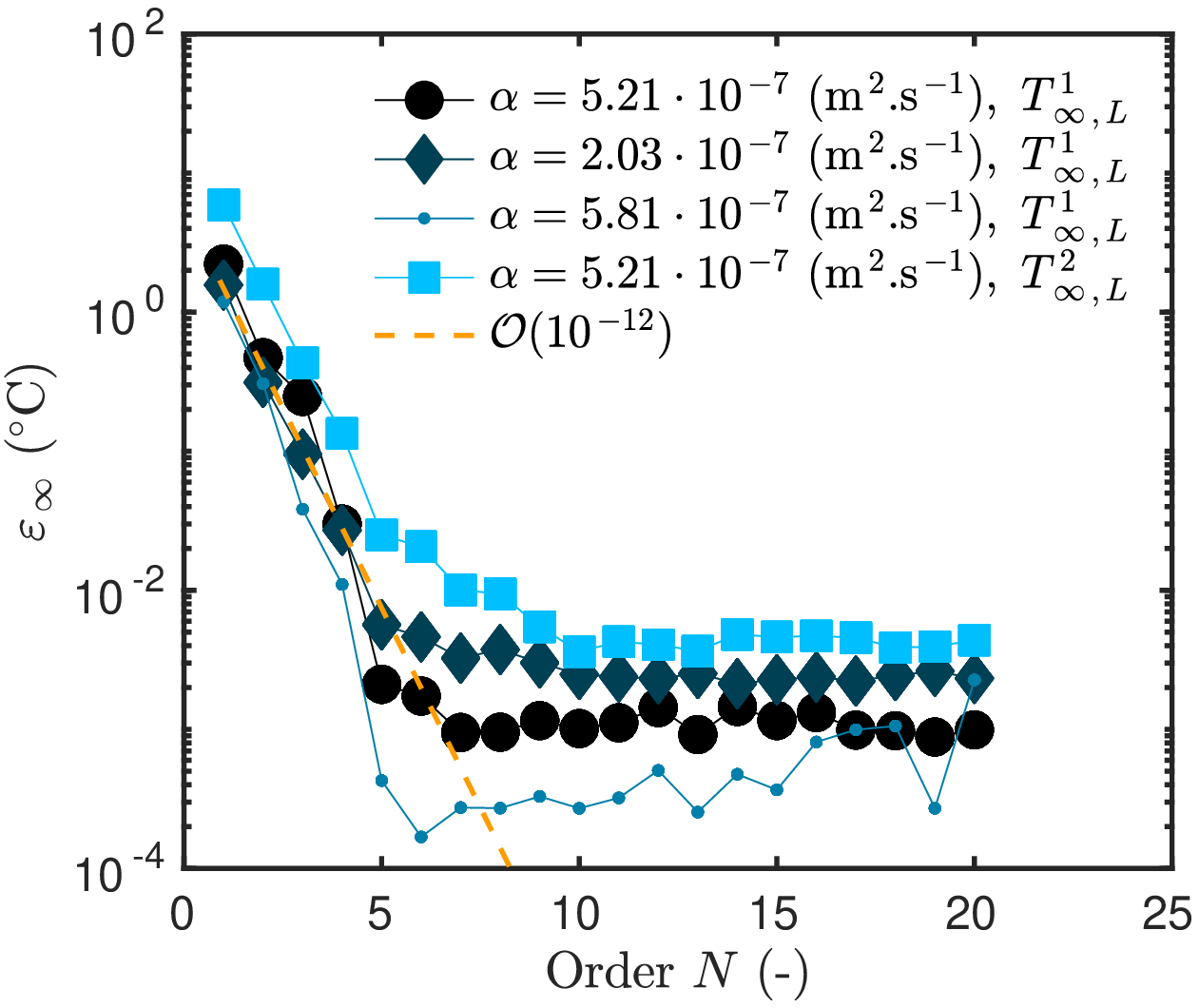}}
\caption{\revision{Evolution of the error of the MIM ROM model according to the order $N$.}}
\end{figure}

Figures~\ref{fig:e2_fN} and \ref{fig:einf_fN} show the error of the reduced order model with the order $N\,$. After the order $10\,$, the accuracy of the model becomes stable. Thus, a MIM ROM with an order $N$ scaling with $\mathcal{O}(\,5\,)$ enables to cut the computational complexity as well as ensure a sufficient accuracy of the predictions. 

\subsection{Accuracy for direct simulations}

To evaluate the accuracy of the built model, the solutions $y_{\,u}$ and $y_{\,\theta}$ are computed with the MIM ROM and the LOM for $50$ values of $\alpha$ in the interval $\bigl[\,2.03 \,,\, 6.37\,\bigr] \cdot 10^{\,-7} \ \mathsf{m^{\,2}\,.\,s^{\,-1}}$. The following initial and boundary conditions are defined: 
\begin{align*}
T_{\,\ini} & \egal T_{\,\infty\,,\,R} \egal 20 \ \mathsf{^{\,\circ}C} \,, &&
T_{\,\infty\,,\,L} \egal T_{\,\ini} \plus 5 \, \sin \biggl(\, \frac{2 \, \pi}{24 \cdot 3600} \ t \,\biggr) \,.
\end{align*}
The final time is $t_{\,\fin} \egal 24 \ \mathsf{h}$ and the wall of length $L \egal 0.5 \ \mathsf{m}\,$. The points of interest are the same as defined in Eq.~\eqref{eq:point_interests}. The discretisation parameters is $\Delta t \egal 30 \ \mathsf{s}$ for both models. For the LOM, $75$ points of space discretisation are used. 

\begin{figure}[h!]
\centering
\subfigure[\label{fig:Yt_ft_kappa1} $\alpha \egal 3 \e{-7} \ \mathsf{m^{\,2}\,.\,s^{\,-1}} $]{\includegraphics[width=.45\textwidth]{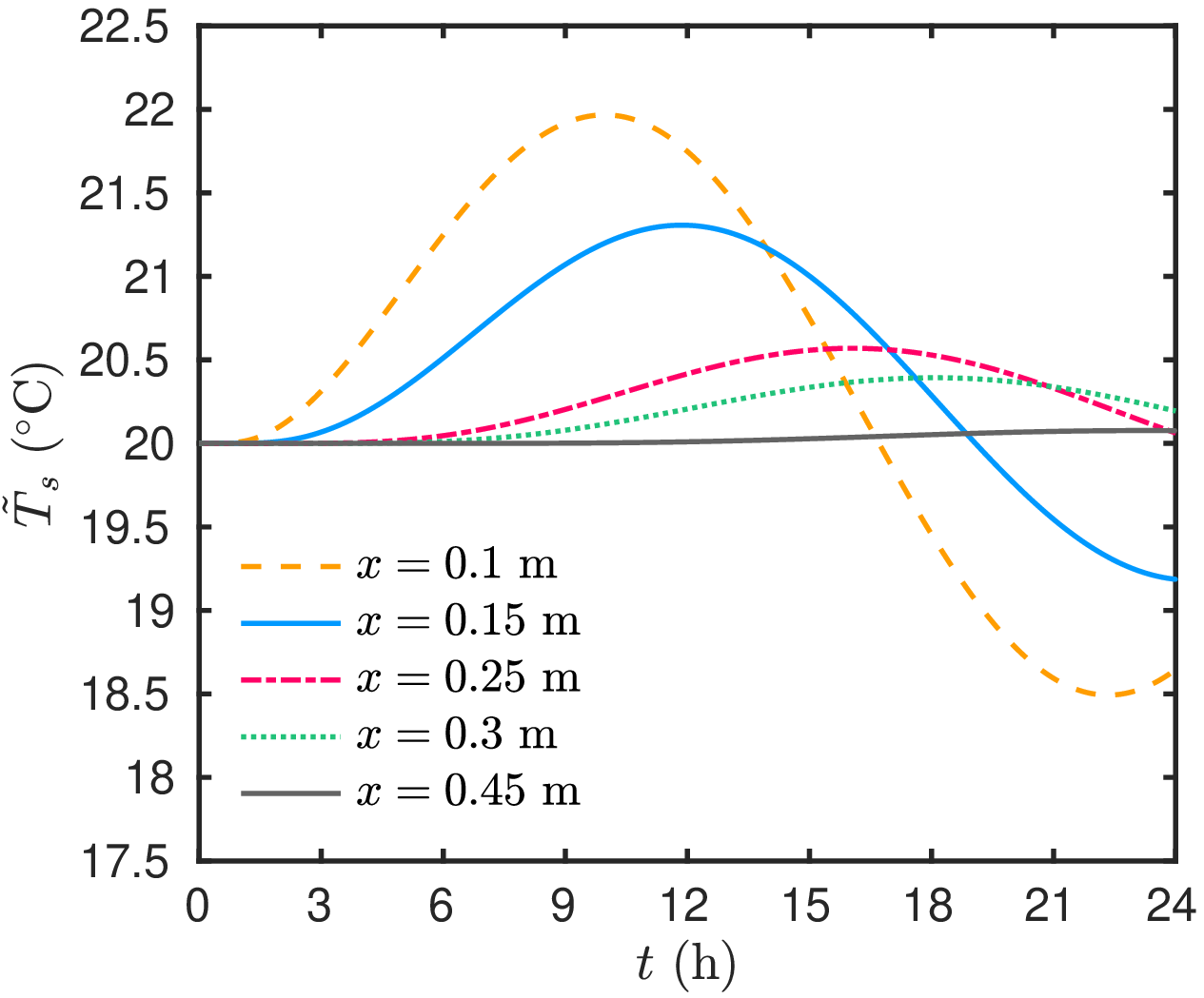}} \hspace{0.2cm}
\subfigure[\label{fig:Ytheta_ft_kappa2} $\alpha \egal 4.5 \e{-7} \ \mathsf{m^{\,2}\,.\,s^{\,-1}} $]{\includegraphics[width=.45\textwidth]{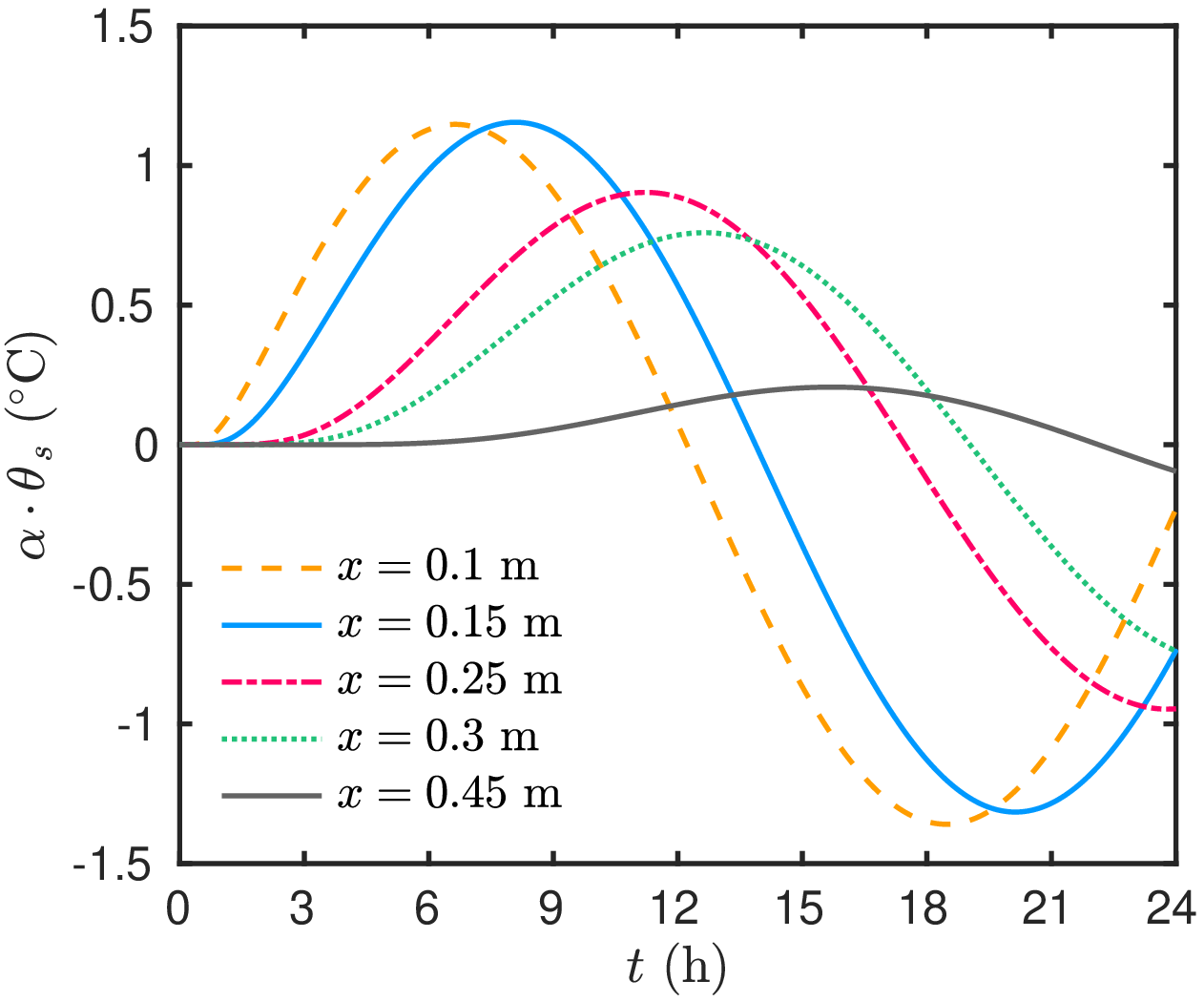}}
\caption{Evolution of the field and the sensitivity at the points of interest for two values of diffusivity.}
\end{figure}

The time evolution of the fields and the sensitivity at the five points of interests is shown in Figures~\ref{fig:Yt_ft_kappa1} and \ref{fig:Ytheta_ft_kappa2} for two values of diffusivity, obtained with the MIM order $5\,$. It highlights that the model enables to simulate the whole dynamic of the observable outputs for thermal diffusivity different from the one used in the learning step.

\begin{figure}[h!]
\centering
\subfigure[\label{fig:e2mimO5_T_fkappa} for $y_{\,u}$]{\includegraphics[width=.45\textwidth]{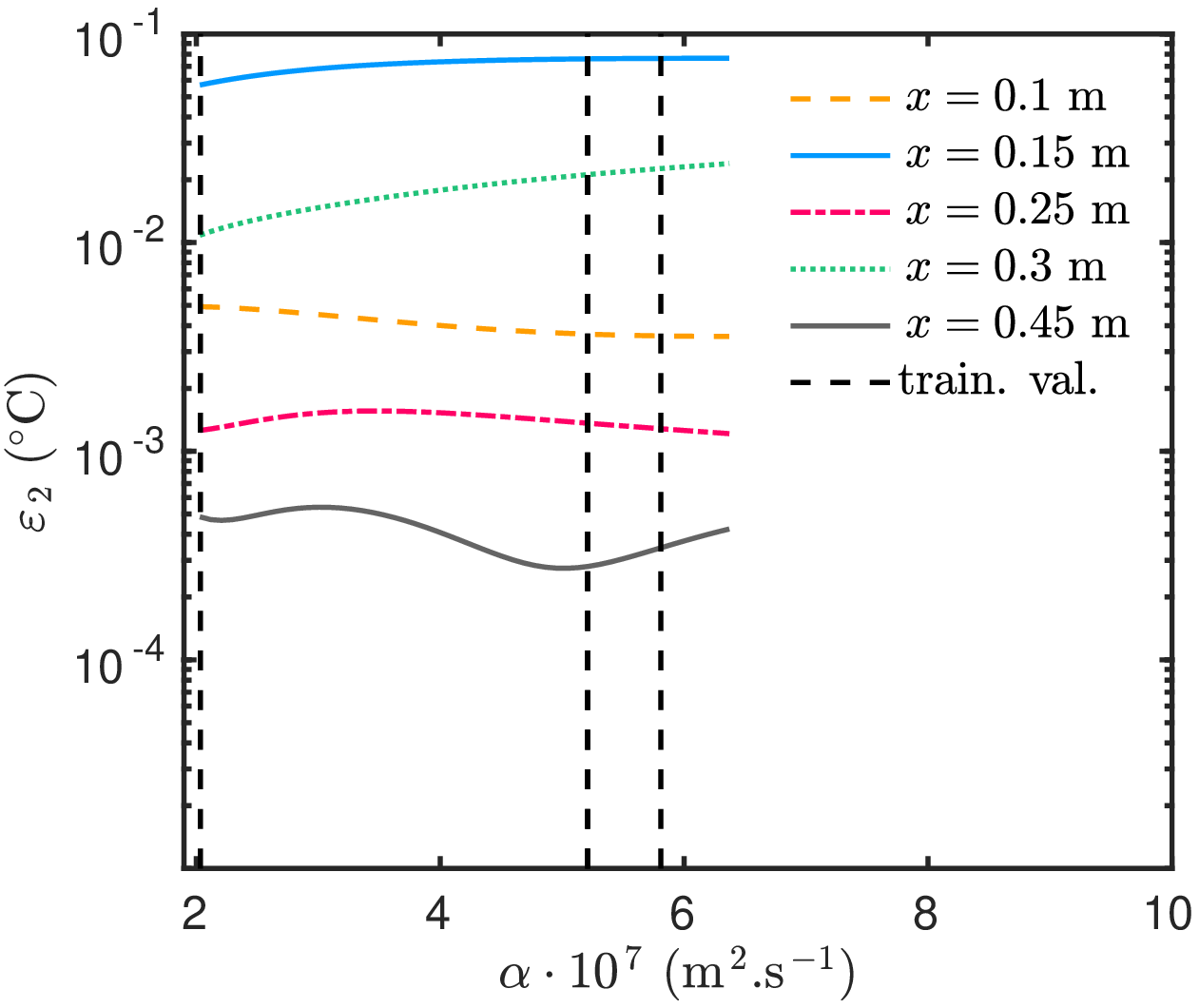}} \hspace{0.2cm}
\subfigure[\label{fig:e2mimO5_Teta_fkappa} for $y_{\,\theta}$]{\includegraphics[width=.45\textwidth]{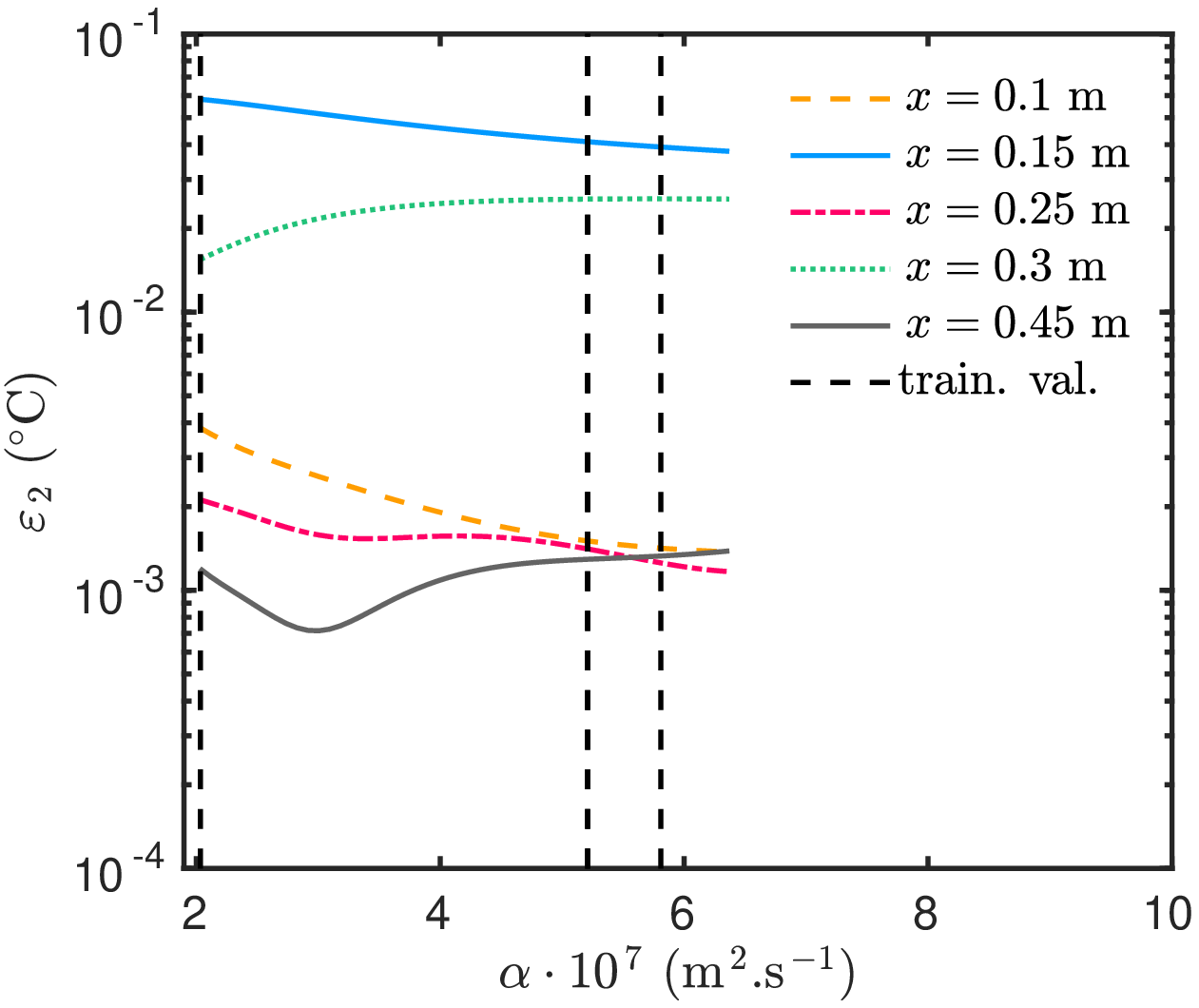}} \\
\subfigure[\label{fig:e2mimO17_T_fkappa} for $y_{\,u}$]{\includegraphics[width=.45\textwidth]{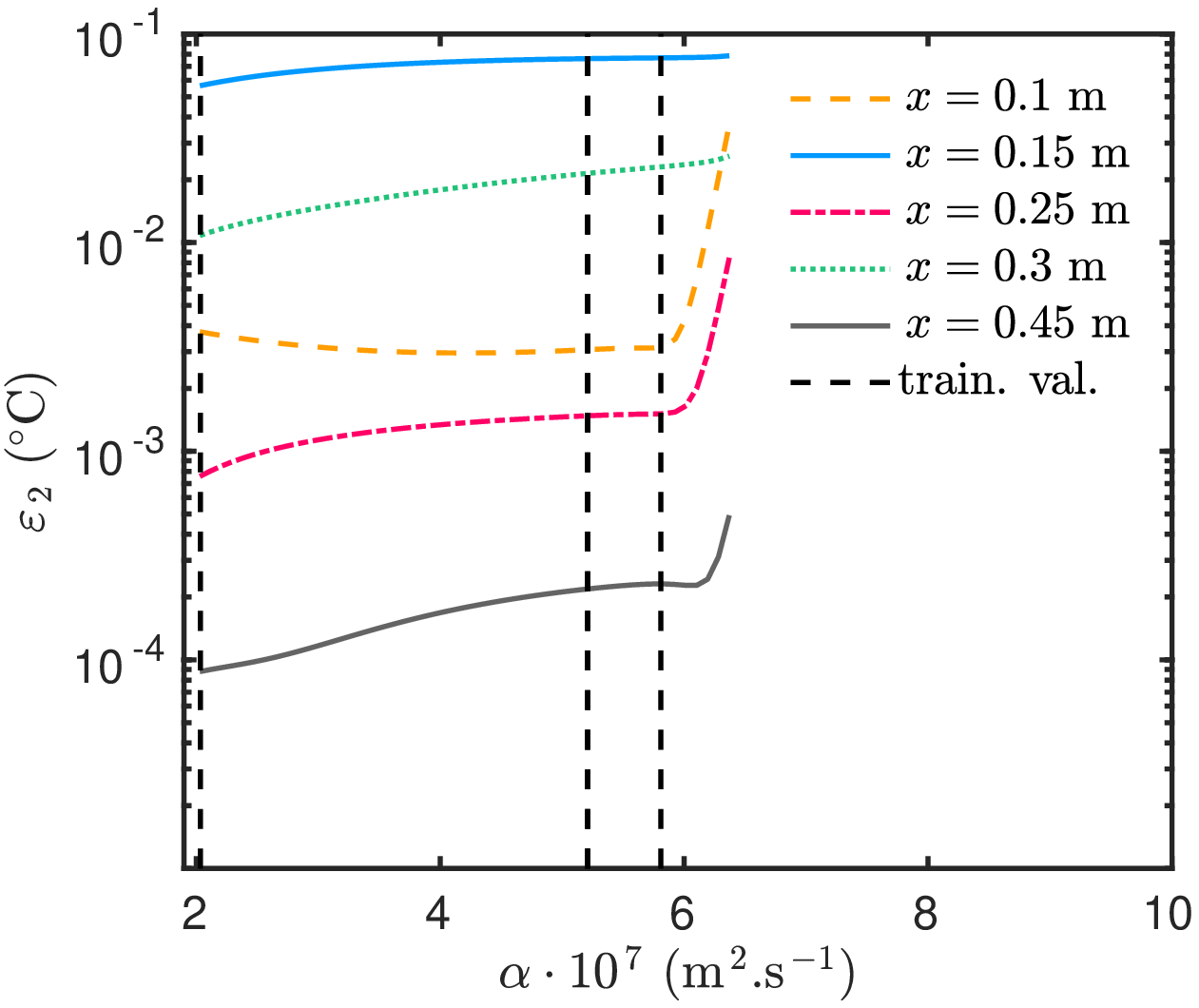}} \hspace{0.2cm}
\subfigure[\label{fig:e2mimO17_Teta_fkappa} for $y_{\,\theta}$]{\includegraphics[width=.45\textwidth]{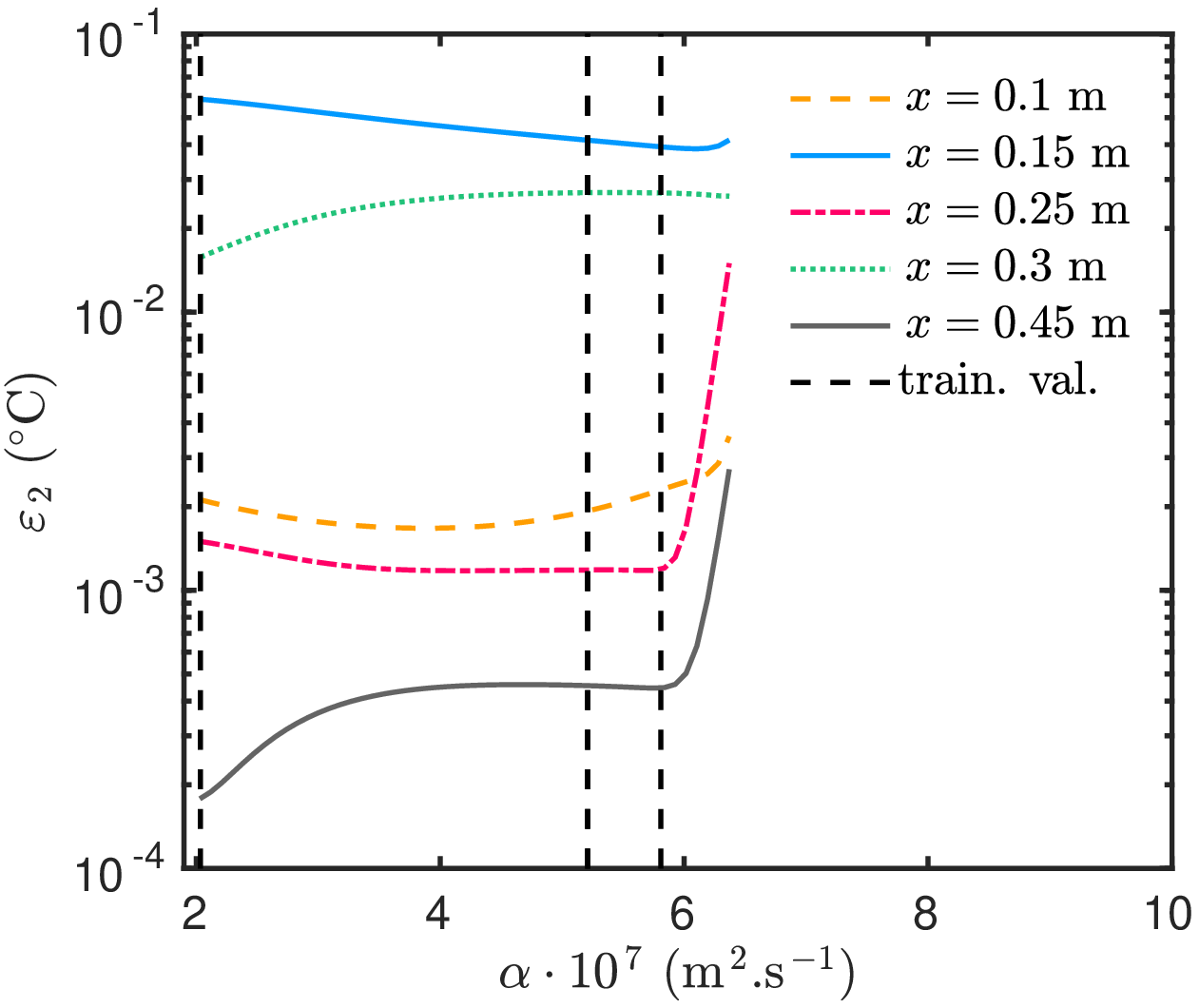}} 
\caption{Variation of the error of the MIM ROM, Orders $5$ \emph{(a,b)} and $17$ \emph{(c,d)}, for the whole interval of thermal diffusivity, for the field \emph{(a,c)} and its sensitivity \emph{(b,d)}.}
\label{fig:e2mim_fkappa}
\end{figure}

Figure~\ref{fig:e2mim_fkappa} shows the variation of the error according to the values of the diffusivity and for two different orders. The ROM of order $5$ has a satisfactory error, lower than $10^{\,-1}$ for both the field and its sensitivity. Moreover, the error is stable for the whole range of thermal diffusivity remaining inside the domain used during the learning step. As noted for $\alpha \, > \, 6 \e{-7} \ \mathsf{m^{\,2} \,.\,s^{\,-1}}\,$, the accuracy of the model is not satisfying anymore. Indeed, this diffusivity is out of the range used during the learning step.

 Table~\ref{tab:model_cpu} provides a synthesis of the effiency of the MIM ROM for a direct simulation. The accuracy of the model is almost similar between order $5$ and $17\,$. In terms of computational time, the MIM ROM of order $5$ enables to cut the computational cost by $87 \%$ compared to the LOM. Since the model reduction method is \emph{a posteriori}, there is an inherent computational cost to build the MIM model. As a synthesis, a learning step composed of $4$ signals is sufficient to build a model of low order and reliable for a wide range of parameter $\alpha$.

\begin{table}
\centering
\caption{Efficiency of the MIM ROM.}
\label{tab:model_cpu}
\setlength{\extrarowheight}{.5em}
\begin{tabular}[l]{@{} c c cc cc cc}
\hline
\hline
Model 
& Order
& \multicolumn{2}{c}{$\varepsilon_{\,\infty}$}
& \multicolumn{2}{c}{$t_{\,\mathrm{cpu}} \ \unit{s}$}
& \multicolumn{2}{c}{$t^{\,\star}_{\,\mathrm{cpu}} \ \unit{-}$}\\
 
& 
& $u_{\,s}$
& $\theta_{\,s}$
& Learning step
& Direct simulation
& Learning step
& Direct simulation\\
LOM
& $75$
& -
& -
& -
& $0.60$
& -
& $1$\\
ROM
& $17$
& $0.08$
& $0.06$
& $2.4$
& $0.10$
& $4$
& $0.16$\\
ROM
& $5$
& $0.07$
& $0.06$
& $2.4$
& $0.08$
& $4$
& $0.13$\\
\hline
\hline 
\end{tabular}
\end{table}

\subsection{Parameter estimation problem}

The reliability of the MIM ROM is now evaluated for solving parameter estimation problem. For this, experimental observations are numerically generated using the LOM for a diffusivity $\alpha^{\,r} \egal 5.21 \ \cdot 10^{\,-7} \ \mathsf{m^{\,2}\,.\,s^{\,-1}}\,$. The initial and boundary conditions are the following: 
\begin{align*}
T_{\,\ini} & \egal T_{\,\infty\,,\,R} \egal 20 \ \mathsf{^{\,\circ}C} \,, &&
T_{\,\infty\,,\,L} \egal T_{\,\ini} \plus 3 \cdot \Biggl(\, 1 \moins \cos \biggl(\, \frac{2 \, \pi}{12 \cdot 3600} \ t \,\biggr) \, \Biggr) \,.
\end{align*}
It can be remarked that the boundary condition at $T_{\,\infty\,,\,L}$ is very different from the one used during the learning step of the MIM ROM~\eqref{eq:learning_step_BC} and \eqref{eq:learning_step_BC_2}. The LOM is computed using $\Delta t \egal 30 \ \mathsf{s}$ and $\Delta x \egal 6.8 \ \mathsf{mm}\,$. Then, the observations are projected at the points of interest considering a time step of $1 \ \mathsf{hour}$ and adding a random normally distributed error. The standard deviation of the measurement is reported in Table~\ref{tab:pep_results}.

\begin{figure}[h!]
\centering
\subfigure[\label{fig:P_fiter}]{\includegraphics[width=.45\textwidth]{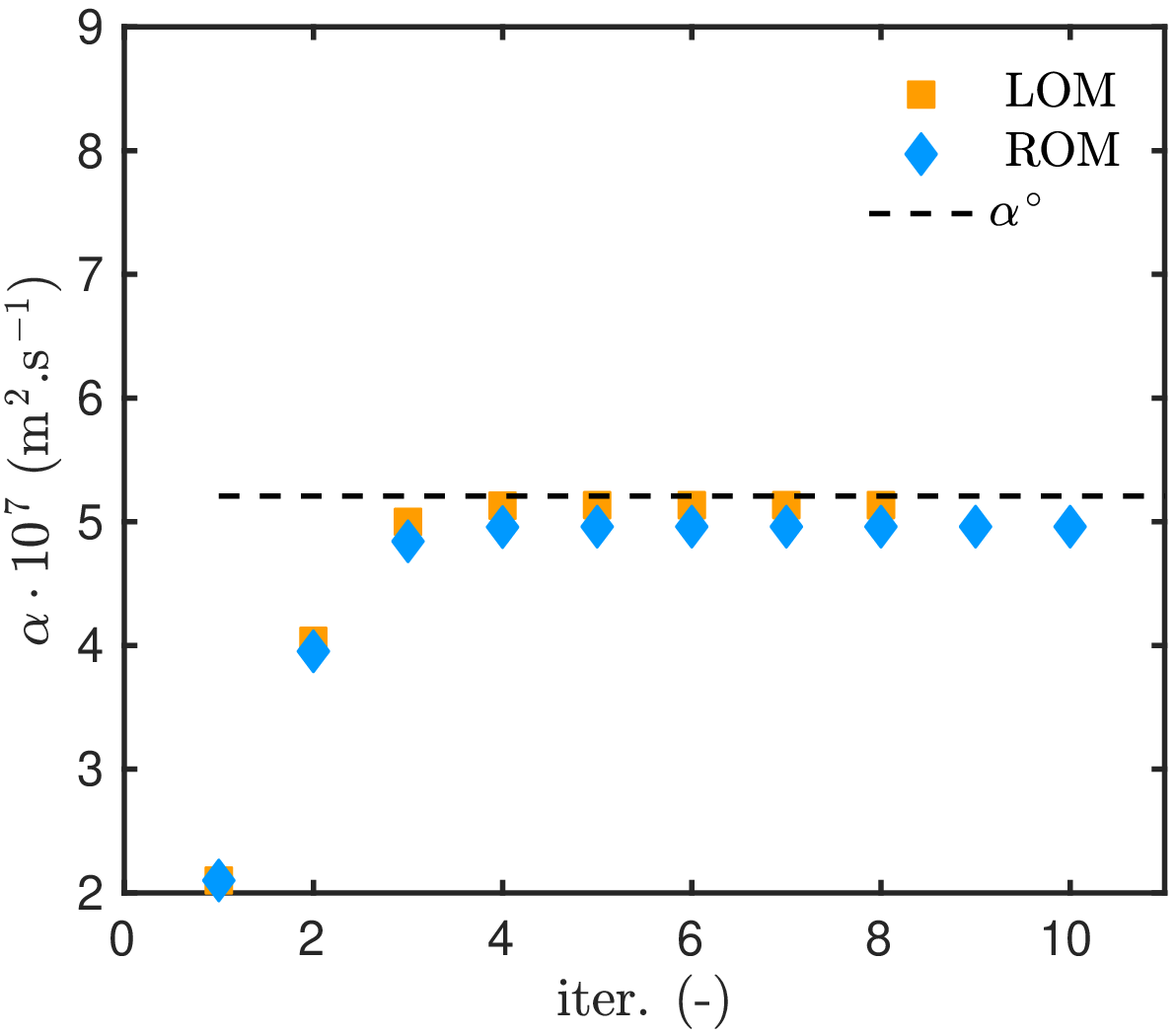}}  \hspace{0.2cm}
\subfigure[\label{fig:gamma2_fiter}]{\includegraphics[width=.45\textwidth]{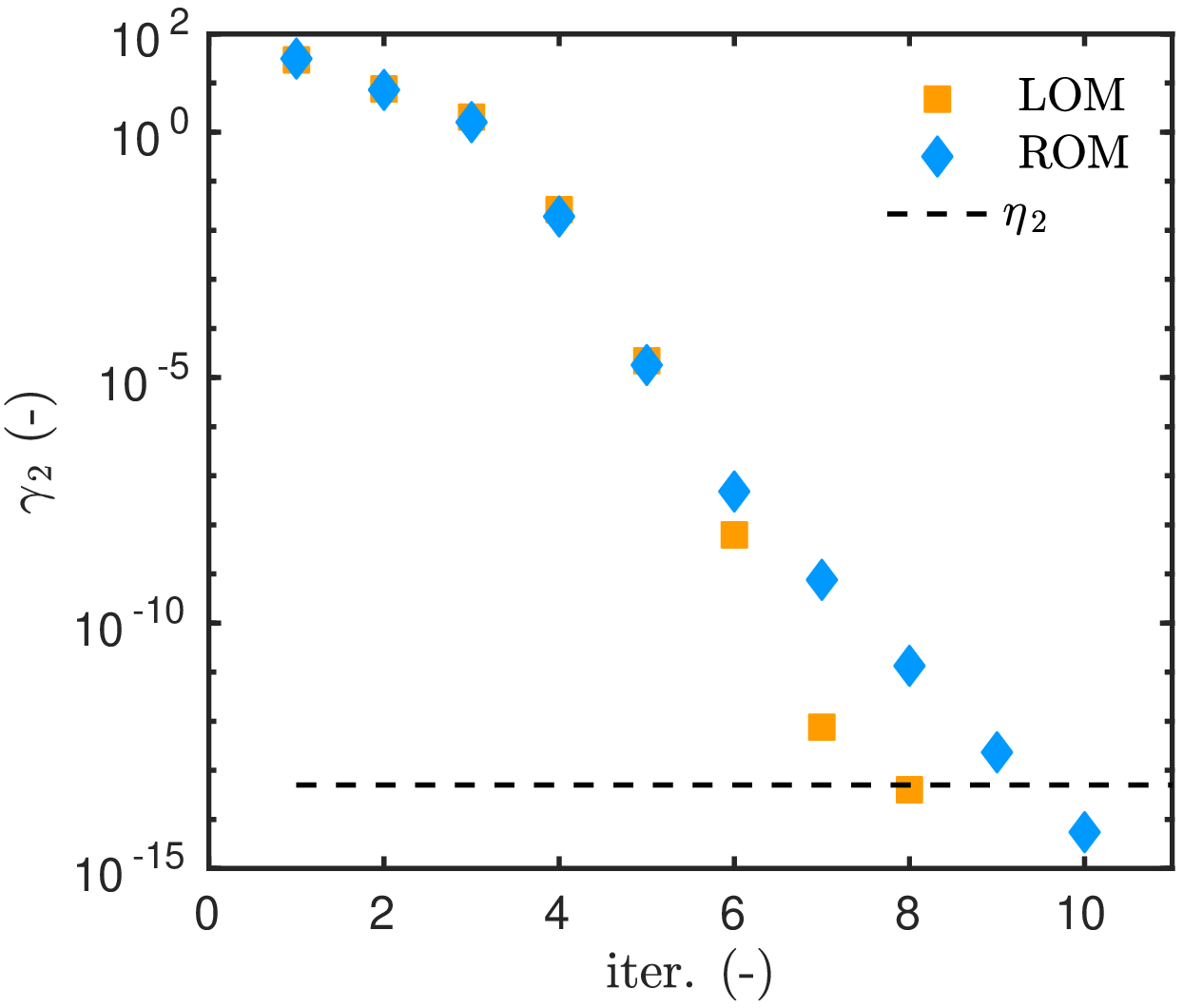}} 
\caption{Evolution of the estimated parameter \emph{(a)} and of the convergence criteria $\gamma_{\,2}$ according to the iteration.}
\end{figure}

\begin{figure}[h!]
\centering
\subfigure[\label{fig:YLOM_ft}]{\includegraphics[width=.45\textwidth]{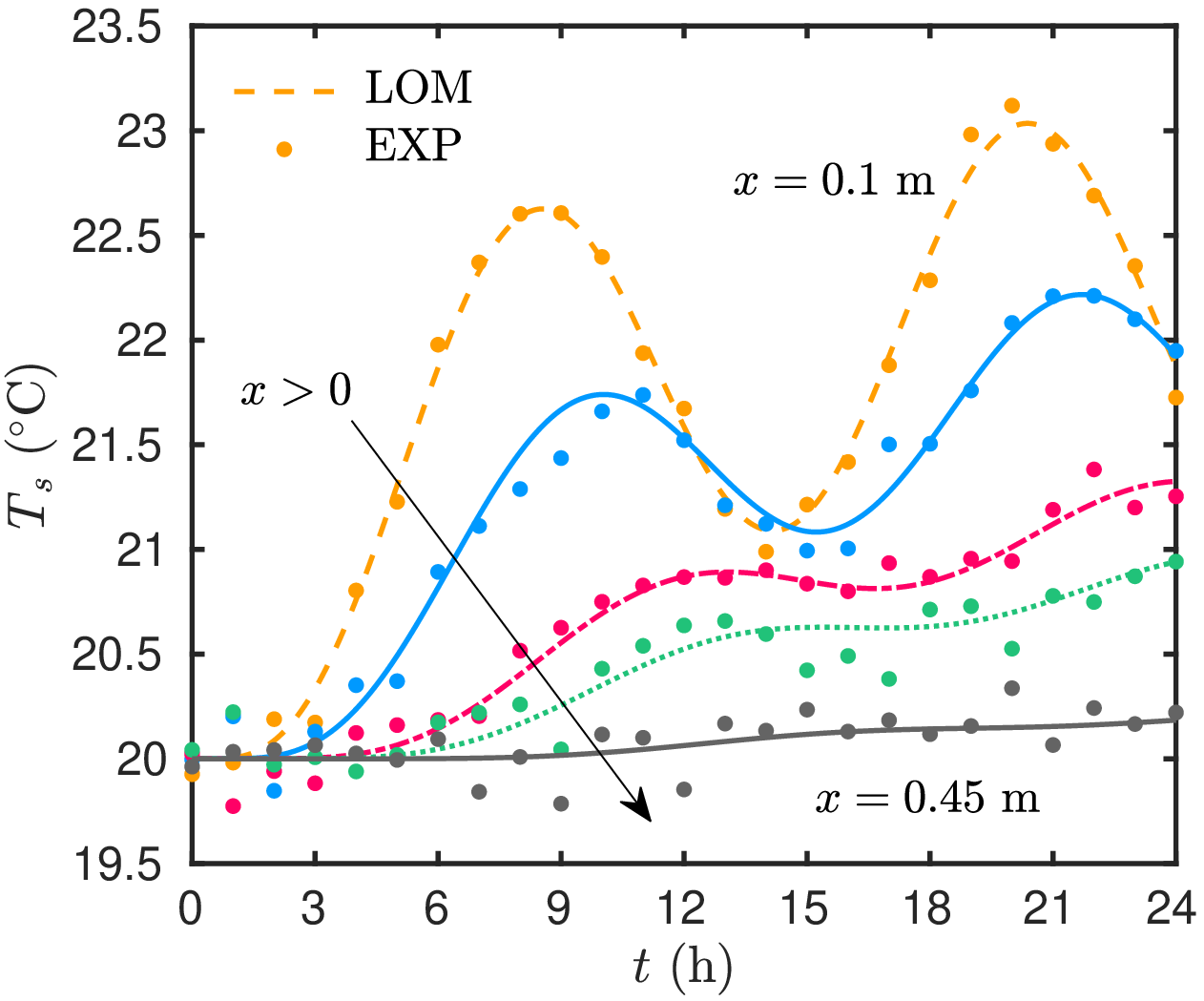}}  \hspace{0.2cm}
\subfigure[\label{fig:YROM_ft}]{\includegraphics[width=.45\textwidth]{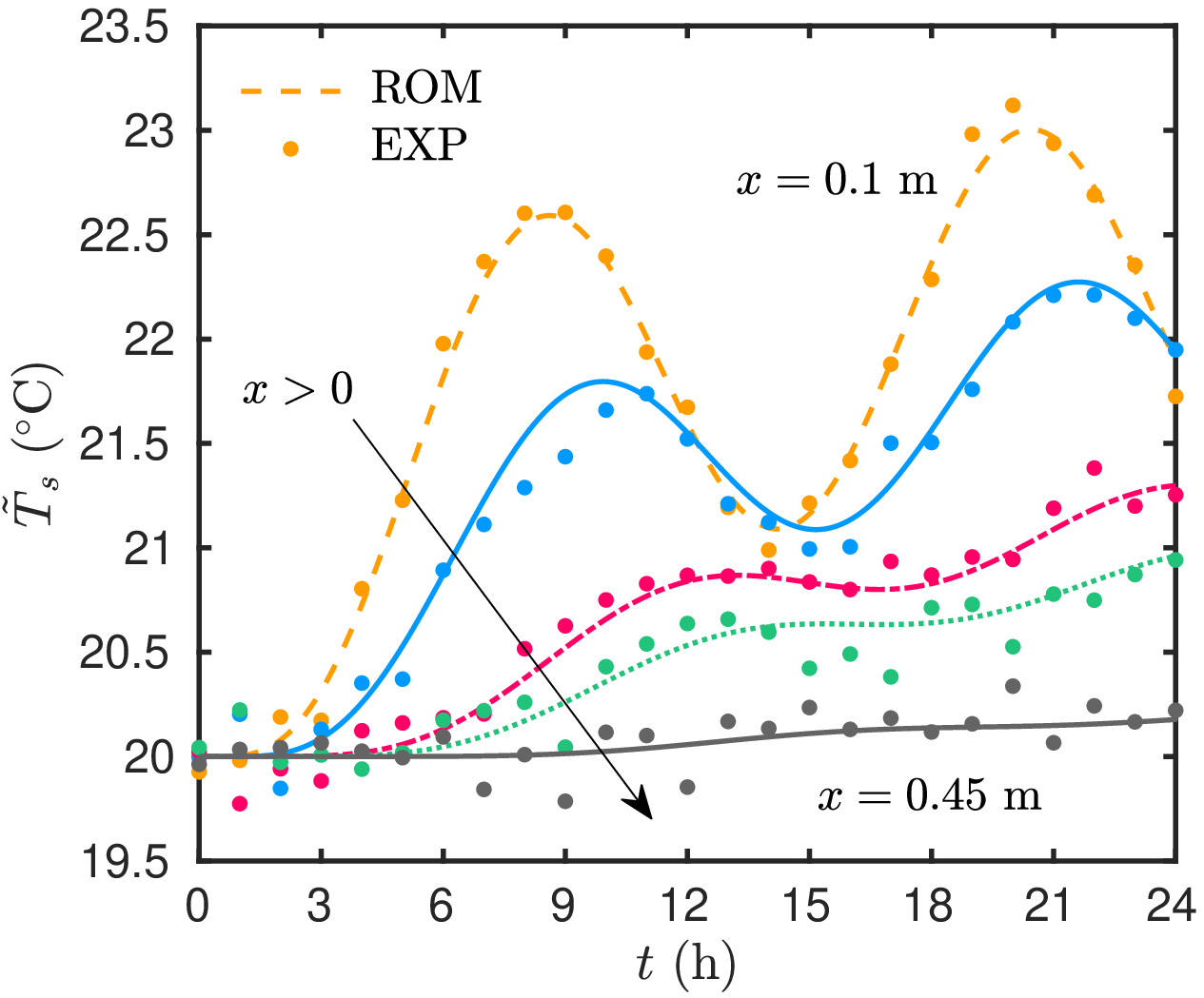}} \\
\subfigure[\label{fig:res_LOM_ft}]{\includegraphics[width=.45\textwidth]{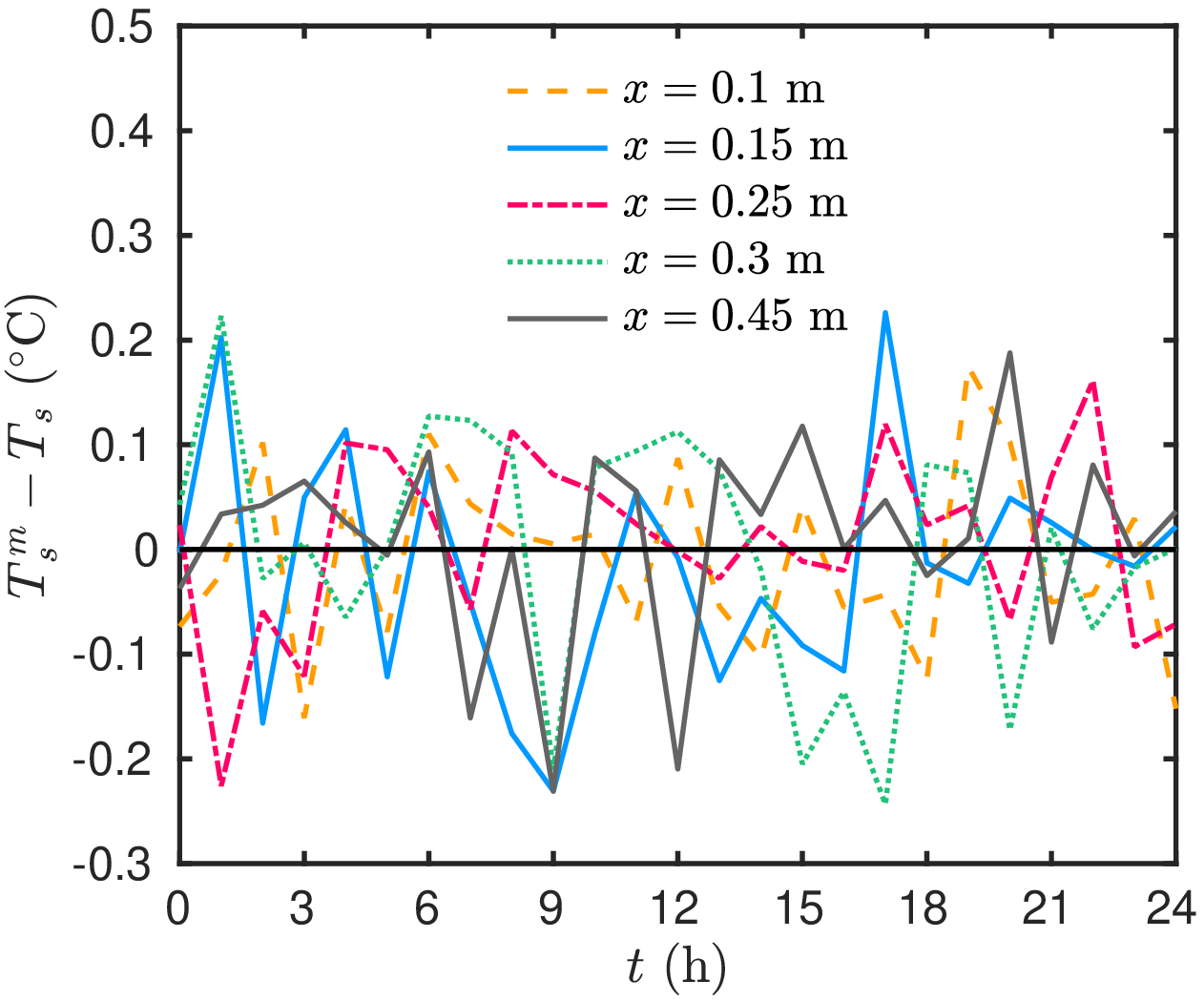}}  \hspace{0.2cm}
\subfigure[\label{fig:res_ROM_ft}]{\includegraphics[width=.45\textwidth]{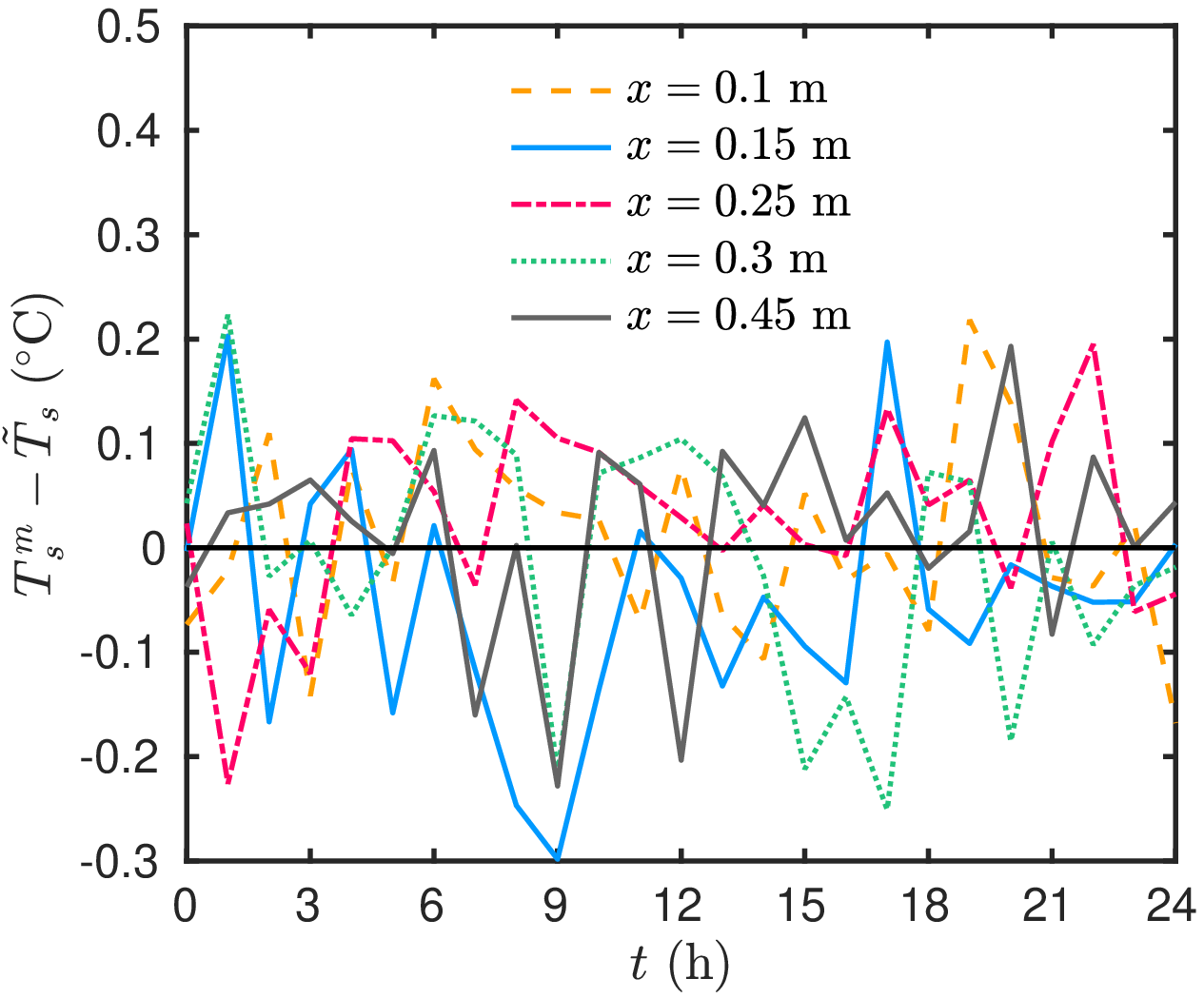}} 
\caption{Comparison of the (simulated) measurement and the model prediction using the estimated parameter for the LOM \emph{(a,c)} and the MIM ROM Order $5$ \emph{(b,d)}.}
\end{figure}

Both the LOM and the MIM ROM of order $5$ are used to solve the parameter estimation problem. The initial \revision{guess} is $\alpha^{\,\apriori} \egal 2.1 \ \cdot 10^{\,-7} \ \mathsf{m^{\,2}\,.\,s^{\,-1}}\,$, corresponding to a different value from those used during the learning step. The stopping criteria are set as $\eta_{\,1} \egal \eta_{\,2} \egal 10^{\,-14}$ and $N_{\,k} \egal 100\,$. The convergence of the algorithm to solve the parameter estimation problem is illustrated in Figures~\ref{fig:P_fiter} and \ref{fig:gamma2_fiter}. The algorithm searches the unknown parameter until the criteria $\gamma_{\,2}$ on the cost function is lower than $10^{\,-14}\,$. It requires $8$ and $10$ iterations when using the LOM and the ROM, respectively. A comparison of the model prediction with the estimated parameters against the measurement is shown in Figures~\ref{fig:YLOM_ft} and \ref{fig:YROM_ft}. A satisfying accuracy is observed for both models. The residuals have no specific signature as illustrated in Figures~\ref{fig:res_LOM_ft} and \ref{fig:res_ROM_ft}. Table~\ref{tab:pep_results} provides a synthesis of the results. The MIM ROM enables to cut the computational time to solve the inverse problem with a satisfying relative error of $4\%$ on the estimated parameter.

\begin{figure}[h!]
\centering
\subfigure[\label{fig:P_fiter_fkappa0}]{\includegraphics[width=.45\textwidth]{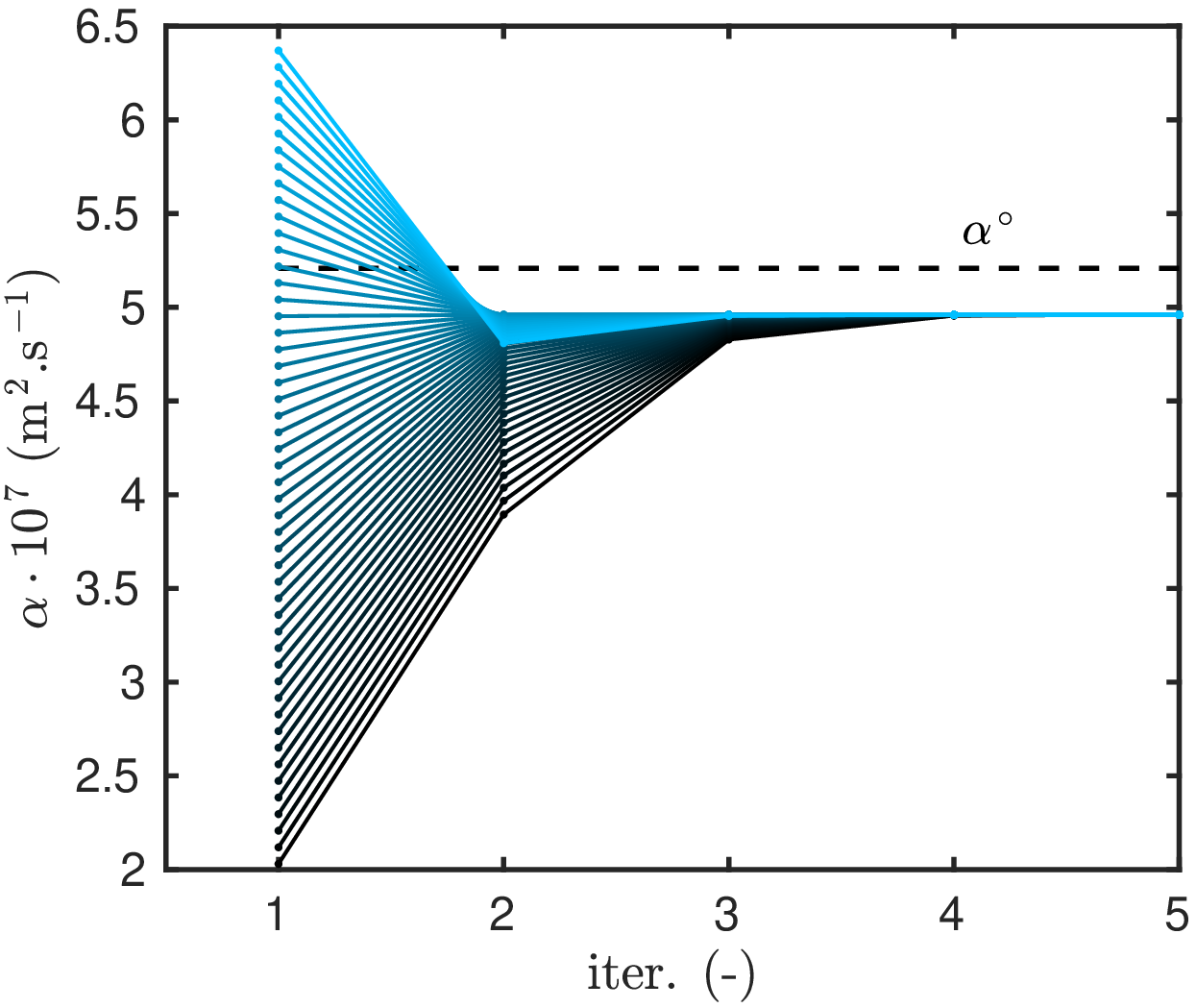}}  \hspace{0.2cm}
\subfigure[\label{fig:cpu_fkappa0}]{\includegraphics[width=.45\textwidth]{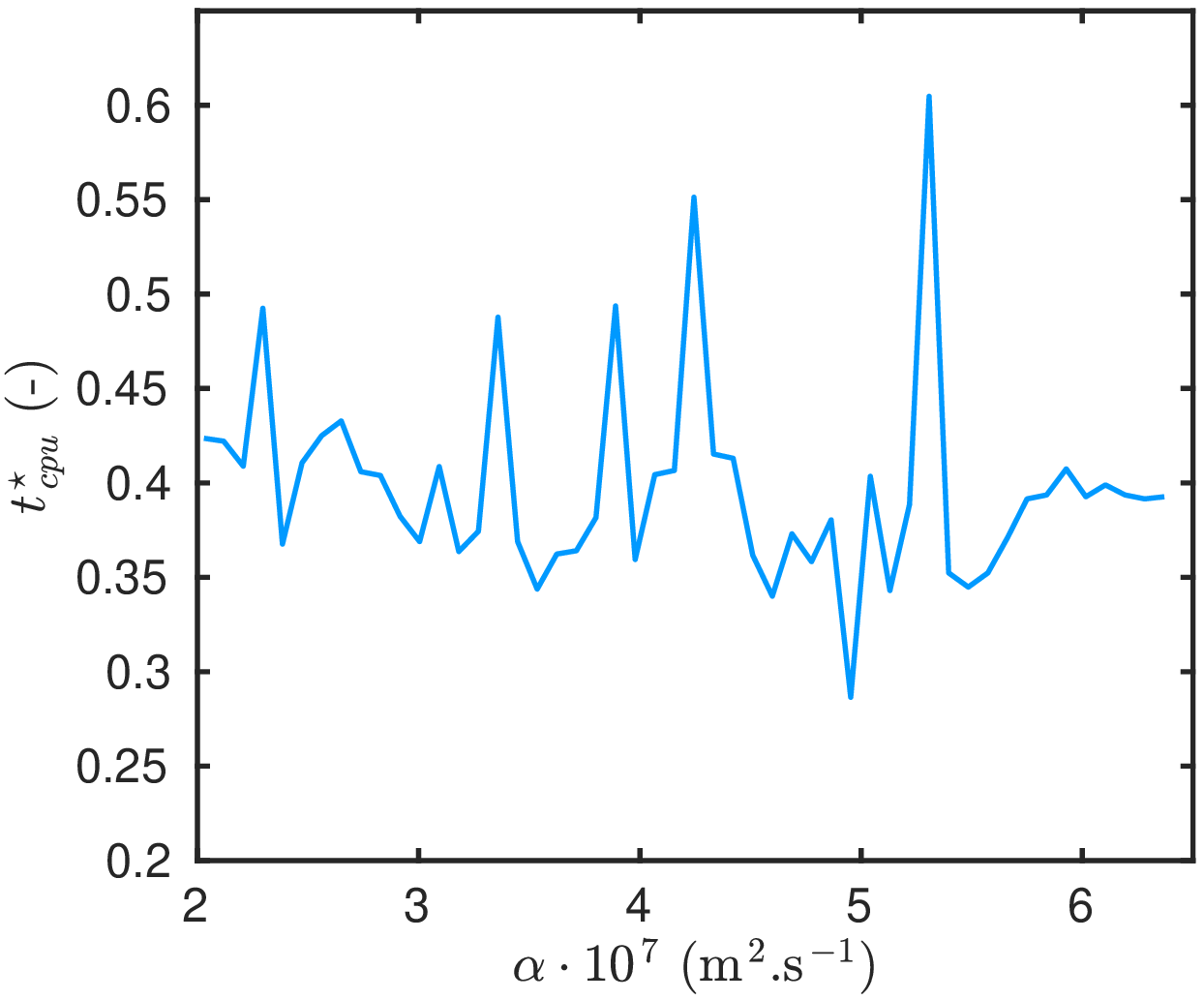}} 
\caption{Evolution of the estimated parameter according to the iteration \emph{(a)} and of the computational time ratio of the algorithm using the MIM ROM Order $5$ \emph{(b)} for $50$ values of initial diffusivity.}
\end{figure}

Further numerical tests are carried for $50$ values of thermal diffusivity. Figure~\ref{fig:P_fiter_fkappa0} shows the evolution of the estimated parameter using the MIM ROM of order $5\,$. The algorithm requires around $5$ iterations to estimate the unknown parameters with an relative accuracy of $4\%\,$. Figure~\ref{fig:cpu_fkappa0} gives the variation of the computational time ratio between the algorithm using the LOM and the MIM ROM. The use of the reduced order model enables to reduce the computational effort by $60\%\,$.

\begin{table}
\centering
\caption{Results of the parameter estimation problem.}
\label{tab:pep_results}
\setlength{\extrarowheight}{.5em}
\begin{tabular}[l]{@{} c ccccc c cc cc}
\hline
\hline

& \multicolumn{5}{c}{\textit{Point of observation} $x_{\,s}\ \unit{m}$}
&
&
&
& \multicolumn{2}{c}{\textit{Error $\varepsilon_{\,r} \ \unit{-}$ on}} \\

& $0.1$
& $0.15$
& $0.25$
& $0.3$
& $0.45$
& 
& \multicolumn{2}{c}{\textit{Computational time}}
& \multicolumn{2}{c}{\textit{parameter estimation}}  \\ 
\textit{Model}
& \multicolumn{5}{c}{\textit{Error $\varepsilon_{\,2}$}}
& \textit{Order}
& $t_{\,\mathrm{cpu}} \ \unit{s}$
& $t^{\,\star}_{\,\mathrm{cpu}} \ \unit{-}$
& $\alpha^{\,\apr}$
& $\alpha^{\,\circ}$ \\ \hline
LOM
& $0.059$
& $0.092$
& $0.073$
& $0.10$
& $0.073$
& $75$
& $4$
& $1$
& $-0.59$
& $-0.01$ \\
ROM
& $0.068$
& $0.11$
& $0.083$
& $0.10$
& $0.074$
& $5$
& $0.9$
& $0.23$
& $-0.59$
& $-0.04$ \\ \hline
& \multicolumn{5}{c}{\textit{Measurement uncertainty} $\unit{^{\,\circ}C}$} \\
& $0.94$
& $0.70$
& $0.46$
& $0.30$
& $0.13$ \\
\hline
\hline 
\end{tabular}
\end{table}

\section{Solving the parameter estimation problem for a real case study}
\label{sec:real_case}

Now the methodology illustrated in Figure~\ref{fig:methodo} is applied to a realistic case study presented in the next section. Then, the reduced sequence of three days is determined using the OED methodology. 

\subsection{Description of the case study}

\begin{figure}[h!]
\begin{center}
\subfigure[\label{fig:photo_int1}]{\includegraphics[width=.45\textwidth]{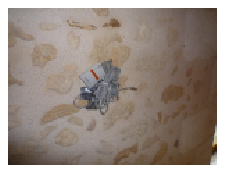}} \hspace{0.2cm}
\subfigure[]{\includegraphics[width=.45\textwidth]{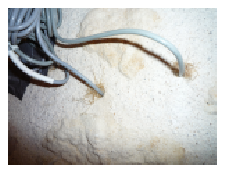}} \\
\subfigure[\label{fig:photo_ext}]{\includegraphics[width=.45\textwidth]{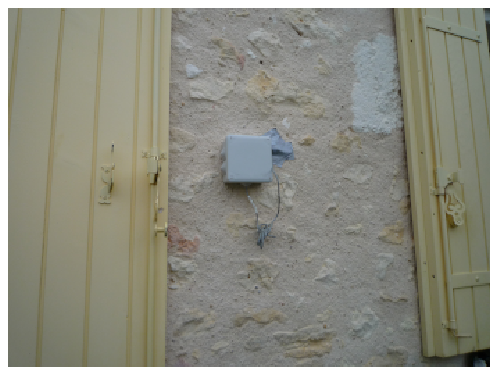}} \hspace{0.2cm}
\subfigure[\label{fig:design}]{\includegraphics[width=.45\textwidth]{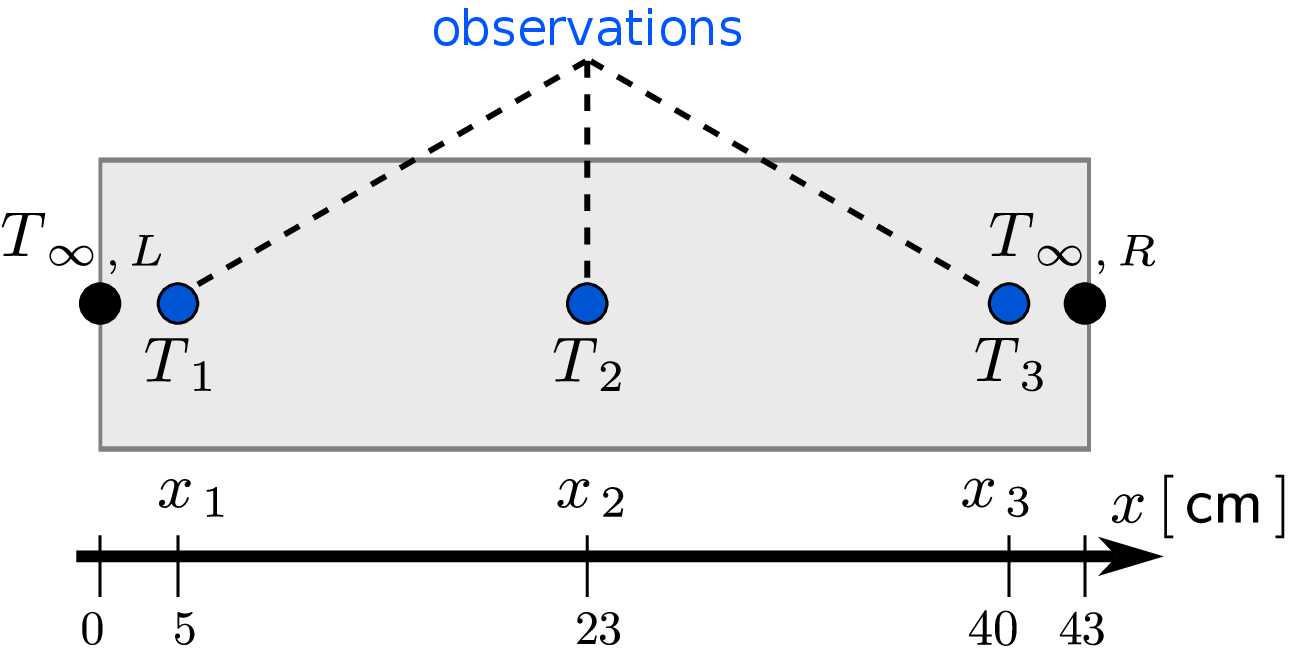}}
\caption{Illustration of the experimental design with picture of the inside sensors \emph{(a,b)}, the outside sensors \emph{(c)} and the experimental design \emph{(d)}.}
\end{center}
\end{figure}

The studied building is an old house, built at the end of the XIX$^{\,th}$ century and located in Saint Julien de Crempse, France. Further information on the architectural aspect of the building can be found in \cite{Cantin_2010}. The investigated wall is oriented East and has a length of $L \egal 43 \ \mathsf{cm} \,$. \revision{The wall is composed of a mixture of lime stones and clay mortar as shown in the picture Figure~\ref{fig:photo_ext}. Due to the vernacular traditional architecture, the stones arises from local around the house.} There is no information on the exact position of the stones. \revision{The issue is to estimate the global thermal diffusivity $\alpha $ of the wall by assuming an equivalent homogenous model \cite{Gori_2018}.} Note that the structural identifiability of this parameter has been demonstrated in \cite{Berger_2019}. For this aim, the wall are monitored using five intrusive calibrated sensors HOBO TMC-$6$-HA as shown in Figures~\ref{fig:photo_int1} to \ref{fig:photo_ext}. Three sensors are installed inside the wall by obliquely drilling a whole of $13 \ \mathsf{mm}$. In the hole, the contact between the material and the sensor is ensured by a thermal conductive paste. The hole is then filled using an insulation material. The sensor located at the inside and outside surfaces are protected by insulation materials from incident radiation. The location of the sensor is set to assume that the heat transfer occurs mainly from the outside to the inside. An illustration of the experimental design is given in Figure~\ref{fig:design}.

\begin{figure}[h!]
\centering
\subfigure[\label{fig:TLR_ft}]{\includegraphics[width=.9\textwidth]{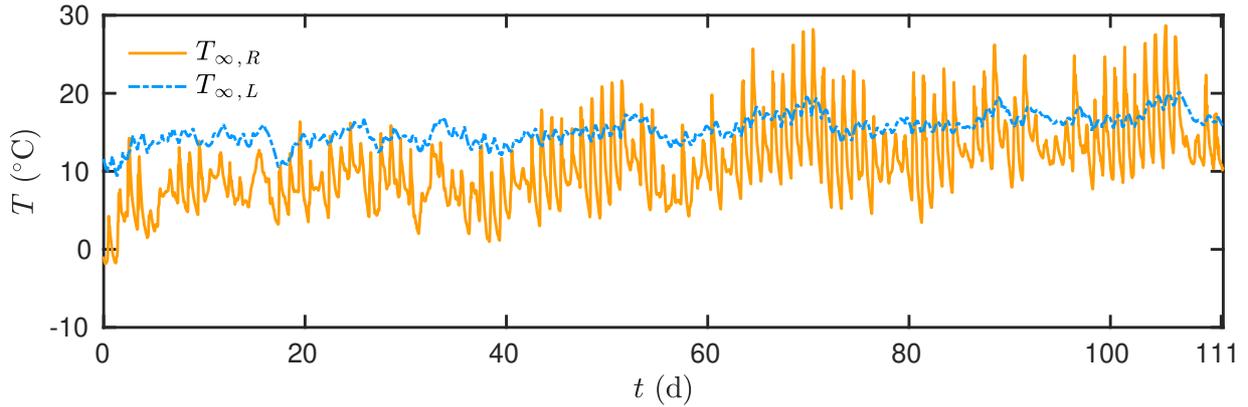}} \\
\subfigure[\label{fig:Tmeas_ft}]{\includegraphics[width=.9\textwidth]{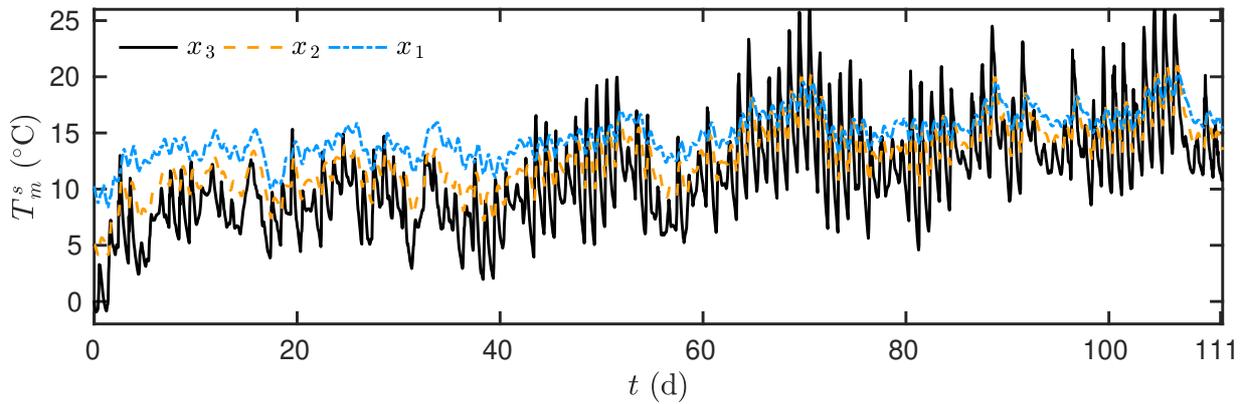}} 
\caption{Time variations of the boundary conditions \emph{(a)} and the measured observation data \emph{(b)}. For the sake of clarity, the measurement uncertainty is not presented.}
\end{figure}

\begin{figure}[h!]
\centering
\subfigure[\label{fig:pdf_sigM}]{\includegraphics[width=.45\textwidth]{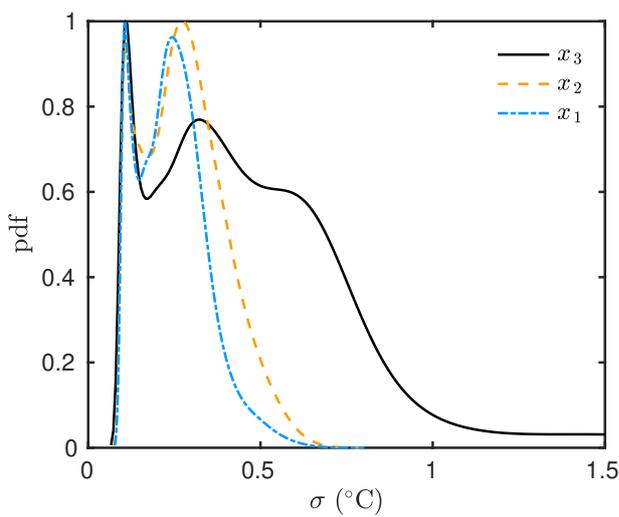}} \hspace{0.2cm}
\subfigure[\label{fig:Tini_fx}]{\includegraphics[width=.45\textwidth]{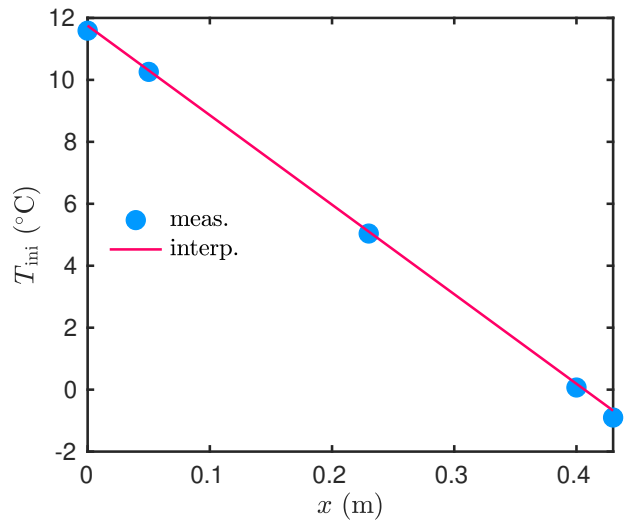}} 
\caption{Probability density function of the uncertainty measurement of the observations \emph{(a)} and initial condition of the problem \emph{(b)}.}
\end{figure}

The monitoring is carried out from January until April 2009. Experimental measures of temperature are available at $x \egal \bigl\{\,5 \,,\, 23 \,,\, 40 \,\bigr\} \ \mathsf{cm} \,$, as illustrated in Figure~\ref{fig:Tmeas_ft}. Thus, $N_{\,s} \egal 3$ are hold. It is not known if the sensors are located in the stone or in the mortar. The temperature $ T_{\,\infty\,,\,L}$ and $ T_{\,\infty\,,\,R}$ are also measured and given in Figure~\ref{fig:TLR_ft}. Measures are available for $\Omega_{\,t} \egal \bigl[\, 0 \,,\, 2670 \,\bigr] \ \mathsf{h} \,$, corresponding to four months. The measurement time step is $1 \ \mathsf{h}\,$. The experimental uncertainties $\sigma$is computed using \cite{Taylor_1970}:
\begin{align*}
\sigma \egal \sqrt{\sigma_{\,m}^{\,2} \plus \sigma_{\,x}^{\,2} } \,,
\end{align*}
where $\sigma_{\,m} \egal 0.1 \ \mathsf{^{\,\circ}C}$ is the sensor measurement uncertainty and $\sigma_{\,x}$ is the sensor position uncertainty. Due to experimental design based on drilling a large wall, the latter can be important. It is computed according to:
\begin{align*}
\sigma_{\,x} \egal \pd{T}{x}\,\biggl|_{\,x \egal x_{\,s}} \ \delta_{\,x} \,,
\end{align*}
where $\delta_{\,x} \egal 2 \ \mathsf{cm}\,$ is the uncertainty on the sensor position. The partial derivative $\pd{T}{x}$ is computed using the numerical model. Figure~\ref{fig:pdf_sigM} shows the probability of the measurement uncertainty at the three points of observation. The uncertainty increases for the sensor located near the outside air conditions. For the initial conditions, a first order polynomial is fitted using the experiment:
\begin{align*}
T_{\,\ini} \,:\, x \,\longmapsto \, T_{\,0} \cdot \biggl(\, 1 \plus \frac{x}{\ell_{\,1}}  \,\biggr)  \,,
\end{align*}
where $T_{\,0} \egal 11.75 \ \mathsf{^{\,\circ}C}$ and $\ell_{\,1} \egal -0.406 \ \mathsf{m} \,$. As shown in Figure~\ref{fig:Tini_fx}, the error fitting is satisfactory with $\varepsilon_{\,2} \egal 0.09 \ \mathsf{^{\,\circ}C}^{\,2}\,$. The \emph{a priori} knowledge from standards gives $\alpha^{\,\apr} \egal 2.8 \cdot 10^{\,-7} \ \mathsf{m^{\,2}\,.\,s^{\,-1}}\,$. \revision{In addition, Table~\ref{tab:wall_description} gives the diffusivity of the different elements composing the wall.}

\begin{table}
\centering
\caption{\revision{Wall stratigraphy.}}
\label{tab:wall_description}
\setlength{\extrarowheight}{.5em}
\begin{tabular}[l]{@{} c c c}
\hline
\hline
\textit{Layer}
& \textit{Thermal diffusivity } $\unit{,m^{\,-2} \,.\, s^{\,-1}}$
& \textit{Thickness} $\unit{cm}$ \\
\hline
Lime stones & $2.7 \cdot 10^{\,-7} $ & $\mathcal{O}(\,20\,)$\\
Mortar & $5.5 \cdot 10^{\,-7} $ & $\mathcal{O}(\,4\,)$\\
Equivalent homogeneous wall & $2.8 \cdot 10^{\,-7} $ & $43$ \\
\hline
\hline
\end{tabular}
\end{table}

\subsection{Determining the reduced sequence of observations using the OED methodology}

\begin{figure}[h!]
\centering
\subfigure[\label{fig:teta_ft}]{\includegraphics[width=.9\textwidth]{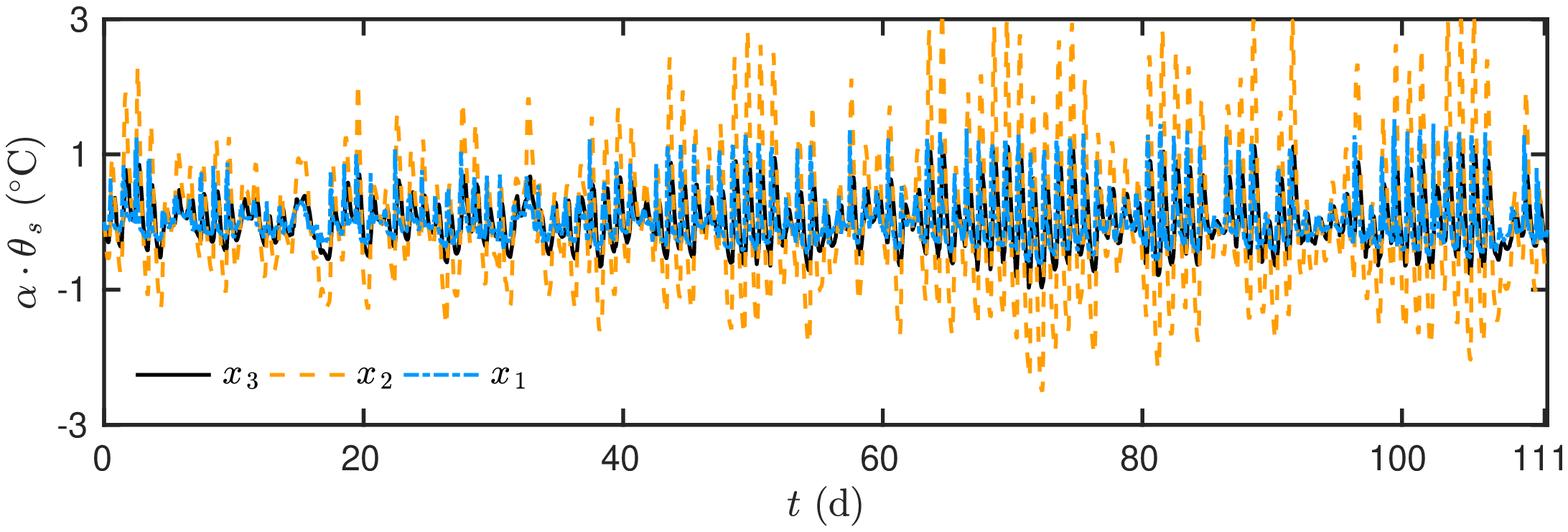}} \\
\subfigure[\label{fig:Psi_fseq}]{\includegraphics[width=.9\textwidth]{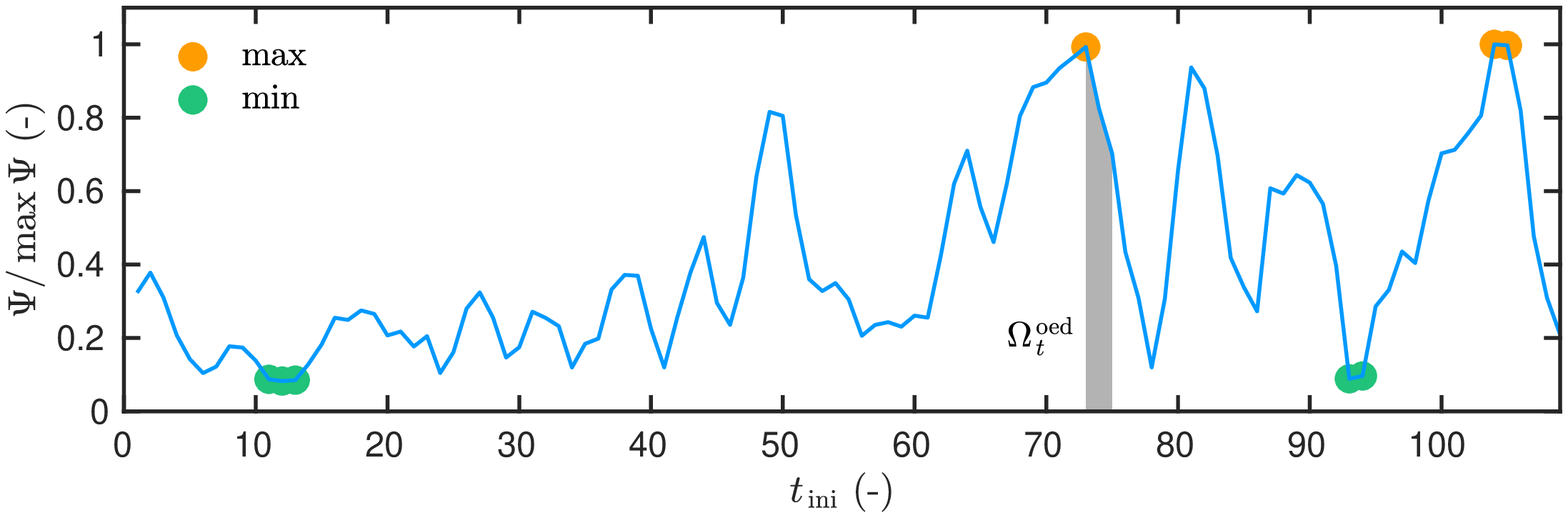}} 
\caption{Time variations of the sensitivity functions of the problem \emph{(a)} and variation of the OED criteria according to the choice of the initial time of the reduced sequence $\Omega_{\,t}^{\,\oed} $ \emph{(b)}.}
\end{figure}

The OED methodology is now carried out to determine the optimal reduced sequence of three days $\Omega_{\,t}^{\,\oed}$ of observations. The sensitivity function at the point of observations $x_{\,s}\,, s \, \in \, \bigl\{\, 1 \,,\,2 \,,\,3 \,\bigr\}$ are computed and shown in Figure~\ref{fig:teta_ft}. Then, the criteria $\Psi$ is evaluated for each sequence of three days in $\Omega_{\,t}\,$. Figure~\ref{fig:Psi_fseq} shows the variation of the dimensionless criteria according to the choice of $\tini\,$. By comparing Figure~\ref{fig:teta_ft} and \ref{fig:Psi_fseq}, it can be remarked that the optimal sequences occurs when the sensitivity of the parameter have high magnitudes of variations. The criteria verifies $\Psi \, \geqslant \, 0.98$ for three initial days. According to these results, the following reduced sequence is adopted for the construction of the MIM ROM and then the estimation of the unknown parameters: 
\begin{align*}
\Omega_{\,t}^{\,\oed} \egal \bigl[\, 72 \,,\, 75\,\bigr] \ \mathsf{d} \,.
\end{align*}

\begin{figure}[h!]
\centering
\includegraphics[width=.45\textwidth]{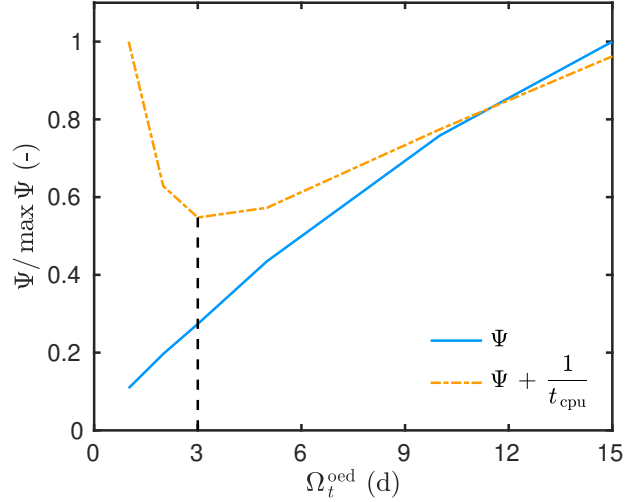}
\caption{\revision{Variation of the OED criteria according to the length of the reduced sequence $\Omega_{\,t}^{\,\mathrm{oed}}$ and the computational time of the MIM ROM.}}
\label{fig:Psi_fseq_fcpu}
\end{figure}

\revision{The chosen monitoring time of $3$ days arises from a compromise among several opposite criteria. Indeed, it could be reduced compared to traditional methods. From an pure inverse problem point of view, the longest is the measurement period, the highest is the accuracy of the parameter estimation. This can be proven by looking at the variation of the OED criteria $\Psi$ according to the length of the sequence in Figure~\ref{fig:Psi_fseq_fcpu}. The criteria is monotonously increasing according to $\Omega_{\,t}^{\,\mathrm{oed}}\,$. However, one has to deal with experimental constraints such as the acceptation of the building occupants. Furthermore, the longer is the measurement period, the higher is the computational time of the direct model and as a consequence the inverse problem algorithm. Figure~\ref{fig:Psi_fseq_fcpu} shows the variation the criteria $\Psi$ added to the inverse of the computational cost of the algorithm. For this case, a good compromise is in three days of monitoring. Note that the influence of the length of the measurement period is investigated empirically in \cite{Rodler_2019}. For their case study, the authors suggest to carry at least three days of monitoring  to have an accurate estimation.}

\subsection{Building the MIM ROM}

\begin{figure}[h!]
\begin{center}
\subfigure[\label{fig:BCOED_ft}]{\includegraphics[width=.45\textwidth]{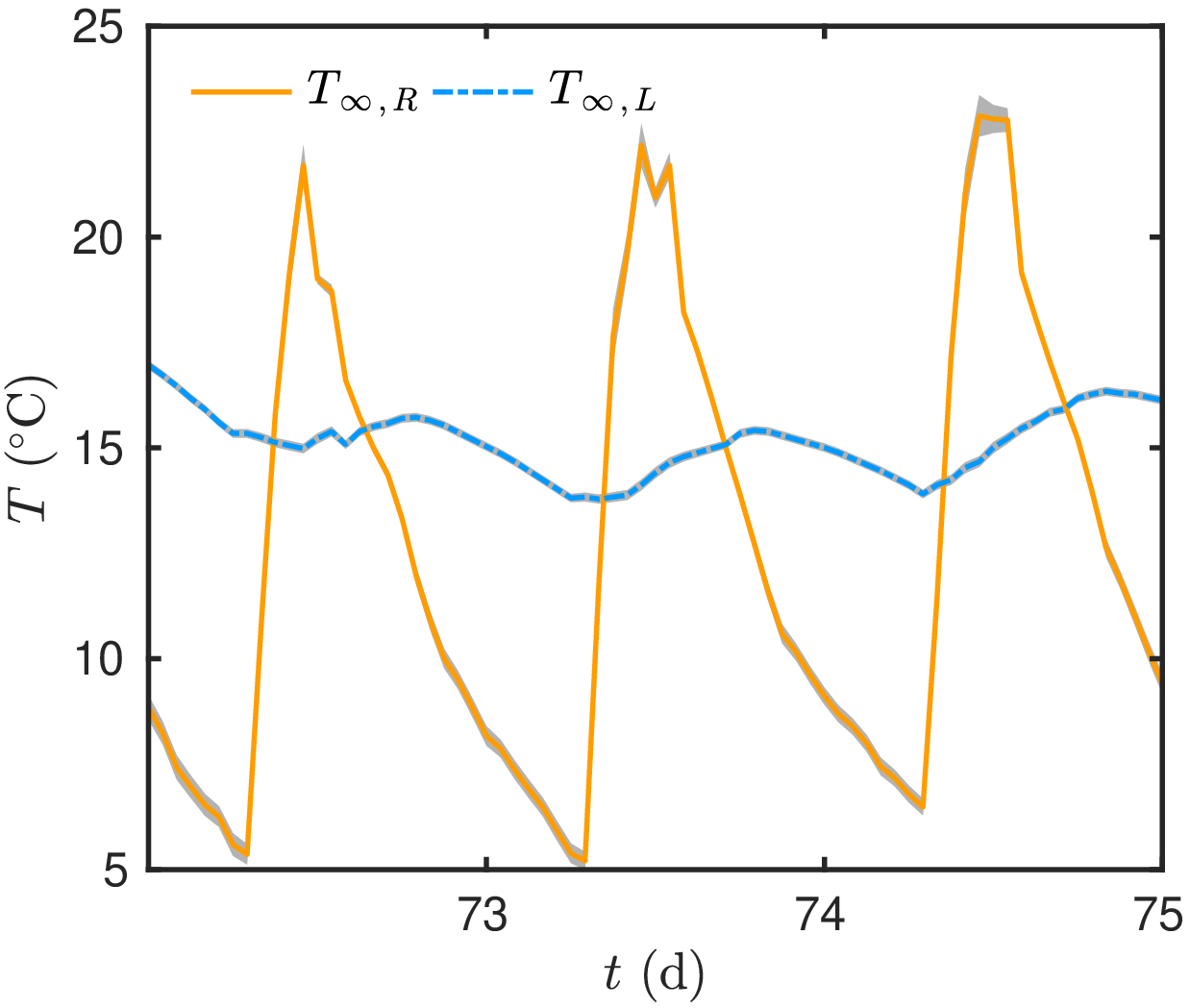}} \hspace{0.2cm}
\subfigure[\label{fig:TiniOED_fx}]{\includegraphics[width=.45\textwidth]{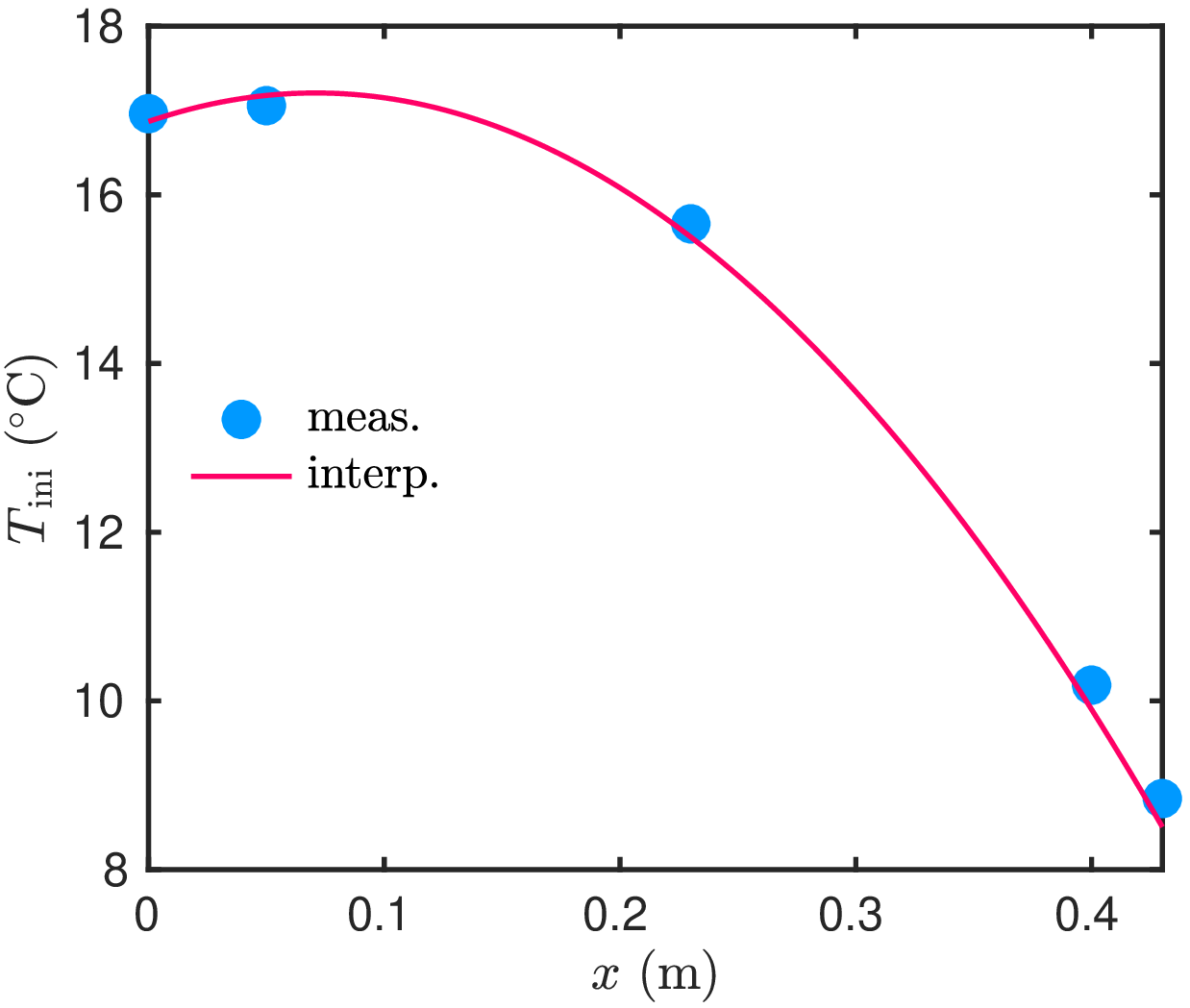}} 
\subfigure[\label{fig:TOED_ft}]{\includegraphics[width=.45\textwidth]{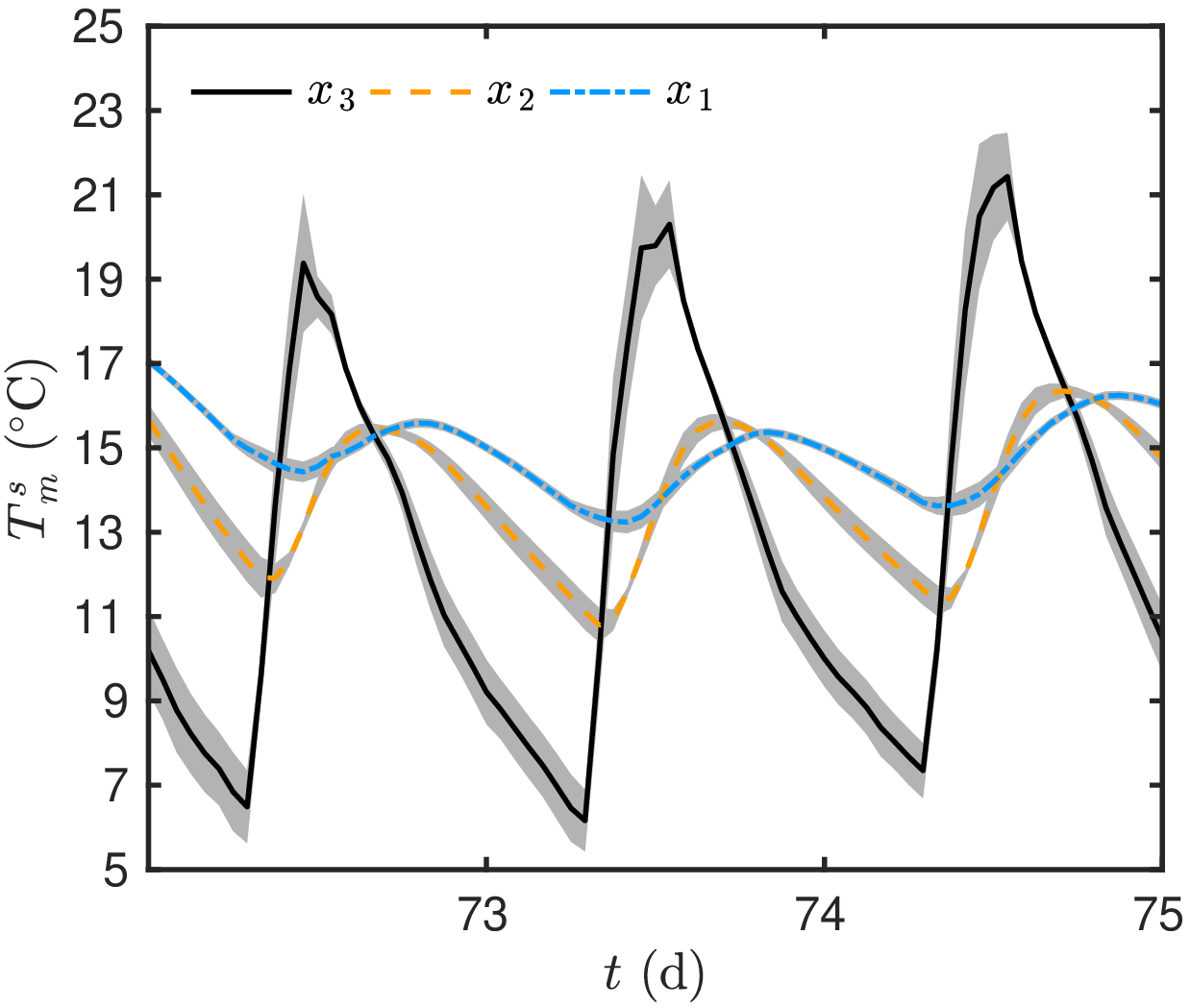}} \hspace{0.2cm}
\subfigure[\label{fig:tetaOED_ft}]{\includegraphics[width=.45\textwidth]{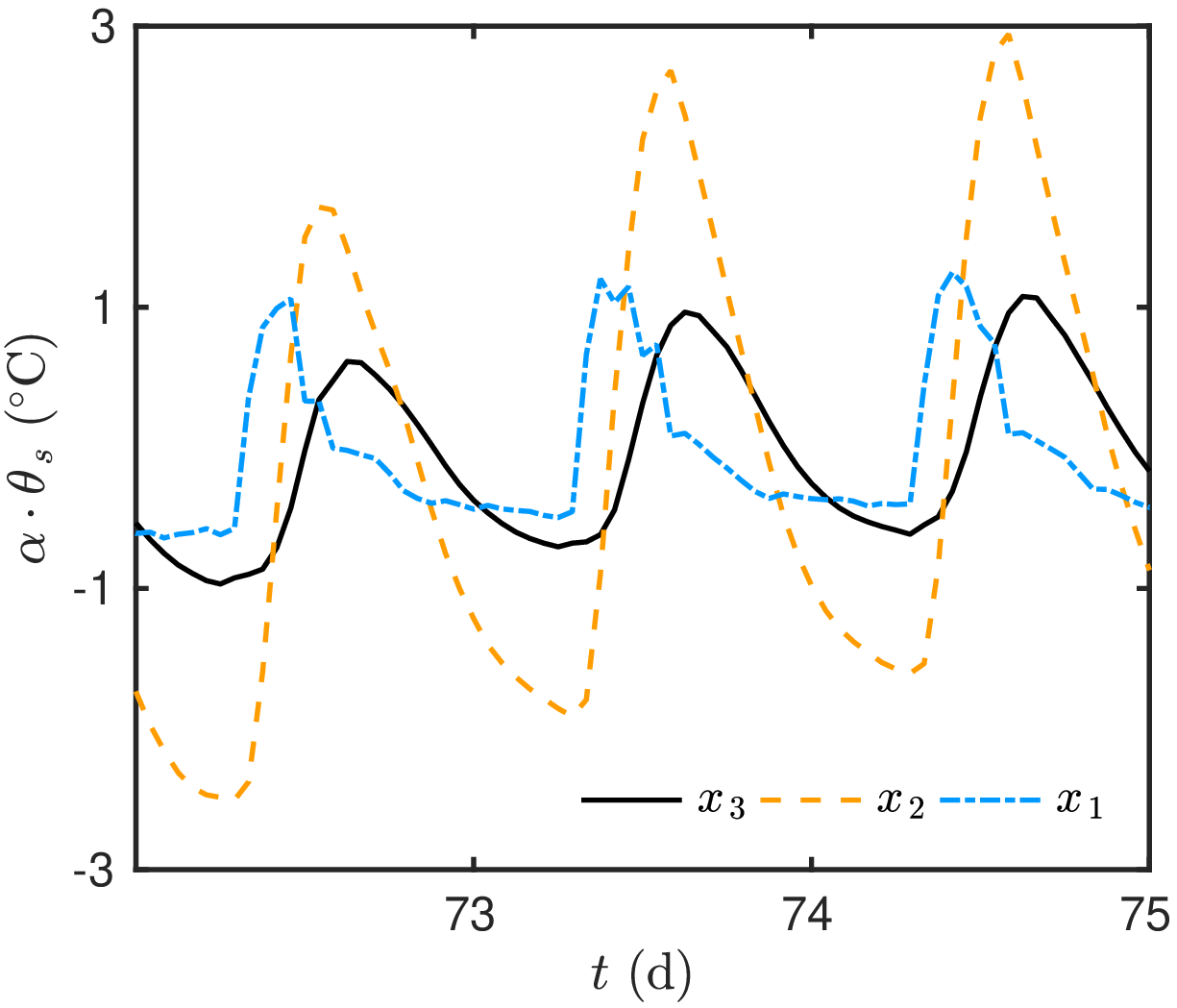}} 
\caption{\revision{Time variations of the boundary conditions \emph{(a)} and and initial condition of the problem \emph{(b)} for the reduced sequence $\Omega_{\,t}^{\,\oed}\,$. Time variation of the measured observation data \emph{(c)} and the sensitivity functions \emph{(d)}. The grey shadow corresponds to the measurement uncertainty.}}
\end{center}
\end{figure}

\begin{figure}[h!]
\centering
\subfigure[\label{fig:case_e2_fN}]{\includegraphics[width=.45\textwidth]{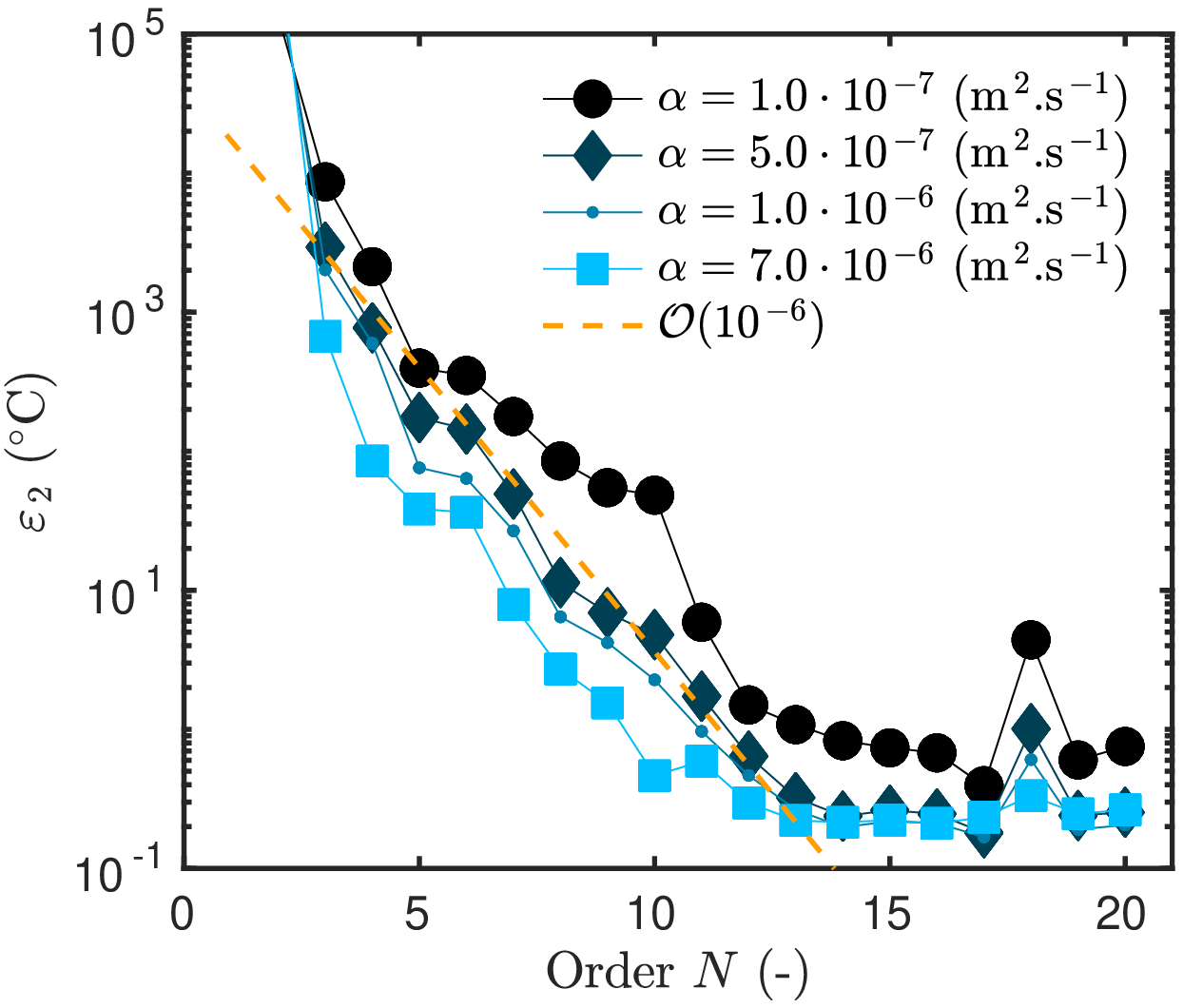}} \hspace{0.2cm}
\subfigure[\label{fig:case_einf_fN}]{\includegraphics[width=.45\textwidth]{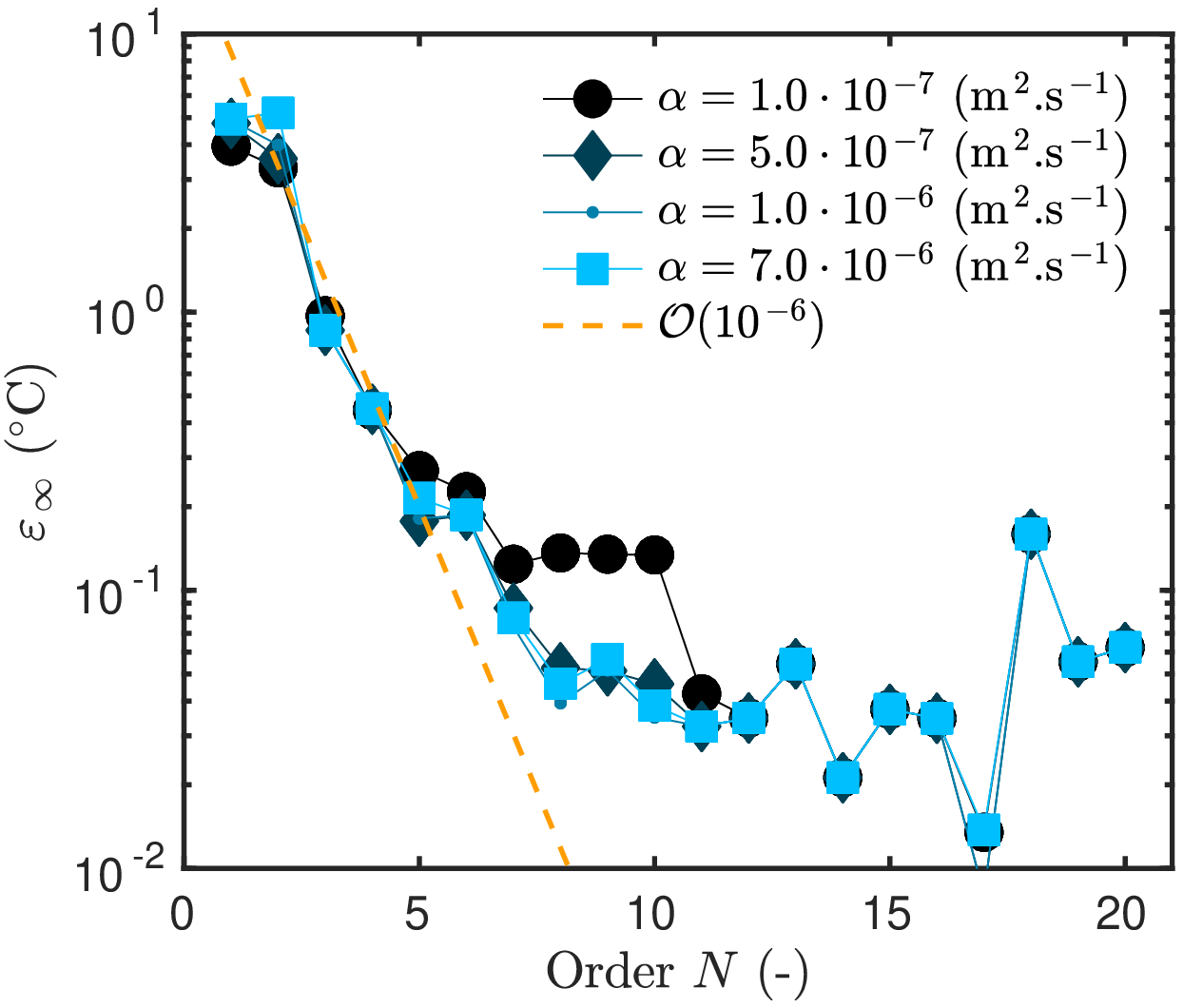}}
\caption{Evolution of the error of the MIM ROM model according to the order $N$ trained over the reduced sequence $\Omega_{\,t}^{\,\oed}\,$.}
\end{figure}

For the reduced sequence $\Omega_{\,t}^{\,\oed}\,$, the boundary conditions are given in Figure~\ref{fig:BCOED_ft}. The initial condition is the following:
\begin{align*}
T_{\,\ini} \,:\, x \,\longmapsto \, T_{\,0} \cdot \Biggl(\, 1 \plus \frac{x}{\ell_{\,1}} \moins \biggl(\, \frac{x}{\ell_{\,2}} \,\biggr)^{\,2} \,\Biggr)  \,,
\end{align*}
where $T_{\,0} \egal 16.87 \ \mathsf{^{\,\circ}C}\,$, $\ell_{\,1} \egal 1.764 \ \mathsf{m}$ and $\ell_{\,2} \egal 0.5 \ \mathsf{m} \,$. The error fitting of the initial condition is satisfactory ($\varepsilon_{\,2} \egal 0.24 \ \mathsf{^{\,\circ}C}^{\,2}$) as shown in Figure~\ref{fig:TiniOED_fx}. The learning step is carrying considering four values of diffusivity $\Omega_{\,\alpha} \egal \bigl\{\, 1 \,,\, 5 \,,\, 10 \,,\, 70  \,\bigr\} \cdot 10^{\,-7} \ \mathsf{m^{\,2} \,.\,s^{\,-1}}\,$. The training is carried until the order $N \egal 20$ taking around $t_{\,\mathrm{cpu}} \egal 3 \ \mathsf{d}\,$. Figures~\ref{fig:case_e2_fN} and \ref{fig:case_einf_fN} shows the evolution of the error according to the order of the model. It can be remarked that a model of order $N \egal 10$ provides a satisfactory accuracy. 

\subsection{Parameter estimation}

\begin{figure}[h!]
\centering
\subfigure[\label{fig:P_fiter_est}]{\includegraphics[width=.45\textwidth]{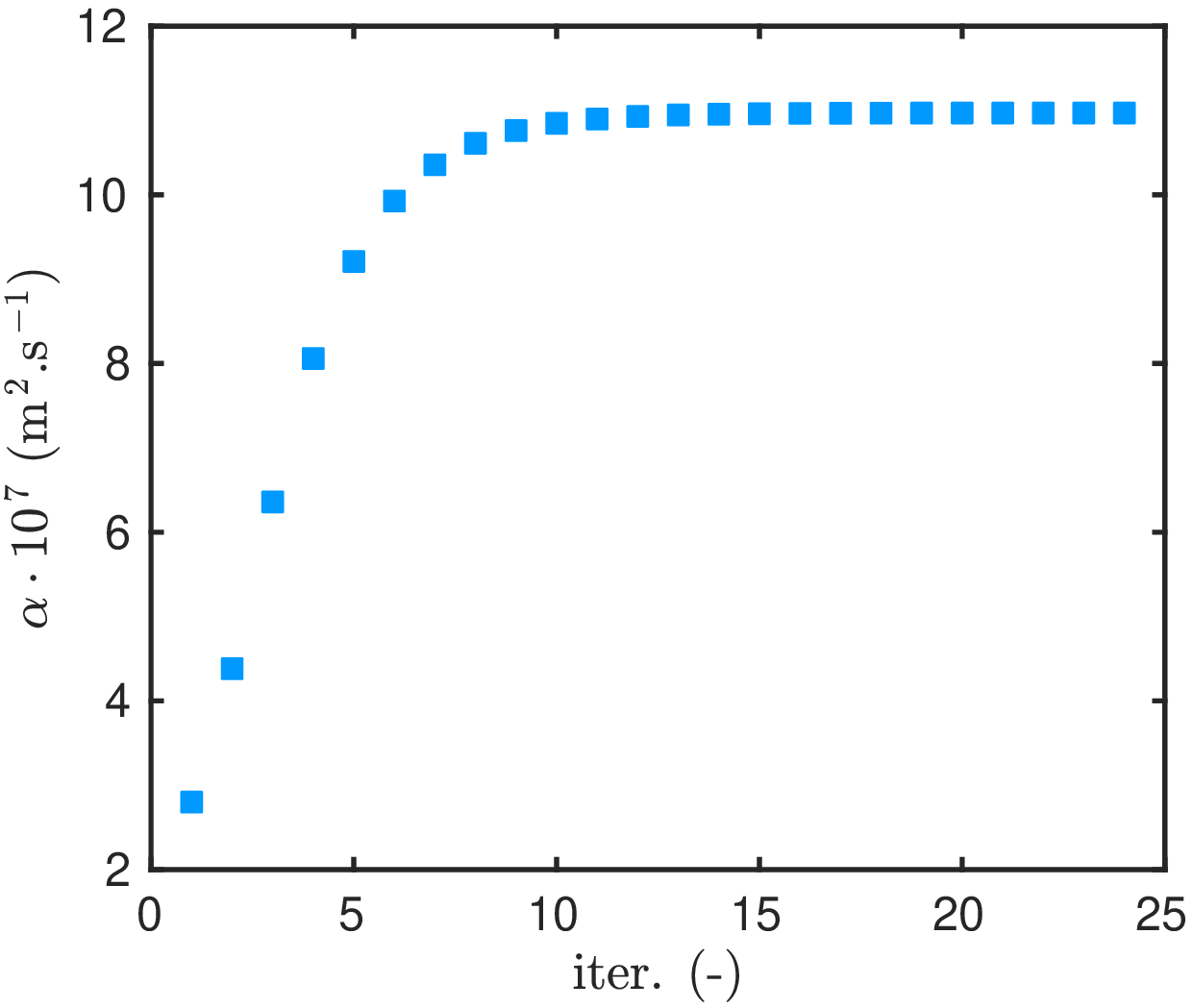}} \hspace{0.2cm}
\subfigure[\label{fig:gamma_fiter}]{\includegraphics[width=.45\textwidth]{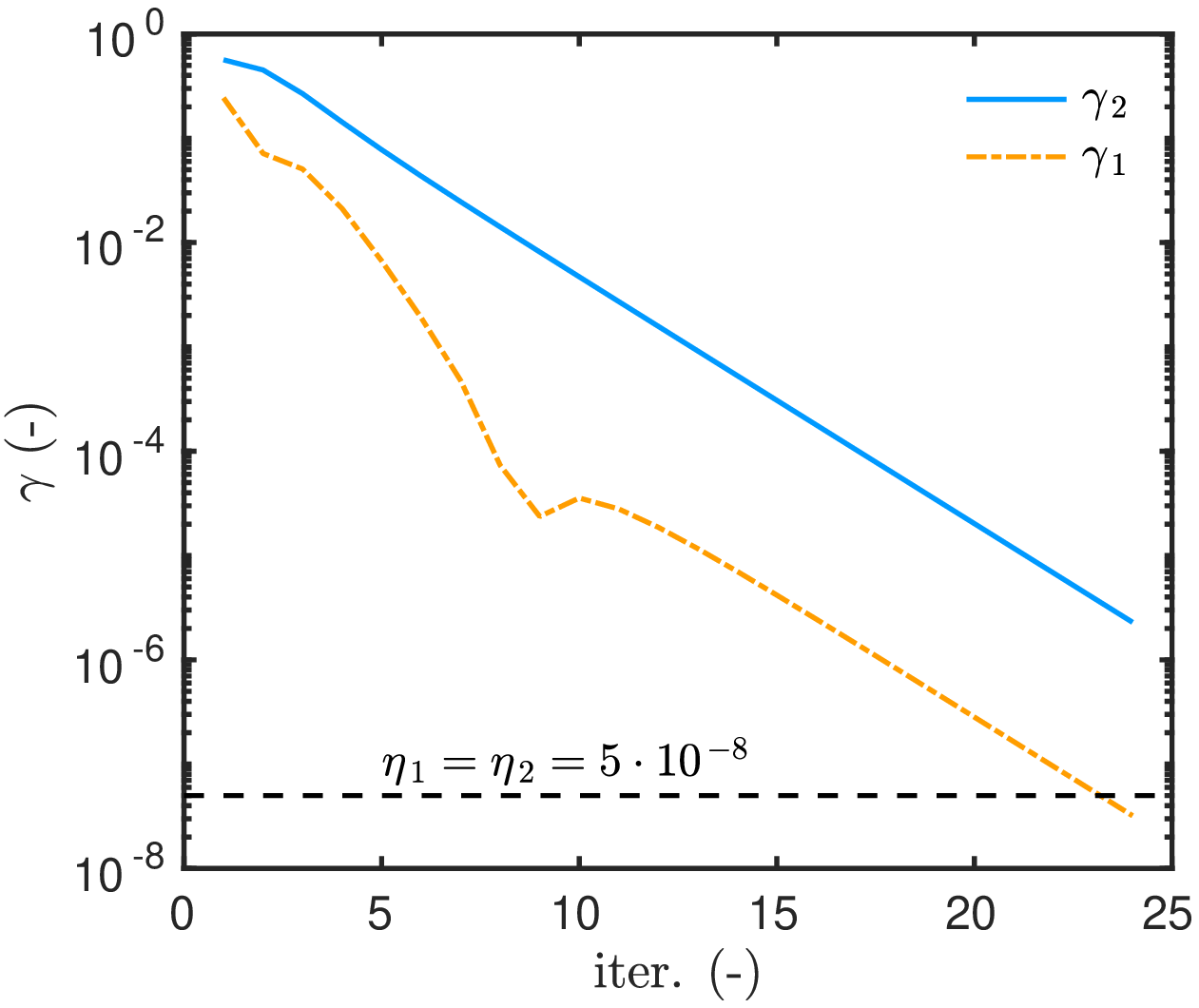}}
\caption{Evolution of the estimated parameter $\alpha$ \emph{(a)} and of the convergence criteria \emph{(b)} according to the iteration number of the algorithm.}
\end{figure}

After the learning step the MIM ROM matrices are known. Thus, the model is now used to estimate the thermal diffusivity of the material. Figure~\ref{fig:tetaOED_ft} shows the variation of the sensitivity functions using the \emph{a priori} value of the diffusivity. The functions have high magnitude of variations ensuring the practical identifiablity of the unknown parameter. The \textsc{Gau}\ss ~algorithm can be employed for the parameter estimation. It uses the MIM ROM of order $N \egal 10$ over the reduced sequence $\Omega_{\,t}^{\,\oed}\,$. The observations are shown in Figures~\ref{fig:TOED_ft}. Figure~\ref{fig:P_fiter_est} shows the evolution of the estimated parameter. The algorithm requires $24$ iterations to satisfy the convergence criteria, which evolution is illustrated in Figure~\ref{fig:gamma_fiter}. In approximately $10$ iterations, the algorithm already founds a trusting value of the thermal diffusivity. Note that the stability of the estimation have been verified by providing several values of initial guess in the algorithm. As presented in Table~\ref{tab:case_pep_results}, the value of the estimated parameter $\alpha^{\,\circ}$ remains stable from $N \egal 7\,$. The computational time to solve the inverse problem is cut by $5$ compared to the LOM. Slightly differences in the computational time are remarked for the MIM ROM due to the number of iterations to converge. The model of order $20$ is more accurate and requires less iterations to retrieve the unknown parameter. \revision{It is important to note that the experimental design relies on embedded sensors inside the wall to obtain the experimental observations. This may be a restriction of the methodology for other case study. Other experimental devices could be used such as heat flux sensor or infrared camera. The methodology to search the optimal experimental design could be extended for such approaches. Furthermore, a MIM ROM needs to be constructed according to the new type of observable field. }

\begin{table}
\centering
\caption{Results of the parameter estimation problem, noting that $\alpha^{\,\apr} \egal 2.8 \e{-6} \ \mathsf{m^{\,2}\,.\,s^{\,-1}}$.}
\label{tab:case_pep_results}
\setlength{\extrarowheight}{.5em}
\begin{tabular}[l]{@{} c c c cc c}
\hline
\hline

& 
& \textit{Estimated parameter} 
& \multicolumn{2}{c}{\textit{Computational time}} 
& \textit{Iterations to} \\ 
\textit{Model}
& \textit{Order}
& $\alpha^{\,\circ} \cdot 10^{\,7} \ \unit{m^{\,2}\,.\,s^{\,-1}}$
& $t_{\,\mathrm{cpu}} \ \unit{s}$
& $t^{\,\star}_{\,\mathrm{cpu}} \ \unit{-}$ 
& \textit{to converged} \\ 
\hline 
LOM
& $100$
& $1.064$ 
& $572$
& $1$
& $19$ \\
ROM
& $5$
& $1.097$
& $114$
& $0.2$ 
& $24$ \\
ROM
& $7$
& $1.064$
& $112$
& $0.19$ 
& $23$ \\
ROM
& $10$
& $1.062$
& $101$
& $0.17$ 
& $21$ \\
ROM
& $15$
& $1.065$
& $102$
& $0.18$ 
& $21$ \\
ROM
& $20$
& $1.067$
& $94$
& $0.16$
& $19$  \\
\hline
\hline 
\end{tabular}
\end{table}

\subsection{Reliability of the model}

\begin{figure}[h!]
\centering
\subfigure[$x_{\,1}$ \label{fig:YL1_ft}]{\includegraphics[width=.45\textwidth]{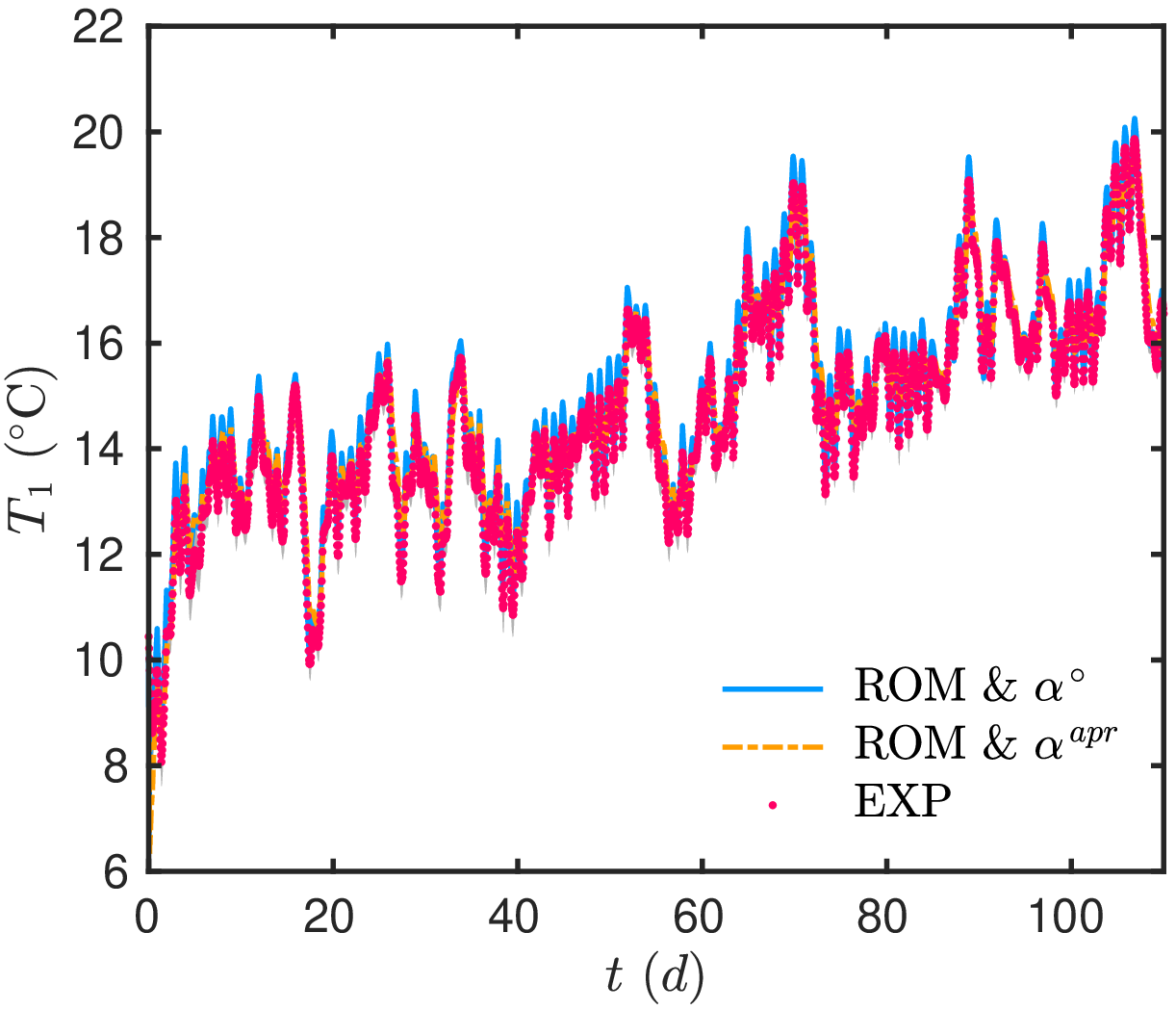}} \hspace{0.2cm}
\subfigure[$x_{\,1}$ \label{fig:pdfL1}]{\includegraphics[width=.45\textwidth]{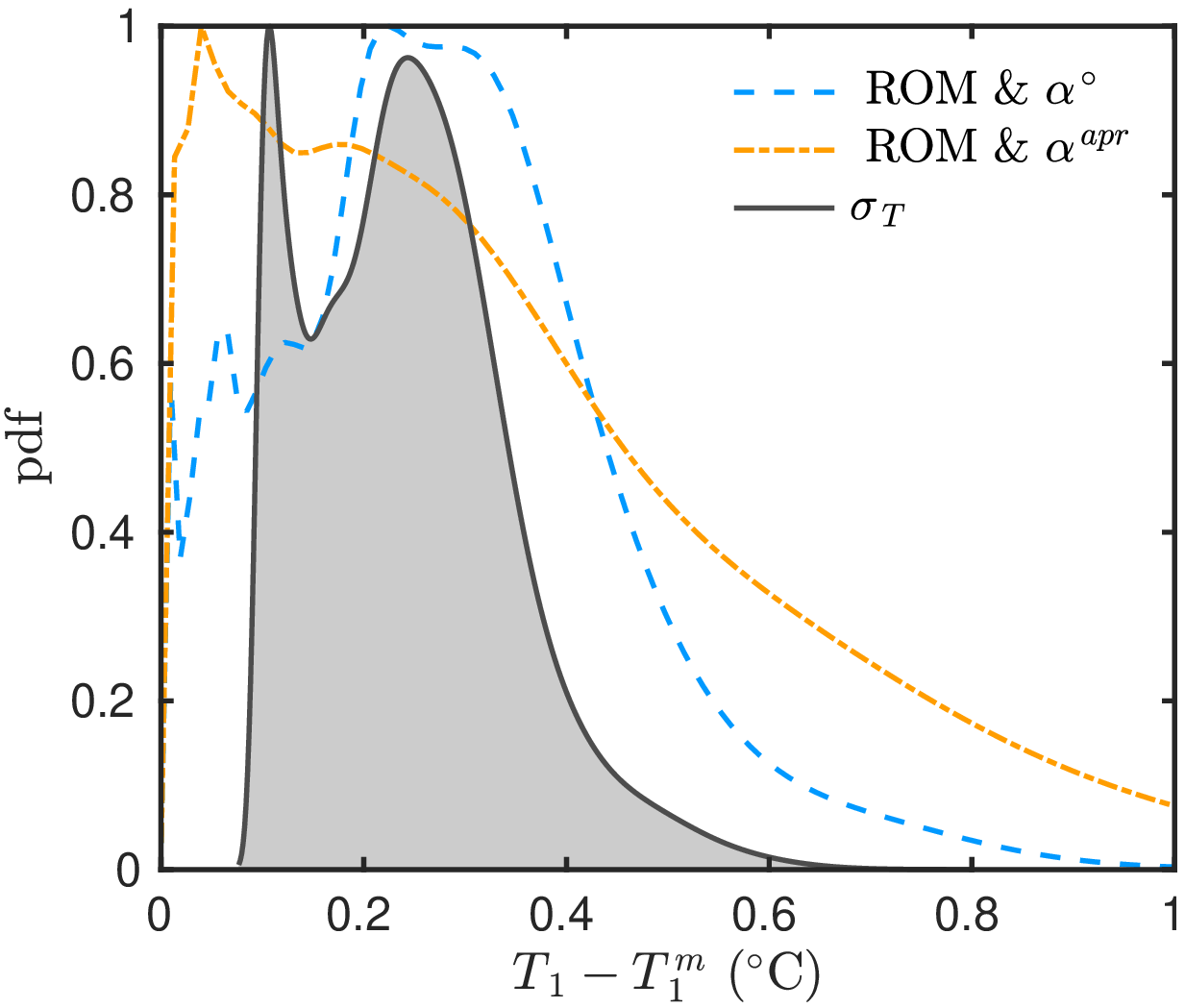}} \\
\subfigure[$x_{\,2}$ \label{fig:YL2_ft}]{\includegraphics[width=.45\textwidth]{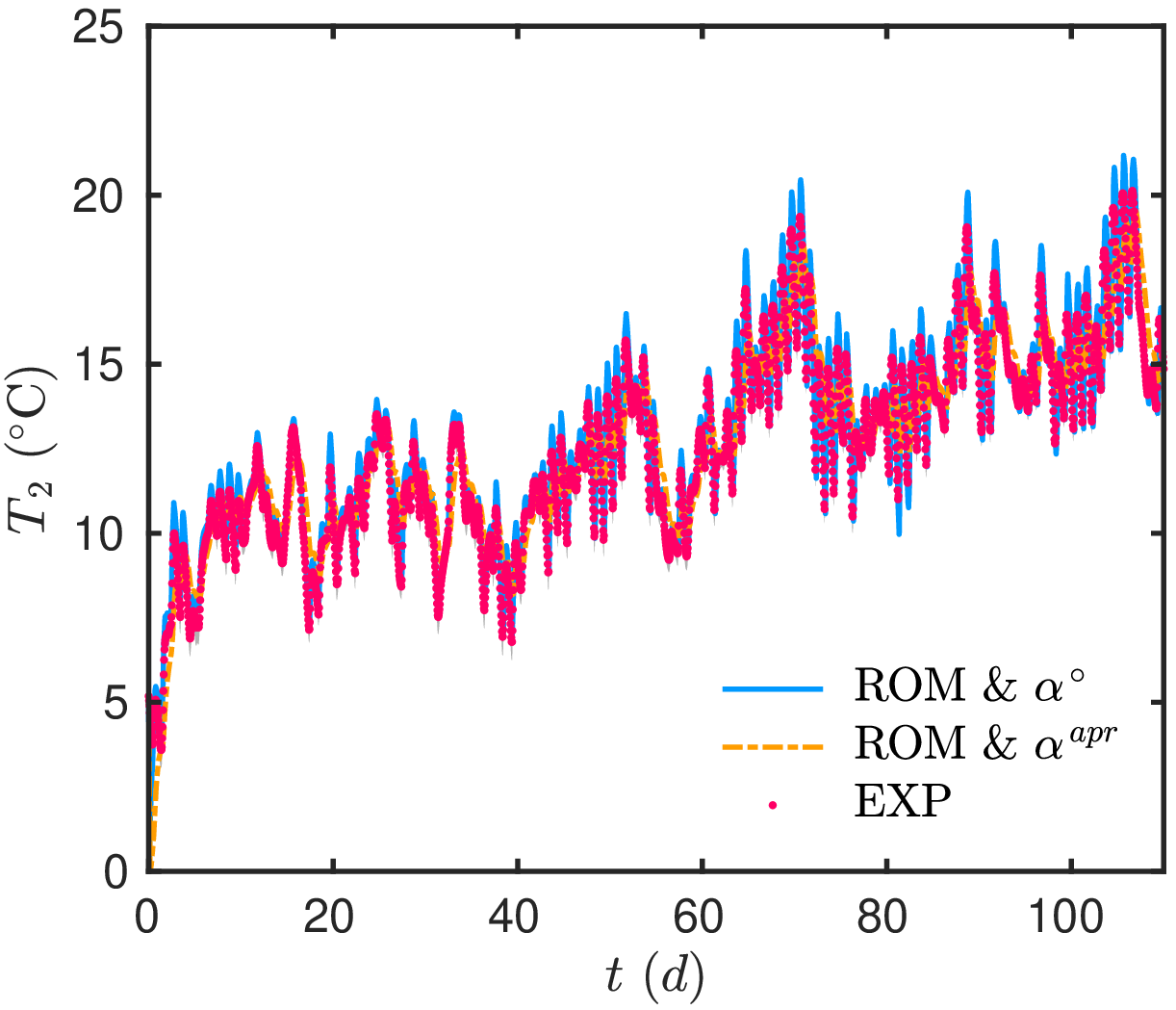}} \hspace{0.2cm}
\subfigure[$x_{\,2}$ \label{fig:pdfL2}]{\includegraphics[width=.45\textwidth]{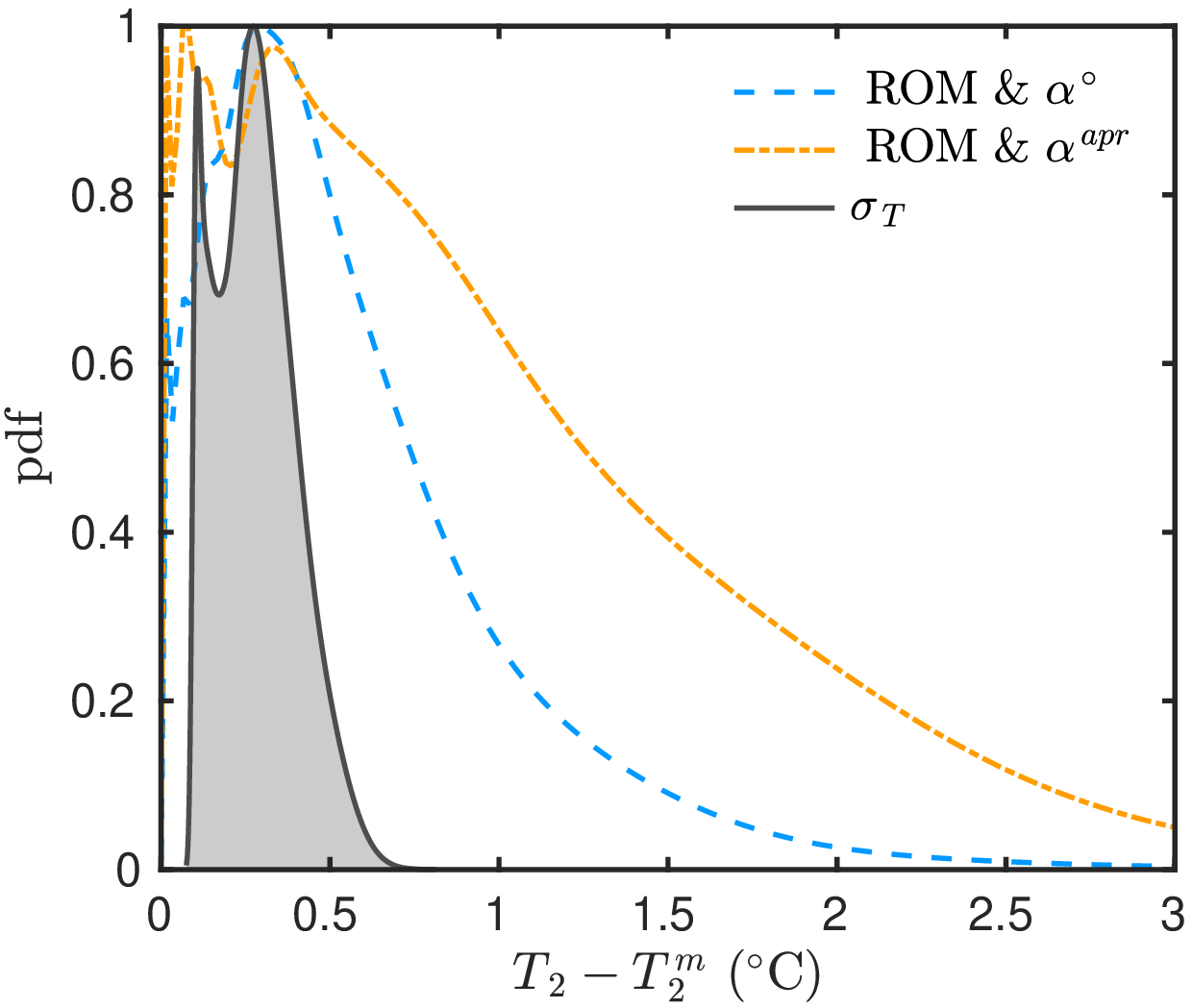}} \\
\subfigure[$x_{\,3}$ \label{fig:YL3_ft}]{\includegraphics[width=.45\textwidth]{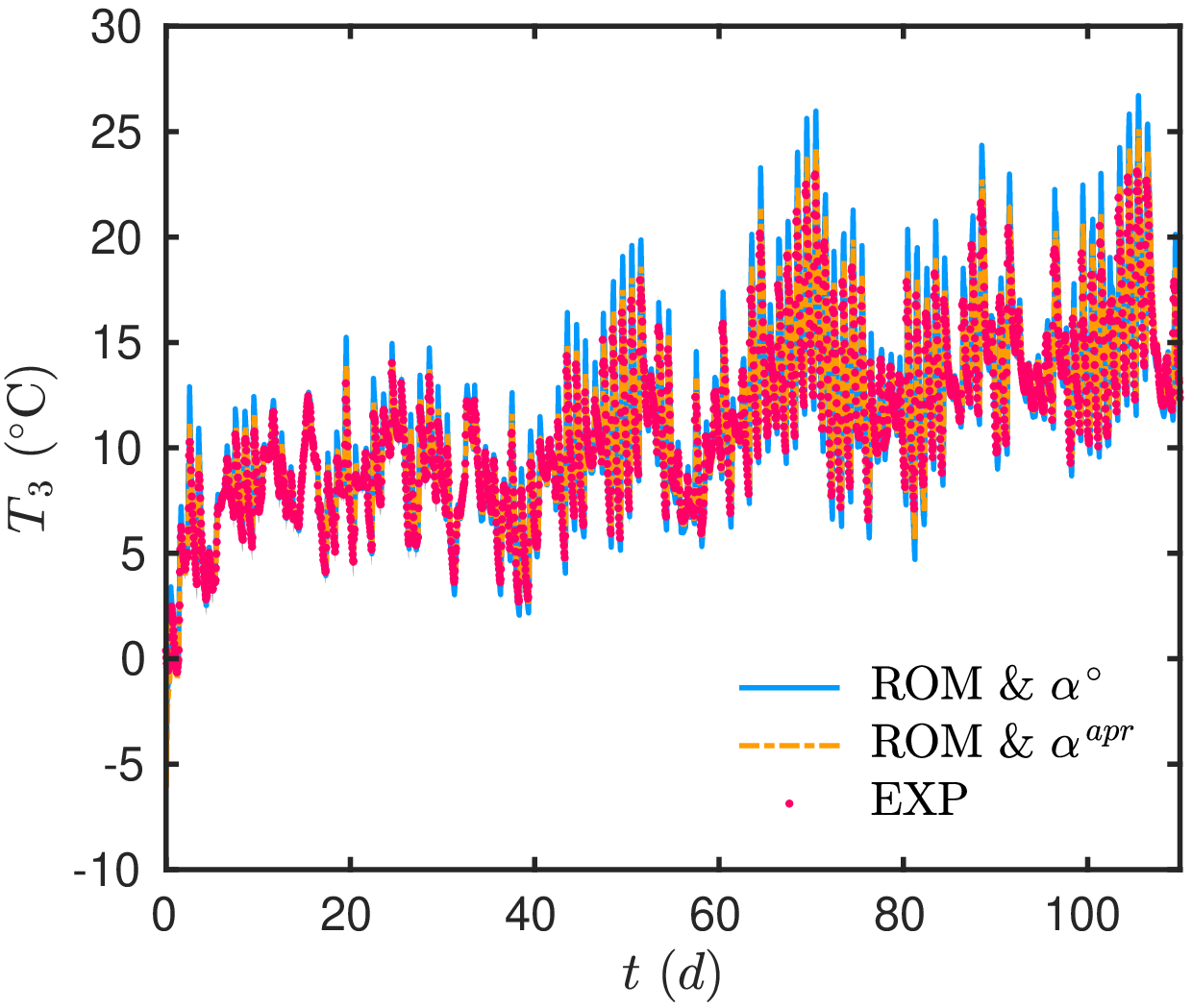}} \hspace{0.2cm}
\subfigure[$x_{\,3}$ \label{fig:pdfL3}]{\includegraphics[width=.45\textwidth]{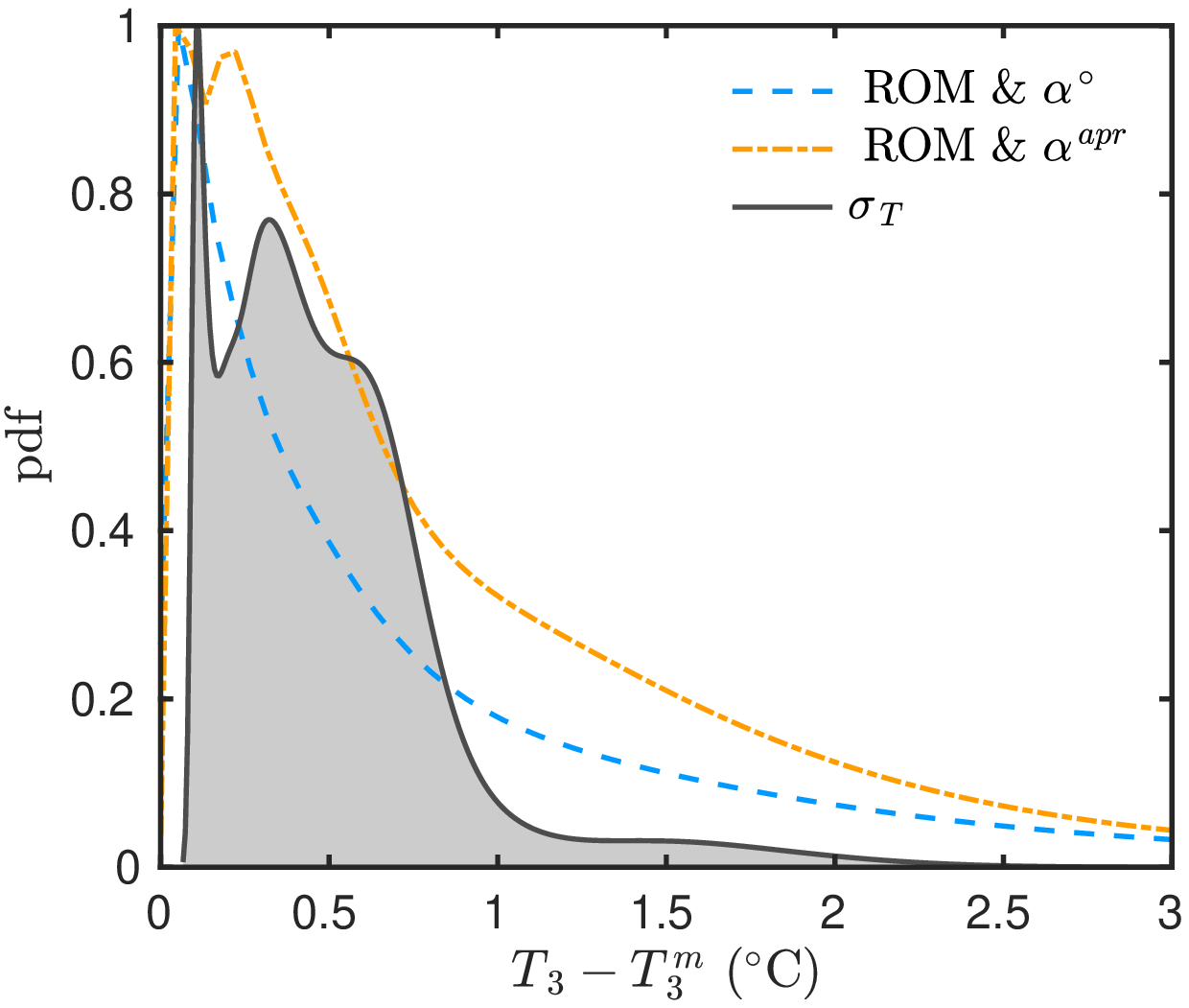}} 
\caption{\revision{Comparison of the numerical predictions for the last week \emph{(a,c,e)} and probability of the error between the predictions and the experimental observations for the whole period $\Omega_{\,t}$ \emph{(b,d,f)}.}}
\end{figure}

\begin{figure}[h!]
\centering
\subfigure[\label{fig:jQ_ft}]{\includegraphics[width=.45\textwidth]{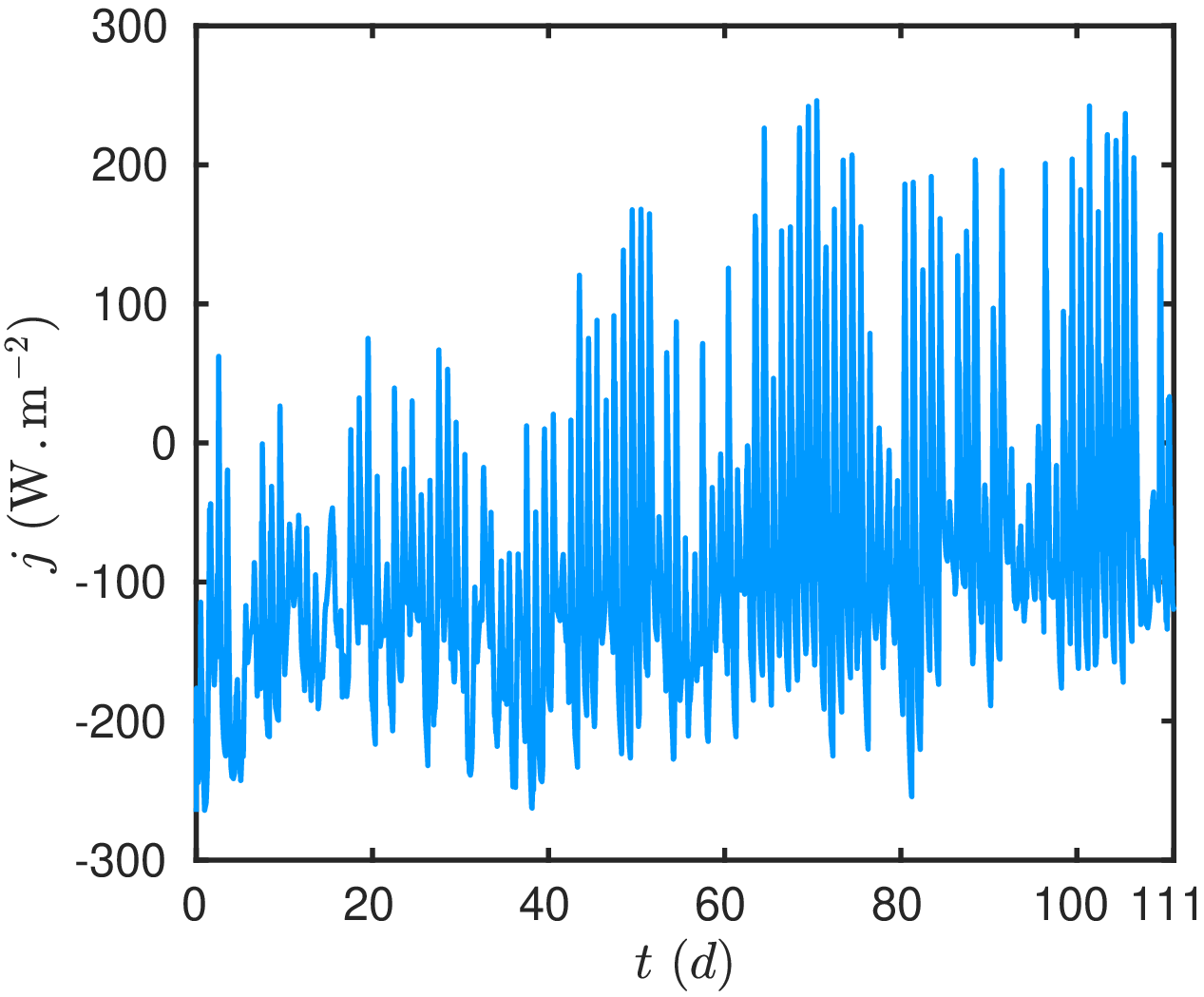}} \hspace{0.2cm}
\subfigure[\label{fig:E_fkappa}]{\includegraphics[width=.45\textwidth]{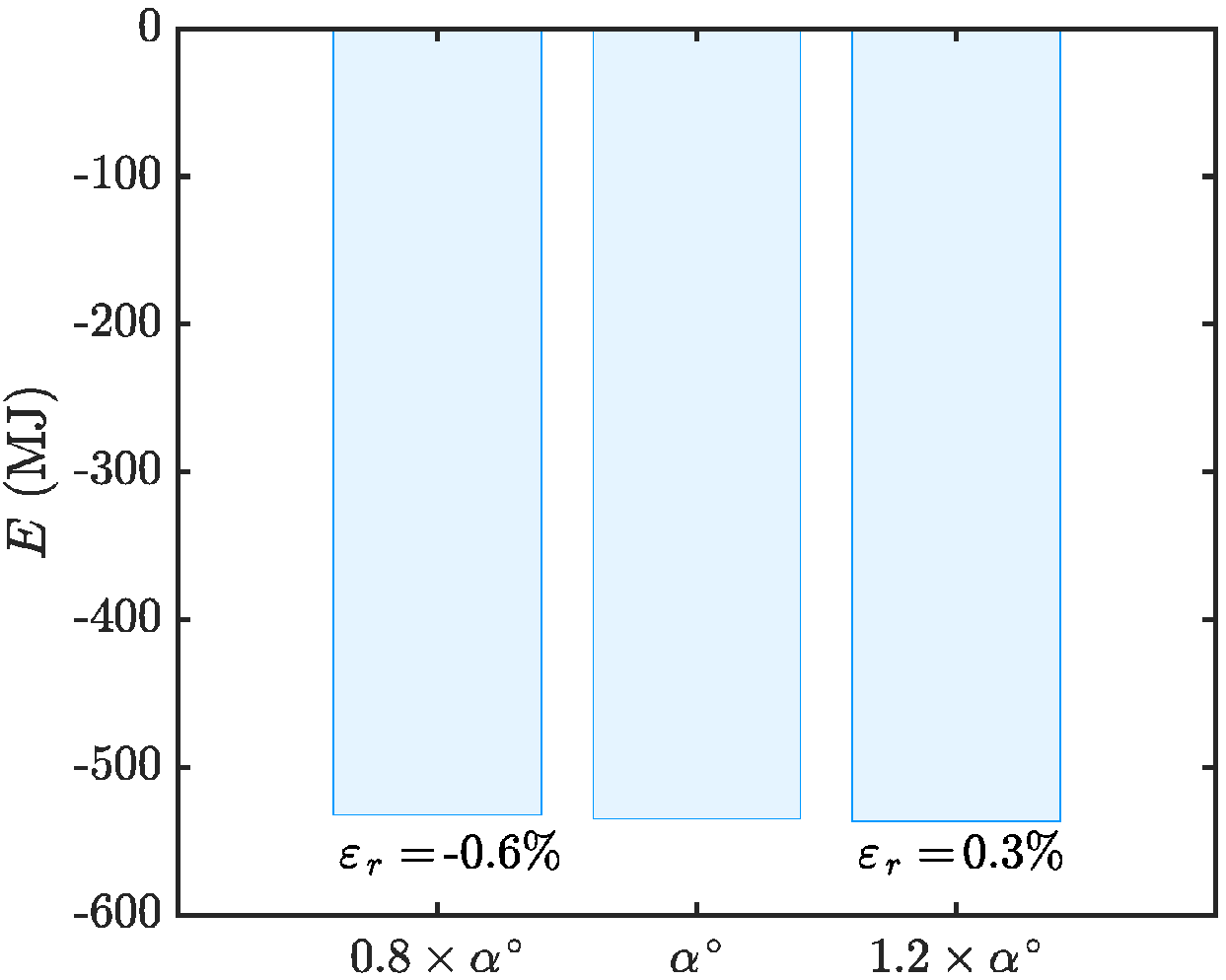}} 
\caption{Time evolution of the inside heat flux \emph{(a)} and variation of the thermal loads according to small perturbation of $\alpha^{\,\circ}$ \emph{(b)}.}
\end{figure}

To evaluate the reliability of the model with the estimated thermal diffusivity $\alpha^{\,\circ} \egal 1.06 \e{-6} \ \mathsf{m^{\,2}\,.\,s^{\,-1}}\,$, the numerical predictions computed with the MIM ROM $N \egal 10$ are compared to the experimental observations at each point $x_{\,s}\,$, for the whole sequence $\Omega_{\,t}\,$. Figures~\ref{fig:YL1_ft},\ref{fig:YL2_ft} and \ref{fig:YL3_ft} show this juxtaposition for the last week of the measurement. The model using \emph{a priori}  parameter lacks of accuracy to represent the physical phenomena. The model with the estimated parameter has a better reliability. However, for the point $x_{\,3}\,$, both models faces some discrepancy with the experimental observations. It can be noted a certain delay between the prediction and the observation at the point $x_{\,2}\,$. Figures~\ref{fig:pdfL1},\ref{fig:pdfL2} and \ref{fig:pdfL3} give the probability of the error with observation over the whole sequence of measurement. The error is compared with the probability of the measurement. The prediction of the model are more precised using the estimated parameter. Indeed the probability has a smaller standard deviation. However, at the point $x_{\,3}\,$, errors of larger magnitude occur for probabilities of both models. It reveals that some phenomena are not considered in the model and the mathematical formulation should be improved by considering for instance heterogeneous diffusivity in the wall. To verify the reliability of the proposed model, the inside heat flux $j$ and the thermal loads $E$ are computed according to: 
\begin{align*}
j \, \,\eqdef \, - k \ \pd{T}{x}\,\biggr|_{\,x \egal L} \,, \qquad
E \, \,\eqdef \,  \int_{\,\Omega_{\,t}} \, j_{\,q} \ \mathrm{d}\tau  \,.
\end{align*}
Figure~\ref{fig:jQ_ft} shows the variation of the thermal flux. It has important magnitude inducing high thermal loads as noted in Figure~\ref{fig:E_fkappa}. The thermal loads are computed for perturbations of $20 \% $ of the estimated diffusivity. It can be noted that the relative error on the loads remains stable around $5\%\,$. It confirms that the approximations introduced from the whole methodology (uncertainty measurement, reduced order model or estimation on a reduced sequence) are minor compared to the ones in the mathematical model itself.

\section{Conclusion}

The uncertain thermal properties of existing building walls can be determined by solving parameter estimation problems. However, such problems have important computational cost for at least two reasons. The inverse problem algorithm requires to solve the direct problem several times. In addition, the experimental measurement are carried on-site implying simulations over several month under climatic variations. These complexities increase the computational requirements.

To answer this issue, an innovative methodology is proposed based on two concepts. First, a reduced order model based on MIM method is used to reduce significantly the computational time of the direct problem without loosing accuracy. The methodology is presented in Section~\ref{sec:numerical_model}. The reduced model is based on a state space representation where the matrices are built during a learning step. The latter is based on a minimization procedure between the predictions of the reduced and complete models. An interesting point is that the model is built to compute the field of interest and its sensitivity to the unknown parameter. The sensitivity is known straightforwardly using the matrices of the reduced model. 

The second concept is the optimal experiment design methodology, described in Section~\ref{sec:pep}. It is employed to determine a reduced sequence of observations. Three advantages are enhanced with this approach. First, it reduces the inherent computational cost of \emph{a posteriori} model reduction methods since the learning step is carried for a reduced sequence. Then, the chosen sequence is optimal to estimate the parameter with accuracy. In addition, the inverse problem is solved only for a few days of observations. It reduced again the computational effort to retrieve the parameter. The solution of the inverse problem is obtained with the \textsc{Gau}\ss ~algorithm since the reduced order model computes the field and its sensitivity.

A first case study is proposed in Section~\ref{sec:validation_MIM} to validate the MIM model reduction method. This step is important to evaluate the efficiency in terms of accuracy and computational cuts. The MIM model is built with a signal based on $4$ values of thermal diffusivity. Then, the MIM model shows a very satisfying accuracy to simulate the direct problem over a wide range of diffusivity. Regardless the inherent cost of the learning step, the model cut by $5$ the computational cost of the direct problem. The model is also evaluated in the framework of inverse problem with simulated experimental observations. The unknown parameter is estimated with an error lower than the measurement uncertainty and a reduced computational time. 

After this simple validation case, the whole methodology is applied to a more realistic one. The issue is to estimate the thermal diffusivity of an old building wall. The latter is monitored during four months with three sensors drilled inside. With the OED methodology, a sequence of three days is identified as optimal. Then, the MIM ROM is built for the reduced sequence and a signal composed of $4$ values of thermal diffusivity. Then, a model of order $10$ is chosen to solve the inverse problem. The estimated diffusivity is three times higher than the one provided by standards. In terms of computational efficiency, the parameter is estimated with an algorithm $5$ times faster. Eventually, the reliability of the model is evaluated by comparing the predictions of the MIM model with the experimental observations of the complete period. It highlights that the global methodology is efficient to calibrate the model with a reduced computational effort. 

Future works should focus on improving the definition of the mathematical model by dealing with anisotropic thermal diffusivity in the building wall. In addition, to evaluate the real efficiency of the methodology, the \textsc{Gau}\ss ~algorithm combined with the MIM ROM should be installed in embedded system for fast estimation of building material properties.

\section*{Acknowledgments}

BK thanks Johnathan Gerardin for the fruitful discussions on MIM.

\clearpage

\section*{Nomenclature and symbols}

\begin{tabular}{|cll|}
\hline
\multicolumn{3}{|c|}{\emph{Physical parameters}} \\ \hline
\multicolumn{3}{|c|}{Latin letters} \\ 
\revision{$c$} & \revision{volumetric heat capacity} & \revision{$\unit{J\,.\,m^{\,-3}\,.\,K^{\,-1}}$} \\
$E$ & thermal loads & $\unit{J}$ \\
$j$ & heat flux & $\unit{W\,.\,m^{\,-2}}$ \\
\revision{$k$} & \revision{thermal conductivity} & \revision{$\unit{W\,.\,m^{\,-1} \,.\, K^{\,-1}}$} \\
$L$ & length & $\unit{m}$ \\
$t$ & time & $\unit{s}$ \\
$x$ & position & $\unit{m}$ \\
$T$ & temperature & $\unit{K}$ \\
\hline
\multicolumn{3}{|c|}{Greek letters} \\
$\alpha$ & thermal diffusivity & $\unit{m^{\,2}\,.\,s^{\,-1}}$ \\
$\delta$ & sensor position uncertainty  & $\unit{m}$ \\
$\sigma$ & measurement uncertainty  & $\unit{K}$ \\
$\Omega_{\,\alpha}$ & thermal diffusivity interval & $\unit{m^{\,2}\,.\,s^{\,-1}}$ \\
$\Omega_{\,t}$ & time interval & $\unit{s}$ \\
$\Omega_{\,t}^{\,\oed}$ & optimal time interval & $\unit{s}$ \\
$\varepsilon \,,\,\varepsilon_{\,2} $ & error & $\unit{K}$ \\
$\varepsilon_{\,r}$ & relative error & $\unit{-}$\\
\hline
\end{tabular}

\hspace{2cm}

\begin{tabular}{|cll|}
\hline
\multicolumn{3}{|c|}{\emph{Mathematical notations}} \\
\hline
\multicolumn{3}{|c|}{Latin letters} \\
$\mA \,,\, \mB \,,\, \mC \,,\, \mF \,,\, \mG \,,\, $ & \multirow{2}{*}{matrices} & \\
$\mH \,,\, \mQ \,,\, \mU \,,\, \mX \,,\, \mY$ &  & \\
$\mathcal{F}$ & \textsc{Fisher} matrix & \\
$\Fo$ & \textsc{Fourier} number & \\
$J$ & cost function & \\
$p$ & unknown parameter & \\
$u$ & dimensionless temperature & \\
$N_{\,r}$ & order of the reduced model & \\
$N_{\,s}$ & number of sensors & \\
\hline
\multicolumn{3}{|c|}{Greek letters} \\
$\Delta x$ & space mesh & \\
$\Delta t$ & time step & \\
$\eta_{\,1}\,,\,\eta_{\,2}$ & tolerance value & \\
$\gamma_{\,1}\,,\,\gamma_{\,2}$ & convergence criteria & \\
$\Pi$ & measurement plan &  \\
$\Psi$ & D-optimum criteria &  \\
$\theta$ & dimensionless sensitivity & \\
$\varepsilon \,,\,\varepsilon_{\,2}\,,\,\varepsilon_{\,r} $ & error &  \\
\hline
\multicolumn{3}{|c|}{Subscripts and superscripts} \\
$\apr$ & \emph{a priori} parameter &  \\
$\mathrm{cpu}$ & computational cost & \\
$\fin$ & final &  \\
$\ini$ & initial &  \\
$m$ & measurement &  \\
$\min$ & minimal value &  \\
$\mathrm{ref}$ & reference value &  \\
$rom$ & reduced order model &  \\
$r$ & real value &  \\
$s$ & sensor &  \\
$x$ & position &  \\
$\circ$ & estimated parameter &  \\
$\star$ & dimensionless value &  \\
$\infty$ & boundary &  \\
\hline
\end{tabular}

%
%

\bibliographystyle{unsrt}  
\bibliography{references}

\end{document}